\documentclass[english,3p]{elsarticle}
\usepackage[T1]{fontenc}
\usepackage[utf8]{inputenc}
\usepackage{float}
\usepackage{booktabs}
\usepackage{units}
\usepackage{textcomp}
\usepackage{amsmath}
\usepackage{amsthm}
\usepackage{amssymb}
\usepackage{stmaryrd}
\usepackage{graphicx}

\makeatletter

\providecommand{\tabularnewline}{\\}
\floatstyle{ruled}
\newfloat{algorithm}{tbp}{loa}
\providecommand{\algorithmname}{Algorithm}
\floatname{algorithm}{\protect\algorithmname}

\numberwithin{equation}{section}
\numberwithin{figure}{section}

\usepackage{algpseudocode,algorithm,algorithmicx}
\usepackage{tikz}
\usetikzlibrary{shapes,arrows,calc,decorations.pathreplacing}

\@ifundefined{showcaptionsetup}{}{%
 \PassOptionsToPackage{caption=false}{subfig}}
\usepackage{subfig}
\makeatother

\usepackage{babel}
\begin{document}

\title{The Moving Discontinuous Galerkin Method with Interface Condition
Enforcement for the Simulation of Hypersonic, Viscous Flows}

\author{Eric J. Ching, Andrew D. Kercher, Andrew Corrigan}
\address{Laboratories for Computational Physics and Fluid Dynamics,  U.S. Naval Research Laboratory, 4555 Overlook Ave SW, Washington, DC 20375}

\begin{abstract}
The moving discontinuous Galerkin method with interface condition
enforcement (MDG-ICE) is a high-order, $r$-adaptive method that treats
the grid as a variable and weakly enforces the conservation law, constitutive
law, and corresponding interface conditions in order to implicitly
fit high-gradient flow features. In this paper, we develop an optimization
solver based on the Levenberg-Marquardt algorithm that features an
anisotropic, locally adaptive penalty method to enhance robustness
and prevent cell degeneration in the computation of hypersonic, viscous
flows. Specifically, we incorporate an anisotropic grid regularization
based on the mesh-implied metric that inhibits grid motion in directions
with small element length scales, an element shape regularization
that inhibits nonlinear deformations of the high-order elements, and
a penalty regularization that penalizes degenerate elements. Additionally,
we introduce a procedure for locally scaling the regularization operators
in an adaptive, elementwise manner in order to maintain grid validity.
We apply the proposed MDG-ICE formulation to two- and three-dimensional
test cases involving viscous shocks and/or boundary layers, including
Mach 17.6 hypersonic viscous flow over a circular cylinder and Mach
5 hypersonic viscous flow over a sphere, which are very challenging
test cases for conventional numerical schemes on simplicial grids.
Even without artificial dissipation, the computed solutions are free
from spurious oscillations and yield highly symmetric surface heat-flux
profiles.
\end{abstract}
\begin{keyword}
Discontinuous Galerkin method; Interface condition enforcement; MDG-ICE;
Implicit shock fitting; Anisotropic curvilinear $r$-adaptivity; Mesh
adaptation
\end{keyword}
\maketitle
\global\long\def\middlebar{\,\middle|\,}%
\global\long\def\cof{\operatorname{cof}}%
\global\long\def\det{\operatorname{det}}%
\global\long\def\adj{\operatorname{adj}}%
\global\long\def\average#1{\left\{  \!\!\left\{  #1\right\}  \!\!\right\}  }%
\global\long\def\expnumber#1#2{{#1}\mathrm{e}{#2}}%
 \newcommand*{\horzbar}{\rule[.5ex]{2.5ex}{0.5pt}}\renewcommand{\algorithmicrequire}{\textbf{Input:}} \renewcommand{\algorithmicensure}{\textbf{Output:}}

\global\long\def\revisionmath#1{\textcolor{red}{#1}}%

\makeatletter \def\ps@pprintTitle{  \let\@oddhead\@empty  \let\@evenhead\@empty  \def\@oddfoot{\centerline{\thepage}}  \let\@evenfoot\@oddfoot} \makeatother

\let\svthefootnote\thefootnote\let\thefootnote\relax\footnotetext{\\ \hspace*{65pt}DISTRIBUTION STATEMENT A. Approved for public release. Distribution is unlimited.}\addtocounter{footnote}{-1}\let\thefootnote\svthefootnote

\section{Introduction\label{sec:intro}}

The moving discontinuous Galerkin finite element method with interface
condition enforcement (MDG-ICE) is an implicit shock fitting method
capable of handling complex shock dynamics~\citep{Cor18,Ker20,Ker20_LS}.
The method is a unique variation of the well-known discontinuous Galerkin
(DG) method~\citep{Ree73,Coc00}. Specifically, neighboring elements
are not coupled through interfacial, single-valued numerical fluxes;
instead, the conservation law and interface conditions (known as the
generalized Rankine-Hugoniot jump conditions~\citep{Maj12}) are
directly discretized, and the grid is treated as a variable. By simultaneously
solving for the flow field and discrete geometry, MDG-ICE is able
to compute highly accurate high-order solutions \emph{without artificial
dissipation} as the grid points are automatically adjusted to fit
shocks and resolve smooth regions of the flow with sharp gradients.
This has significant advantages over traditional shock capturing approaches,
such as artificial viscosity and limiting; the former introduces low-order
errors into high-order approximations that can excessively smear discontinuous
and high-gradient features (and even cause movement of such features
that may then contaminate prediction of other flow structures~\citep{Chi18}),
while the latter can obstruct iterative convergence, fail to provide
sufficient stabilization, and be used only for certain approximation
orders and/or element types. Furthermore, since MDG-ICE adapts the
grid to satisfy the weak form, grid interfaces are naturally repositioned
to fit a priori unknown shocks with arbitrary topology, overcoming
key limitations of explicit shock fitting methods~(\citep{Mor02,Sal09,Sal11}).
Note that in the inviscid case, shocks are fit exactly along grid
interfaces, while in the viscous setting, high-aspect-ratio cells
form to resolve viscous shocks, which are sharp (yet smooth) features,
via anisotropic curvilinear $r$-adaptivity. In previous work, MDG-ICE
was shown to achieve not only extremely sharp, oscillation-free viscous-shock
profiles, but also significantly higher accuracy than standard DG
schemes in boundary-layer problems~\citep{Ker20}. Another form of
implicit shock fitting is the high-order implicit shock tracking (HOIST)
framework developed by Zahr and Persson~\citep{Zah18} and improved
in~\citep{Zah20,Shi22,Hua22,Hua23}, which also treats the discrete
geometry as a variable while retaining a standard discontinuous Galerkin
method. High-quality solutions to inviscid flows with discontinuities
on coarse grids have been achieved using HOIST. Additonally, a variant
of MDG-ICE has been developed by Luo et al.~\citep{Luo21}. 

\begin{figure}[ht]
\centering{}\includegraphics[clip,width=0.8\textwidth]{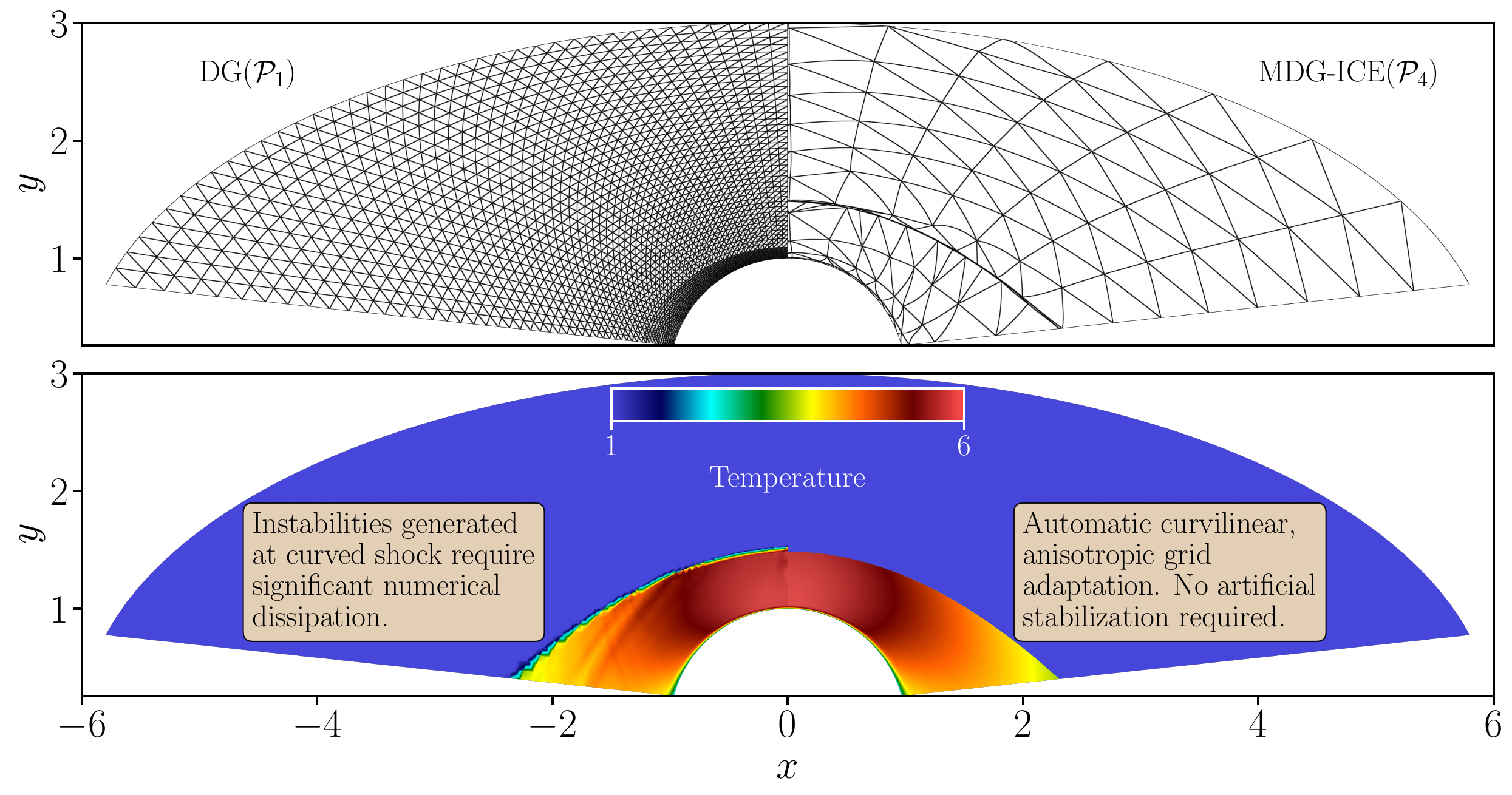}
\caption{Comparison of DG($\mathcal{P}_{1}$) (with artificial viscosity) (left)
and MDG-ICE($\mathcal{P}_{4}$) (right) solutions to two-dimensional
Mach 5 flow over a cylinder at $\mathrm{Re}=10^{4}$~\citep{Ker20}.
The corresponding meshes are also displayed.}
\label{fig:viscous_bow_shock_Temperature_p4_Mach_05_Re_00010000}
\end{figure}

Figure~\ref{fig:viscous_bow_shock_Temperature_p4_Mach_05_Re_00010000}
presents the temperature fields and corresponding meshes for hypersonic
viscous flow over a half-cylinder at Mach 5 and Reynolds number $10^{4}$
on two-dimensional triangular grids. A DG($\mathcal{P}_{1}$) solution
(left), where $\mathcal{P}_{p}$ denotes the space of polynomials
of total degree $p$ on simplicial elements, with artificial viscosity
is compared with an isoparametric MDG-ICE($\mathcal{P}_{4}$) solution
(right) without any additional stabilization~\citep{Ker20}. The
corresponding meshes are also displayed. Unlike in the MDG-ICE solution,
the shock in the DG solution is noticeably smeared, and spurious oscillations
are visible both at the shock and in the shock layer. The resulting
surface heating profile in the MDG-ICE solution is highly symmetric,
which, despite the relatively simple flow conditions, is nevertheless
encouraging since surface heat-flux predictions in state-of-the-practice
finite-volume simulations of external hypersonic flows are known to
deterioriate considerably when using simplicial grids~\citep{Nom04,Gno04}.
In particular, finite-volume schemes are extremely sensitive to misalignment
between the grid and the shock and boundary layer, which generates
errors in vorticity and entropy that propagate downstream and negatively
impact the accuracy of computed surface quantities. Shock-tailored
quadrilateral/hexahedral grids are typically needed to obtain heat-flux
profiles that do not exhibit noticeable nonphysical asymmetries~\citep{Can09}.
Constructing such grids for large-scale geometries often requires
significant time and user effort~\citep{Nom04,Can09}, which is exacerbated
when performing parametric studies.

\begin{figure}[ht]
\centering{}\includegraphics[clip,width=0.8\textwidth]{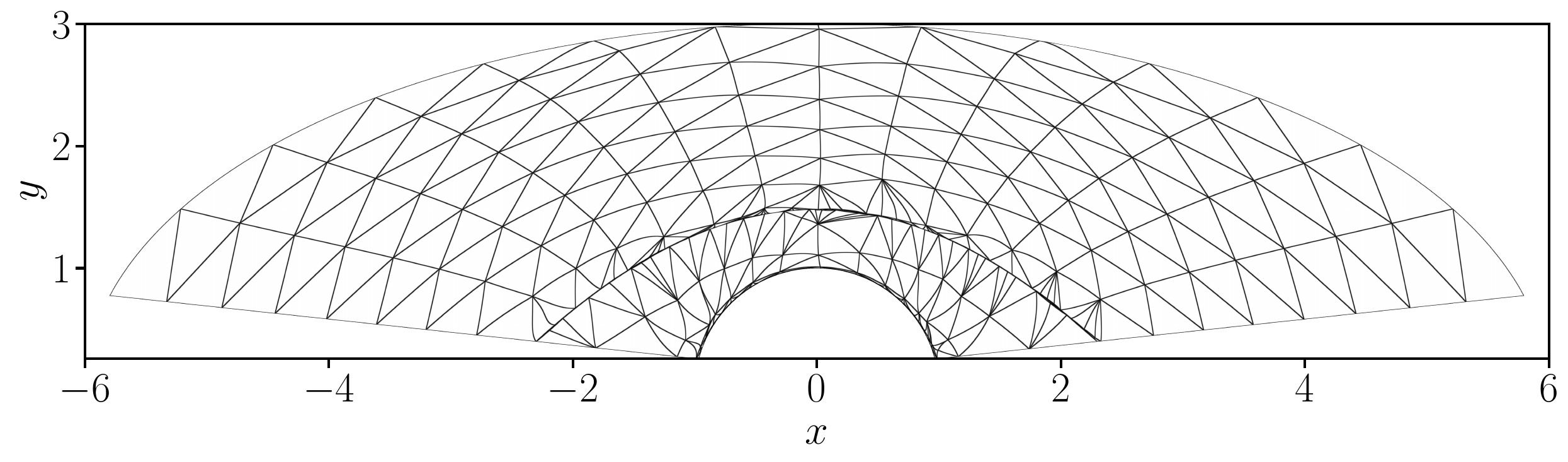}
\caption{Final mesh for the isoparametric MDG-ICE($\mathcal{P}_{4}$) solution
to two-dimensional Mach 5 flow over a cylinder at $\mathrm{Re}=10^{5}$~\citep{Ker20}.}
\label{fig:viscous_bow_shock_Mesh_p4_Mach_05_Re_00100000_00162550}
\end{figure}

In contrast, implicit shock-fitting methods, such as MDG-ICE, automatically
produce shock-aligned grids as part of the solution process, thereby
mitigating the need for shock-tailored grids composed of cubiod elements,
which significantly reduces the burden associated with generating
a suitable grid for high-speed flow calculations. However, a major
difficulty encountered in MDG-ICE calculations is frequent cell degeneration
(i.e., the determinant of the geometric Jacobian becomes negative),
particularly as curved, high-aspect-ratio cells form to resolve sharp
viscous features. Figure~\ref{fig:viscous_bow_shock_Mesh_p4_Mach_05_Re_00100000_00162550}
displays the final mesh for an MDG-ICE solution to the same cylinder
problem, but with a higher Reynolds number~\citep{Ker20}. Degenerate
cells were treated via longest-edge refinement, leading to the generation
of ``sliver'' elements that introduce unnecessary degrees of freedom.
Furthermore, although adequate for this relatively simple flow, in
problems of moderately greater complexity, this strategy either requires
an inordinate number of refinements or, worse, simply fails to recover
a valid grid. Given this significant bottleneck precluding the application
of MDG-ICE to more complicated, larger-scale configurations, the primary
objective of this work is to develop an optimization solver that improves
the robustness of the Levenberg-Marquardt algorithm employed in~\citep{Ker20}
and is capable of producing high-order approximations of hypersonic
viscous flows in two and three dimensions without introducing low-order
errors associated with artificial stabilization. The key feature of
the solver is an anisotropic, locally adaptive penalty technique.
In particular, we leverage the mesh-implied metric commonly employed
in output-based anisotropic mesh adaptation~\citep{Fid07,Oli08},
which encodes information about the local element size and orientation
(even on curved, anisotropic grids), to create an anisotropic grid
regularization that inhibits grid motion in directions with small
element length scales. This is similar to the inverse-volume scaling
introduced by Zahr et al.~\citep{Zah20} and employed in~\citep{Ker20_LS},
but with element anisotropy explicitly taken into account. Furthermore,
we incorporate additional regularization operators and develop an
adaptive elementwise regularization strategy that locally adjusts
regularization parameters as needed to maintain grid validity. We
also employ increment limiting as described in~\citep{Cez15}, where
the increment at each iteration is dynamically reduced to mitigate
excessive changes in pressure and/or density. The improved solver,
paired with a full parallelization of our MDG-ICE implementation,
enables consideration of significantly more challenging hypersonic
test cases. We specifically employ simplicial grids that are initially
coarse with respect to the final high-gradient features in order to
demonstrate the potential of MDG-ICE to greatly alleviate the burden
of mesh generation on the user. Finally, we note that although incorporating
artificial dissipation into the formulation in some capacity (e.g.,
applied only during intermediate iterations or minimally in the final
solution) may be beneficial, the present study aims to aggressively
test the underlying MDG-ICE formulation without the aid of additional
stabilization and, as a secondary goal, identify in what ways stabilization
would be most useful; incorporation of such dissipation mechanisms
is left for future work.

The remainder of this paper is organized as follows. Sections~\ref{sec:governing-equations}
and~\ref{sec:mdg-ice-review} briefly summarize the governing equations
considered in this study and the basic MDG-ICE formulation, respectively.
The improved optimization solver, which is the primary contribution
of this work, is then detailed in Section~\ref{sec:nonlinear-solver}.
Section~\ref{sec:results}presents solutions for a variety of challenging
test cases, including two- and three-dimensional hypersonic viscous
flows over blunt bodies in which surface heat flux is a target quantity.
We conclude with final remarks and recommendations for future work.

\section{Governing equations\label{sec:governing-equations}}

Let $\Omega\subset\mathbb{R}^{d}$ denote the domain, which can be
either a space-time domain, $\Omega\subset\mathbb{R}^{d=d_{x}+1}$,
or a spatial domain, $\Omega\subset\mathbb{R}^{d=d_{x}}$. Consider
the following nonlinear conservation law governing the vector of $m$
state variables, $y:\Omega\rightarrow\mathbb{R}^{m}$: 
\begin{equation}
%\frac{\partialy}{\partialt}+\frac{\partial}{\partialx}\left(\frac{y^{2}}{2}\right)=\frac{\partial}{\partialx}\left(\mu\frac{\partialy}{\partialx}\right),
\nabla\cdot\mathcal{F}(y,\nabla_{x}y)=0\text{ in }\Omega,\label{eq:nonlinear-conservation-law}
\end{equation}
where $\mathcal{F}:\mathbb{R}^{m}\times\mathbb{R}^{m\times d_{x}}\rightarrow\mathbb{R}^{m\times d}$
is the flux and $\nabla_{x}(\cdot)$ denotes the spatial gradient,
\begin{equation}
\nabla_{x}y=\left(\frac{\partial y}{\partial x_{1}},\ldots,\frac{\partial y}{\partial x_{d_{x}}}\right).\label{eq:spatial-gradient}
\end{equation}
In the case of a space-time domain (i.e., $d=d_{x}+1$), the space-time
flux is given by 
\begin{equation}
\mathcal{F}(y,\nabla_{x}y)=\left(\mathcal{F}_{1}^{x}(y,\nabla_{x}y),\ldots,\mathcal{F}_{d_{x}}^{x}(y,\nabla_{x}y),y\right),
\end{equation}
where $\mathcal{F}^{x}:\mathbb{R}^{m}\times\mathbb{R}^{m\times d_{x}}\rightarrow\mathbb{R}^{m\times d_{x}}$
is the spatial flux, defined as the difference between the convective
flux and the viscous flux,
\[
\mathcal{F}^{x}(y,\nabla_{x}y)=\mathcal{F}^{c}(y)-\mathcal{F}^{v}(y,\nabla_{x}y).
\]
The divergence operator in Equation~(\ref{eq:nonlinear-conservation-law})
is then the space-time divergence operator, given by 
\begin{equation}
\nabla\cdot\mathcal{F}(y,\nabla_{x}y)=\nabla_{x}\cdot\mathcal{F}^{x}(y,\nabla_{x}y)+\frac{\partial}{\partial t}y.
\end{equation}
In the case of a spatial domain (i.e., $d=d_{x}$), the flux is simply
the spatial flux, and the divergence operator is the spatial divergence
operator. In this study, we consider the viscous Burgers equation
and the compressible Navier-Stokes equations.

\subsection{One-dimensional viscous Burgers equation\label{subsec:Burgers}}

The one-dimensional viscous Burgers equation involves a single state
variable, $y:\Omega\rightarrow\mathbb{R}^{1}$. The convective and
viscous fluxes are given by 
\begin{equation}
\mathcal{F}^{c}\left(y\right)=\frac{1}{2}y^{2},\quad\mathcal{F}^{v}\left(y,\nabla_{x}y\right)=\mu\nabla_{x}y,
\end{equation}
where $\mu\in\mathbb{R}_{+}$ is the viscosity. The spatial flux is
then written as 
\begin{equation}
\mathcal{F}^{x}\left(y,\nabla_{x}y\right)=\left(\frac{1}{2}y^{2}-\mu\nabla_{x}y\right),\label{eq:spatial-burgers-flux}
\end{equation}

\subsection{Compressible Navier-Stokes equations\label{subsec:CNS}}

The vector of state variables, $y:\Omega\rightarrow\mathbb{R}^{m}$,
where $m=d_{x}+2$, is given by 
\begin{equation}
y=\left(\rho,\rho v_{1},\ldots,\rho v_{d_{x}},\rho E\right),\label{eq:navier-stokes-state}
\end{equation}
where $\rho:\Omega\rightarrow\mathbb{R}_{+}$ is the density, $v=\left(v_{1},\ldots,v_{d_{x}}\right)$
is the velocity vector, with $v:\mathbb{R}^{m}\rightarrow\mathbb{R}^{d_{x}}$,
and $E:\mathbb{R}^{m}\rightarrow\mathbb{R}_{+}$ is the specific total
energy. The $i$th spatial component of the convective flux is written
as
\begin{equation}
\mathcal{F}_{i}^{c}\left(y\right)=\left(\rho v_{i},\rho v_{i}v_{1}+P\delta_{i1},\ldots,\rho v_{i}v_{d_{x}}+P\delta_{id_{x}},\rho Hv_{i}\right),\label{eq:ns-convective-flux-spatial-component}
\end{equation}
where $P:\mathbb{R}^{m}\rightarrow\mathbb{R}_{+}$ is the pressure,
$\delta_{ij}$ is the Kronecker delta, and $H=\left(\rho E+P\right)/\rho$
is the specific total enthalpy, with $H:\mathbb{R}^{m}\rightarrow\mathbb{R}_{+}$.
In this work, we assume a calorically perfect gas, such that 
\begin{equation}
P=\left(\gamma-1\right)\left(\rho E-\frac{1}{2}\sum_{i=1}^{d_{x}}\rho v_{i}v_{i}\right),\label{eq:pressure}
\end{equation}
where $\gamma$ is the specific heat ratio, set to $1.4$. Note that
the assumption of a single-species calorically perfect gas is appropriate
for evaluating the capability of a numerical method to handle sharp,
high-gradient features such as strong shocks and boundary layers,
although more complex gas and transport models are typically employed
in realistic hypersonic applications~\citep{Gno99}.

The $i$th spatial component of the viscous flux is written as
\begin{equation}
\mathcal{F}_{i}^{\nu}\left(y,\nabla_{x}y\right)=\left(0,\tau_{1i},\ldots,\tau_{d_{x}i},\sum_{j=1}^{d_{x}}\tau_{ij}v_{j}-q_{i}\right),\label{eq:navier-stokes-viscous-flux-spatial-component}
\end{equation}
where $q:\mathbb{R}^{m}\times\mathbb{R}^{m\times d_{x}}\rightarrow\mathbb{R}^{d_{x}}$
is the heat flux and $\tau:\mathbb{R}^{m}\times\mathbb{R}^{m\times d_{x}}\rightarrow\mathbb{R}^{d_{x}\times d_{x}}$
is the viscous stress tensor. The heat flux is expanded as $q=-\kappa\nabla_{x}T$,
where $\kappa:\mathbb{R}^{m}\rightarrow\mathbb{R}_{+}$ is the thermal
conductivity and $T:\mathbb{R}^{m}\rightarrow\mathbb{R}_{+}$ is the
temperature, calculated as $T=P/(\rho R)$, with $R=287$ denoting
the specific gas constant. The $i$th spatial component of the viscous
stress tensor is given by

\noindent 
\begin{equation}
\tau_{i}=\mu\left(\frac{\partial v_{1}}{\partial x_{i}}+\frac{\partial v_{i}}{\partial x_{1}}-\delta_{i1}\frac{2}{3}\sum_{j=1}^{d_{x}}\frac{\partial v_{j}}{\partial x_{j}},\ldots,\frac{\partial v_{d_{x}}}{\partial x_{i}}+\frac{\partial v_{i}}{\partial x_{d_{x}}}-\delta_{id_{x}}\frac{2}{3}\sum_{j=1}^{d_{x}}\frac{\partial v_{j}}{\partial x_{j}}\right),\label{eq:navier-stokes-viscous-stress-tensor-component}
\end{equation}

\noindent where $\mu:\mathbb{R}^{m}\rightarrow\mathbb{R}_{+}$ is
the dynamic viscosity obtained from Sutherland's law,
\begin{equation}
\mu=\mu_{0}\frac{T_{0}+C}{T+C}\left(\frac{T}{T_{0}}\right)^{\frac{3}{2}}.\label{eq:sutherlands-law}
\end{equation}
$u_{0}\in\mathbb{R}_{+}$ is the reference dynamic viscosity, and
$T_{0}\in\mathbb{R}_{+}$ and $C\in\mathbb{R}_{+}$ are reference
temperatures. Nondimensionalizing Equation~(\ref{eq:sutherlands-law})
using freestream quantities, denoted $\left(\cdot\right)_{\infty}$,
yields 
\begin{equation}
\mu^{*}=\frac{1+C^{*}}{T^{*}+C^{*}}\left(T^{*}\right)^{\frac{3}{2}},\label{eq:sutherlands-law-nondimensional}
\end{equation}
where $\mu^{*}=\mu/\mu_{\infty}$, $T^{*}=T/T_{\infty}$, and $C^{*}=C/T_{\infty}$.
In this study, we use $C=110$ K and $T_{\infty}=293.15$ K. The
thermal conductivity is computed as
\begin{equation}
k=\frac{c_{p}\mu}{\mathrm{Pr}},\label{eq:thermal-conductivity}
\end{equation}
where $\mathrm{Pr}=0.72$ is the Prandtl number and $c_{p}=\gamma R/\left(\gamma-1\right)$
is the specific heat capacity at constant pressure.

Two important nondimensional quantities that characterize compressible,
viscous flows are the Reynolds number, $\mathrm{Re}$, and Mach number,
$\mathrm{Ma}$, defined as

\begin{equation}
\mathrm{Re}=\frac{\rho\lvert v\rvert L}{\mu},\label{eq:reynolds-number}
\end{equation}
where $L\in\mathbb{R}_{+}$ is a characteristic length scale, and
\begin{equation}
\mathrm{Ma}=\frac{\lvert v\rvert}{c},\label{eq:mach-number}
\end{equation}
where $c=\sqrt{\gamma P/\rho}$ is the speed of sound, with $c:\mathbb{R}^{m}\rightarrow\mathbb{R}_{+}$,
respectively.

\section{Moving discontinuous Galerkin method with interface condition enforcement\label{sec:mdg-ice-review}}

In this subsection, we briefly review the MDG-ICE formulation for
a system governed by a set of conservation laws and a generalized
constitutive law. Further details reagarding the formulation and its
application can be found in~\citep{Ker20}, and an alternative least-squares
formulation is described in~\citep{Ker20_LS}.

Let $\Omega$ be partitioned by $\mathcal{T}$, which consists of
cells $\kappa$. Furthermore, we define the set $\mathcal{E}$ of
interfaces such that $\bigcup_{\epsilon\in\mathcal{E}}=\bigcup_{\kappa\in\mathcal{T}}\partial\kappa$.
The normal over each interface $\epsilon\in\mathcal{E}$ is denoted
$n:\epsilon\rightarrow\mathbb{R}^{d}$. For space-time domains, $n_{x}:\epsilon\rightarrow\mathbb{R}^{d_{x}}$
denotes the spatial normal.

\subsection{Strong and weak formulations\label{subsec:Formulation}}

Consider the following conservation law, constitutive law, and associated
interface conditions in strong form: 
\begin{align}
\nabla\cdot\mathcal{F}\left(y,\sigma\right)=0 & \textup{ in }\kappa\qquad\forall\kappa\in\mathcal{T},\label{eq:conservation-strong-viscous}\\
\sigma-G\left(y\right)\nabla_{x}y=0 & \textup{ in }\kappa\qquad\forall\kappa\in\mathcal{T},\label{eq:constitutive-strong-viscous}\\
\left\llbracket n\cdot\mathcal{F}\left(y,\sigma\right)\right\rrbracket =0 & \textup{ on }\epsilon\qquad\forall\epsilon\in\mathcal{E},\label{eq:interface-condition-strong-viscous}\\
\average{G\left(y\right)}\left\llbracket y\otimes n_{x}\right\rrbracket =0 & \textup{ on }\epsilon\qquad\forall\epsilon\in\mathcal{E},\label{eq:interface-condition-state-strong-viscous}
\end{align}
where $\sigma:\Omega\rightarrow\mathbb{R}^{m\times d_{x}}$ is an
auxiliary variable, $G(y)\in\mathbb{R}^{m\times d_{x}\times m\times d_{x}}$
is the homogeneity tensor that satisfies $G(y)\nabla_{x}y=\mathcal{F}^{v}\left(y,\nabla_{x}y\right)=\mathcal{F}_{\nabla_{x}y}^{v}\left(y,\nabla_{x}y\right)\nabla_{x}y$
(assuming the viscous flux is linear with respect to the spatial gradient
of the state), and $\average{\cdot}$ and $\llbracket\cdot\rrbracket$
denote the average and jump operators, respectively. The interface
condition~(\ref{eq:interface-condition-strong-viscous}), which corresponds
to the conservation law~(\ref{eq:conservation-strong-viscous}),
is known as the jump or generalized Rankine-Hugoniot conditions~\citep{Maj12}.
The interface condition~(\ref{eq:interface-condition-state-strong-viscous}),
derived in~\citep{Ker20}, is associated with the constitutive law~(\ref{eq:constitutive-strong-viscous})
and constrains the continuity of the state variable at the interface.

Let the solution spaces $Y$ and $\Sigma$ be the broken Sobolev spaces
\begin{eqnarray}
Y & = & \left\{ y\in\left[L^{2}\left(\Omega\right)\right]^{m\hphantom{\times d_{x}}}\bigl|\forall\kappa\in\mathcal{T},\:\:\hphantom{\nabla_{x}\cdot}\left.y\right|_{\kappa}\in\left[H^{1}\left(\kappa\right)\right]^{m}\right\} ,\label{eq:solution-space-vector}\\
\Sigma & = & \left\{ \sigma\in\left[L^{2}\left(\Omega\right)\right]^{m\times d_{x}}\bigl|\forall\kappa\in\mathcal{T},\left.\nabla_{x}\cdot\sigma\right|_{\kappa}\in\left[L^{2}\left(\Omega\right)\right]^{m}\right\} .\label{eq:solution-space-tensor}
\end{eqnarray}
The MDG-ICE weak formulation is then obtained by integrating Equations~(\ref{eq:conservation-strong-viscous})-(\ref{eq:interface-condition-state-strong-viscous})
against separate test functions: find $\left(y,\sigma\right)\in Y\times\Sigma$
such that 
\begin{align}
0= & \;\;\;\;\sum_{\kappa\in\mathcal{T}}\left(\nabla\cdot\mathcal{F}\left(y,\sigma\right),v_{y}\right)_{\kappa}\nonumber \\
 & +\sum_{\kappa\in\mathcal{T}}\left(\sigma-G\left(y\right)\nabla_{x}y,v_{\sigma}\right)_{\kappa}\nonumber \\
 & -\sum_{\epsilon\in\mathcal{E}}\left(\left\llbracket n\cdot\mathcal{F}\left(y,\sigma\right)\right\rrbracket ,w_{y}\right)_{\epsilon}\nonumber \\
 & -\sum_{\epsilon\in\mathcal{E}}\left(\average{G\left(y\right)}\left\llbracket y\otimes n_{x}\right\rrbracket ,w_{\sigma}\right)_{\epsilon}\qquad\forall\left(v_{y},v_{\sigma},w_{y},w_{\sigma}\right)\in V_{y}\times V_{\sigma}\times W_{y}\times W_{\sigma}.\label{eq:weak-formulation-viscous}
\end{align}
where the test spaces are $V_{y}=\left[L^{2}\left(\Omega\right)\right]^{m}$
and $V_{\sigma}=\left[L^{2}\left(\Omega\right)\right]^{m\times d}$,
with $W_{y}$ and $W_{\sigma}$ defined as the corresponding single-valued
trace spaces. Note that numerical flux functions are not employed
in the current MDG-ICE formulation, although with slight modifications,
it is straightforward to include them~\citep{Cor18}.

To treat the grid as a variable, the weak formulation~(\ref{eq:weak-formulation-viscous})
is transformed from physical to reference space. Let $u:\hat{\Omega}\rightarrow\Omega$
be a continuous, invertible mapping from the reference domain, $\hat{\Omega}$,
to the physical domain, $\Omega$. $\hat{\Omega}$ is assumed to be
partitioned by $\hat{\mathcal{T}}$, such that $\overline{\hat{\Omega}}=\cup_{\hat{\kappa}\in\hat{\mathcal{T}}}\overline{\hat{\kappa}}$.
Furthermore, let $\hat{\mathcal{E}}$ denote the set of interfaces,
$\hat{\epsilon}$, such that $\cup_{\hat{\epsilon}\in\hat{\mathcal{E}}}\hat{\epsilon}=\cup_{\hat{\kappa}\in\hat{\mathcal{T}}}\partial\hat{\kappa}$,
and let $U=\left[H^{1}\left(\hat{\Omega}\right)\right]^{d}$, the
$\mathbb{R}^{d}$-valued Sobolev space over $\hat{\Omega}$. The solution
and test spaces are also now assumed to be defined over reference
space. We then define a provisional state operator, $\tilde{e}:Y\times\Sigma\times U\rightarrow\left(V_{y}\times V_{\sigma}\times W_{y}\times W_{\sigma}\right)^{*}$
for $\left(y,\sigma,u\right)\in Y\times\Sigma\times U$, as 
\begin{align}
\tilde{e}\left(y,\sigma,u\right)=\left(v_{y},v_{\sigma},w_{y},w_{\sigma}\right)\mapsto & \;\;\;\;\sum_{\hat{\kappa}\in\hat{\mathcal{T}}}\left(\left(\cof\left(\nabla u\right)\nabla\right)\cdot\mathcal{F}\left(y,\sigma\right),v_{y}\right)_{\hat{\kappa}}\nonumber \\
 & +\sum_{\hat{\kappa}\in\hat{\mathcal{T}}}\left(\det\left(\nabla u\right)\sigma-G\left(y\right)\left(\cof\left(\nabla u\right)\nabla\right)_{x}y,v_{\sigma}\right)_{\hat{\kappa}}\nonumber \\
 & -\sum_{\hat{\epsilon}\in\hat{\mathcal{E}}}\left(\left\llbracket s\left(\nabla u\right)\cdot\mathcal{F}\left(y,\sigma\right)\right\rrbracket ,w_{y}\right)_{\hat{\epsilon}}\nonumber \\
 & -\sum_{\hat{\epsilon}\in\hat{\mathcal{E}}}\left(\average{G\left(y\right)}\left\llbracket y\otimes s\left(\nabla u\right)_{x}\right\rrbracket ,w_{\sigma}\right)_{\hat{\epsilon}}.\label{eq:state-operator-weak-formulation-reference-viscous}
\end{align}

The state operator, $e:Y\times\Sigma\times U\rightarrow\left(V_{y}\times V_{\sigma}\times W_{y}\times W_{\sigma}\right)^{*}$,
is defined as 
\begin{equation}
e\left(y,\sigma,u\right)=\tilde{e}\left(y,\sigma,b\left(u\right)\right),\label{eq:state-operator-composition}
\end{equation}
which imposes geometric boundary conditions via the geometric projection
operator, $b(u)$.

The state equation in reference space is $e\left(y,\sigma,u\right)=0,$
such that the corresponding weak formulation in reference space is
as follows: find $\left(y,\sigma,u\right)\in Y\times\Sigma\times U$
such that 
\begin{equation}
\left\langle e\left(y,\sigma,u\right),\left(v_{y},v_{\sigma},w_{y},w_{\sigma}\right)\right\rangle =0\qquad\forall\left(v_{y},v_{\sigma},w_{y},w_{\sigma}\right)\in V_{y}\times V_{\sigma}\times W_{y}\times W_{\sigma}.\label{eq:weak-formulation-3-reference-coordinates}
\end{equation}
The solution is therefore given by $\left(y,\sigma,b\left(u\right)\right)\in Y\times\Sigma\times U$.

\subsection{Discretization\label{subsec:Discretization}}

To discretize the weak formulation (\ref{eq:weak-formulation-3-reference-coordinates}),
we choose discrete subspaces $Y_{h}\subset Y$, $\Sigma_{h}\subset\Sigma$,
$U_{h}\subset U$, $V_{y,h}\subset V_{y}$, $V_{\sigma,h}\subset V_{\sigma}$,
$W_{y,h}\subset W_{y}$, and $W_{\sigma,h}\subset W_{\sigma}$ and
define a discrete state operator 
\begin{equation}
e_{h}:Y_{h}\times\Sigma_{h}\times U_{h}\rightarrow\mathbb{R}^{\dim\left(V_{y,h}\times V_{\sigma,h}\times W_{y,h}\times W_{\sigma,h}\right)}.\label{eq:discrete-state-operator}
\end{equation}
For a simplicial grid, 
\begin{eqnarray}
Y_{h} & = & \left\{ y\in Y\middlebar\forall\hat{\kappa}\in\hat{\mathcal{T}},\left.y\right|_{\hat{\kappa}}\in\left[\mathcal{P}_{p}\right]^{m\hphantom{\times d_{x}}}\right\} ,\label{eq:discrete-solution-state-space-simplex}\\
\Sigma_{h} & = & \left\{ \sigma\in\Sigma\middlebar\forall\hat{\kappa}\in\hat{\mathcal{T}},\left.\sigma\right|_{\hat{\kappa}}\in\left[\mathcal{P}_{p}\right]^{m\times d_{x}}\right\} ,\label{eq:discrete-solution-flux-space-simplex}
\end{eqnarray}
where $\mathcal{P}_{p}$ is the space of polynomials spanned by the
monomials $\boldsymbol{x}^{\beta}$ with multi-index $\beta\in\mathbb{N}_{0}^{d}$
satisfying $\sum_{i=1}^{d}\beta_{i}\leq p$. We set $V_{y,h}=Y_{h}$
and $V_{\sigma,h}=\Sigma_{h}$. $W_{y,h}$ and $W_{\sigma,h}$ are
selected to be the corresponding single-valued polynomial trace spaces.
The discrete subspace $U_{h}$ of geometric mappings is defined as
\begin{equation}
U_{h}=\left\{ u\in U\middlebar\forall\hat{\kappa}\in\hat{\mathcal{T}},\left.u\right|_{\hat{\kappa}}\in\left[\mathcal{P}_{p}\right]^{d}\right\} .\label{eq:discrete-shape-space-simplex}
\end{equation}
In this work, the polynomial degrees of $Y_{h}$ and $\Sigma_{h}$
are chosen to be equal; the polynomial degree of $U_{h}$, however,
may be different. The cases in which the polynomial degree of $U_{h}$
is greater than, the same as, and less than the polynomial degrees
of $Y_{h}$ and $\Sigma_{h}$ are referred to as superparametric,
isoparametric, and subparametric, respectively.

\section{Nonlinear solver\label{sec:nonlinear-solver}}

The nonlinear solver developed for this work is based on the Levenberg-Marquardt
method that was previously employed~\citep{Ker20}, in which the
weak formulation is solved iteratively (in nondimensional form) using
unconstrained optimization to minimize the objective function 
\begin{equation}
J\left(y,\sigma,u\right)=\frac{1}{2}\left\Vert e_{h}\left(y,\sigma,u\right)\right\Vert ^{2}\label{eq:objective-function}
\end{equation}
by seeking a stationary point 
\begin{equation}
\nabla J\left(y,\sigma,u\right)=e_{h}'\left(y,\sigma,u\right)^{*}e_{h}\left(y,\sigma,u\right)=0,\label{eq:discrete-residual-stationary}
\end{equation}
where $e_{h}'\left(y,\sigma,u\right)^{*}:\mathbb{R}^{\dim\left(V_{y,h}\times V_{\sigma,h}\times W_{y,h}\times W_{\sigma,h}\right)}\rightarrow Y_{h}\times\Sigma_{h}\times U_{h}$
is the adjoint operator. Given an initialization $\left(y,\sigma,u\right)_{0}$,
the solution is repeatedly updated 
\begin{equation}
\left(y,\sigma,u\right)_{i+1}=\left(y,\sigma,u\right)_{i}+\Delta\left(y,\sigma,u\right)_{i}\qquad i=0,1,2,\ldots\label{eq:discrete-iteration}
\end{equation}
until~(\ref{eq:discrete-residual-stationary}) is satisfied to a
given tolerance. The Levenberg-Marquardt method~\citep{Lev44,Mar63}
is then employed to solve~(\ref{eq:discrete-residual-stationary}),
which yields the increment 
\begin{equation}
\Delta\left(y,\sigma,u\right)=-\left(e_{h}'\left(y,\sigma,u\right)^{*}e_{h}'\left(y,\sigma,u\right)+I_{h,\lambda}\left(y,\sigma,u\right)\right)^{-1}\left(e_{h}'\left(y,\sigma,u\right)^{*}e_{h}\left(y,\sigma,u\right)\right),\label{eq:discrete-increment}
\end{equation}
where where $I_{h,\lambda}\left(y,\sigma,u\right):\left(Y_{h}\times\Sigma_{h}\times U_{h}\right)\times\left(Y_{h}\times\Sigma_{h}\times U_{h}\right)\rightarrow\mathbb{R}$
is a symmetric, positive-definite bilinear form that defines the choice
of regularization 
\begin{equation}
I_{h,\lambda}\left(y,\sigma,u\right)\left(\left(\delta y,\delta\sigma,\delta u\right),\left(v_{y},v_{\sigma},v_{u}\right)\right)=\left(\delta y,\lambda_{y}v_{y}\right)+\left(\delta\sigma,\lambda_{\sigma}v_{\sigma}\right)+\left(\delta u,\lambda_{u}v_{u}\right).\label{eq:regularization-operator}
\end{equation}
with $\lambda_{y}$, $\lambda_{\sigma}$, and $\lambda_{u}$ denoting
nonnegative regularization coefficients for each solution variable.
In practice, $\lambda_{y}$ and $\lambda_{\sigma}$ are set to zero.
In addition to the identity regularization~(\ref{eq:regularization-operator}),
which ensures positive definiteness and prevents large-scale grid
changes, we incorporate a Laplacian-type grid regularization, 
\begin{equation}
I_{h,\lambda}^{\Delta}\left(y,\sigma,u\right)\left(\left(\delta y,\delta\sigma,\delta u\right),\left(v_{y},v_{\sigma},v_{u}\right)\right)=-\left(\nabla\left(b_{h}'\left(u\right)\delta u\right),\lambda_{\Delta u}\nabla\left(b_{h}'\left(u\right)v_{u}\right)\right).\label{eq:regularization-laplacian-old}
\end{equation}
where $\lambda_{\Delta u}\geq0$ is the corresponding regularization
coefficient and $v_{u}\in U_{h}$, which introduces a compressibility
effect into the grid motion. In previous work~\citep{Ker20_LS},
following the approach of Zahr et al.~\citep{Zah20}, the regularization
factor $\lambda_{\Delta u}$ was modified to incorporate a factor
proportional to the inverse of the element volume, which locally stiffens
small elements in an \emph{isotropic} manner. Note that the regularizations~(\ref{eq:regularization-operator})
and~(\ref{eq:regularization-laplacian-old}) are not incorporated
into the objective function~(\ref{eq:objective-function}).

The solver as hitherto described was employed in~\citep{Ker20}.
In the remainder of this section, we introduce enhancements to the
solver that significantly improve its robustness. 

\subsection{Regularization terms\label{subsec:regularization}}

\subsubsection{Anisotropic Laplacian-type grid regularization}

In this work, we modify the regularization~(\ref{eq:regularization-laplacian-old})
to directly account for element anisotropy as

\begin{equation}
I_{h,\lambda}^{\Delta}\left(y,\sigma,u\right)\left(\left(\delta y,\delta\sigma,\delta u\right),\left(v_{y},v_{\sigma},v_{u}\right)\right)=-\left(\nabla\left(\mathcal{H}^{-\alpha/2}b_{h}'(u)\delta u\right),\lambda_{\Delta u}\nabla\left(\mathcal{H}^{-\alpha/2}b_{h}'(u)v_{u}\right)\right),\label{eq:regularization_laplacian-new}
\end{equation}
where $\alpha\geq0$ is a parameter and $\mathcal{H}$, in the two-dimensional
case, is an element-local, $2\times2$, symmetric-positive-definite
transformation matrix given by~\citep{Fid07,Chi21}

\begin{equation}
\mathcal{H}=\mathcal{M}^{-1/2}=V\Sigma V^{T}=\left[\begin{array}{cc}
\rule[-1ex]{0.5pt}{2.5ex} & \rule[-1ex]{0.5pt}{2.5ex}\\
\widehat{e}_{1} & \widehat{e}_{2}\\
\rule[-1ex]{0.5pt}{2.5ex} & \rule[-1ex]{0.5pt}{2.5ex}
\end{array}\right]\begin{bmatrix}h_{1} & 0\\[1ex]
0 & h_{2}
\end{bmatrix}\left[\begin{array}{ccc}
\rule[.5ex]{2.5ex}{0.5pt} & \widehat{e}_{1}^{T} & \rule[.5ex]{2.5ex}{0.5pt}\\
\rule[.5ex]{2.5ex}{0.5pt} & \widehat{e}_{2}^{T} & \rule[.5ex]{2.5ex}{0.5pt}
\end{array}\right].\label{eq:transformation_matrix}
\end{equation}
$\mathcal{M}$ is the metric implied by the mesh~\citep{Fid07},
the columns of $V$ are the (orthonormal) left singular vectors, $\widehat{e}_{1}$
and $\widehat{e}_{2}$, of the geometric Jacobian, $J$, and $\Sigma$
is a diagonal matrix with the singular values, $h_{1}$ and $h_{2}$,
of $J$ along the main diagonal. $\mathcal{H}$ projects the unit
circle to an ellipse with principal directions $\widehat{e}_{1}$
and $\widehat{e}_{2}$ and principal stretching magnitudes $h_{1}$
and $h_{2}$~\citep{Chi21}. Without loss of generality, we assume
that the singular values are ordered such that $h_{1}\leq h_{2}$.
$\mathcal{H}^{-\alpha}$ can be expanded as 
\begin{equation}
\mathcal{H}^{-\alpha}=V\Sigma^{-\alpha}V^{T}=\left[\begin{array}{cc}
\rule[-1ex]{0.5pt}{2.5ex} & \rule[-1ex]{0.5pt}{2.5ex}\\
\widehat{e}_{1} & \widehat{e}_{2}\\
\rule[-1ex]{0.5pt}{2.5ex} & \rule[-1ex]{0.5pt}{2.5ex}
\end{array}\right]\begin{bmatrix}h_{1}^{-\alpha} & 0\\[1ex]
0 & h_{2}^{-\alpha}
\end{bmatrix}\left[\begin{array}{ccc}
\rule[.5ex]{2.5ex}{0.5pt} & \widehat{e}_{1}^{T} & \rule[.5ex]{2.5ex}{0.5pt}\\
\rule[.5ex]{2.5ex}{0.5pt} & \widehat{e}_{2}^{T} & \rule[.5ex]{2.5ex}{0.5pt}
\end{array}\right].\label{eq:transformation_matrix-alpha}
\end{equation}
As such, the modified regularization~(\ref{eq:regularization_laplacian-new})
limits grid motion in directions with small element length scales
while allowing for greater changes in directions with larger length
scales, aiding in the formation of non-degenerate high-aspect-ratio
elements to resolve thin viscous structures.  For simplicity, $\mathcal{H}^{-\alpha}$
is evaluated at the centroid of the given element. The metric described
here is a powerful tool that can be used not only for anisotropic
grid regularization but also for remeshing via a metric-based grid
generator~\citep{Fid07} in order to maintain high-quality grids,
which will be the subject of future work. Moreover, we employ additional
regularization operators that further inhibit element degeneration,
specifically $\mathcal{P}_{1}$ grid regularization and penalty grid
regularization, which are described below. Similar forms of these
two regularization operators were employed in the mesh untangling
algorithm by Toulorge et al.~\citep{Tou13}.

\subsubsection{$\mathcal{P}_{1}$ grid regularization}

To penalize excessive element curvature, the following regularization
term is included in the objective function~(\ref{eq:objective-function}):
\begin{equation}
J_{1}(u)=\frac{1}{2}h_{1}^{\beta}\lambda_{1}\left|\left|u-\Pi_{1}u\right|\right|^{2},\label{eq:p1-regularization}
\end{equation}
where $\lambda_{1}$ is the corresponding regularization coefficient,
$h_{1}$ is obtained from the mesh-implied metric, $\Pi_{1}$ denotes
the projection to a (multi-)linear subspace of $U_{h}$, and $\beta>0$
is a parameter. The $h_{1}^{\beta}$ scaling allows for increased
curvature of elements with small length scales, which is important
for resolving curved shocks.

\subsubsection{Penalty grid regularization}

Finally, a penalty regularization term is included in the objective
function~(\ref{eq:objective-function}):
\begin{equation}
J_{b}\left(u\right)=\frac{1}{2}\lambda_{b}\left|\left|f\left(u\right)\right|\right|^{2},\label{eq:penalty-regularization}
\end{equation}
where $\lambda_{b}$ is the corresponding regularization coefficient
and $f$ is defined as
\begin{equation}
f\left(u\right)=\max\left\{ 0,\mathcal{J}_{b}-\det\left(\nabla u\right)\right\} ,\label{eq:penalty-function}
\end{equation}
with $\mathcal{J}_{b}$ represents a desired lower bound on the Jacobian
determinant (e.g., $10^{-10}$). The regularization~(\ref{eq:penalty-regularization})
penalizes invalid elements (i.e., elements for which the determinant
of the geometric Jacobian is nonpositive). Note that the non-differentiability
of $f\left(u\right)$ at $\det\left(\nabla u\right)=\mathcal{J}_{b}$
is not found to cause any noticeable issues; furthermore, it can easily
be made differentiable by squaring the RHS of~(\ref{eq:penalty-function}).
In previous work~\citep{Cor18}, barrier grid regularization was
found to frequently lead to solver stagnation, although such regularization
may nevertheless be worthy of future investigation.

\subsubsection{Global scaling}

In this work, the regularization coefficients $\lambda_{u}$, $\lambda_{\Delta u}$,
$\lambda_{1}$, and $\lambda_{b}$ are scaled by the residual magnitude,
$\left\Vert e_{h}\left(y,\sigma,u\right)\right\Vert $. Early in the
simulation, when the residual is large and spurious transients may
appear in the solution, the higher regularization prevents excessive
grid changes that would otherwise occur. As the solution converges
and these transients disappear, the lower regularization encourages
greater adjustments to the grid to facilitate resolution of high-gradient
features. To prevent rank deficiency of the linear system at low residuals,
we select a small, positive number as a lower bound on $\lambda_{u}$
(e.g., $10^{-7}$).

\subsection{Adaptive, elementwise regularization\label{subsec:adaptive-elementwise-regularization}}

Even with manual tuning, the aforementioned global, static scaling
of the regularization terms is not sufficiently robust for preventing
cell degeneration, especially as highly anisotropic, curved elements
form to resolve multidimensional, high-gradient features. Local grid
operations, whether topology-preserving (e.g., vertex smoothing~\citep{Ala14})
or topology-changing (e.g., edge refinement, collapse~\citep{Loh08},
and swapping~\citep{Ala14}), may be beneficial if used sparingly;
however, such operations often fail to recover valid cells and will
inhibit iterative convergence of the solver since they are not intrinsic
to the formulation (i.e., they are applied as a ``postprocessing''
step at the end of the iteration). This obstruction of convergence
is exacerbated in the viscous setting, wherein high-aspect-ratio elements
are gradually compressed to better resolve thin viscous structures.
Mesh deformation techniques have also been developed to improve mesh
quality and untangle invalid elements, typically in the context of
a posteriori high-order mesh generation. Examples include elasticity-based
methods~\citep{Xie13,Mox16,Mar21}, in which a set of elasticity
equations is solved to obtain a mesh-displacement field, and mesh
optimization algorithms~\citep{Tou13,Gar15}, which seek to minimize
element distortion and constrain invalid Jacobians via regularization
operators similar to~(\ref{eq:p1-regularization}) and~(\ref{eq:penalty-regularization}).
Although more reliable for recovering valid elements and potentially
useful for occasionally ``resetting'' the grid in case convergence
begins to stall, these mesh-deformation techniques can be expensive
and again disrupt solver convergence, especially because they need
to be engaged frequently in the presence of curved, high-gradient
features. Incorporating regularization operators directly into the
objective function~(\ref{eq:objective-function}) eliminates, or
at least mitigates, obstruction of iterative convergence, but, as
previously mentioned, global, static scaling of the regularization
terms does not provide sufficient robustness.

In light of the above, we propose a strategy for maintaining grid
validity that can reliably recover valid cells while avoiding disruption
of solver convergence. In particular, when a cell becomes invalid,
the iteration is restarted (i.e., the solution is ``rolled back'')
and the regularization coefficients are automatically adjusted in
an elementwise fashion until the solver can proceed with a valid grid.
In this work, $\lambda_{\Delta u}$, $\lambda_{1}$, and $\lambda_{b}$
are dynamically scaled; however, if the aspect ratio is high (e.g.,
$h_{2}/h_{1}>10$), then $\lambda_{1}$ is left unchanged in order
to facilitate resolution of thin, curved features using curved cells.
The regularization coefficients are then successively decreased to
their base values, unless the element once again becomes invalid.
The rollback of the solution is crucial for ensuring a valid grid
at every iteration. Although our formulation does not necessarily
diverge in the presence of (slightly) negative Jacobian determinants,
maintaining grid validity at every iteration is important for the
following reasons: (a) allowing the solver to proceed with invalid
cells can result in pollution of solution accuracy and lower-quality
grids that may drive the solution towards a less-optimal local minimum;
(b) in our experience, once the solver proceeds with negative Jacobian
determinants, it can be extremely difficult to eliminate them; and
(c) if negative Jacobian determinants become too large in magnitude,
the solver can easily diverge or at least produce clearly erroneous
results. Furthermore, although the intermittent scaling of the penalty
grid regularization can cause abrupt increases in the objective function
(specifically, $J_{b}$), this is not a major concern since we are
ultimately interested in minimizing $\left\Vert e_{h}\left(y,\sigma,u\right)\right\Vert $.
This adaptive, elementwise regularization strategy is found to be
critical for preventing cell degeneration in the vicinity of viscous
shocks and boundary layers at hypersonic flow conditions. With such
high-gradient features, zero to three rollbacks per iteration are
typically required to preserve grid validity. Note that if the solution
is rolled back, only the regularization terms for the corresponding
cells need to be recomputed. The implementation details of this strategy
will be provided in Section~\ref{subsec:line-search}.

\subsection{Increment limiting\label{subsec:increment-limiting}}

The consideration of very strong shocks without artificial dissipation
often leads to undershoots and overshoots in pressure and other quantities
during intermediate iterations. To mitigate these instabilities, which
can otherwise cause solver divergence, we employ an increment-limiting
strategy used in standard implicit-DG solvers~\citep{Cez15}, in
which the increment is scaled by a factor no greater than unity such
that the maximum change in pressure and density at the integration
points is less than a user-specified fraction, denoted $\mathsf{f}$
(e.g., $\mathsf{f}=10$\%). The implementation details will be provided
in Section~\ref{subsec:line-search}. 

We have also experimented with projecting the state in troubled elements
to a first-order approximation ($\mathcal{P}_{0}$), but this can
hinder iterative convergence by creating a cycle in which first-order
projections are followed by eventual reappearance of undershoots/overshoots,
which then causes additional first-order projections, and so on. Applying
sophisticated limiters that, for example, nominally preserve order
of accuracy to troubled elements during intermediate iterations may
also be useful, but is not considered in this work.

\subsection{Line search\label{subsec:line-search}}

The locally adaptive regularization and increment limiting discussed
in Sections~\ref{subsec:adaptive-elementwise-regularization} and~\ref{subsec:increment-limiting},
respectively are incorporated into a simple line search method, as
presented in Algorithm~\ref{alg:Line-search-algorithm}. It should
be noted that decreasing $\left\Vert e_{h}\left(y,\sigma,u\right)\right\Vert $
and maintaining grid validity (i.e., minimizing $J_{b}$) are often
at odds with each other since penalizing small Jacobian determinants
can hinder resolution of high-gradient features, especially those
which are curved. Penalty regularization is thus activated only when
necessary and in a local, gradual fashion in order to avoid interfering
with reduction of $\left\Vert e_{h}\left(y,\sigma,u\right)\right\Vert $
and potential convergence towards a less optimal local minimum. The
local regularization coefficients $\lambda_{\Delta u}^{\kappa}$,
$\lambda_{1}^{\kappa}$, and $\lambda_{b}^{\kappa}$ are then successively
decreased every few steps towards their base values and $\mathcal{J}_{b}^{\kappa}$
is reset, provided that the given element remains valid. Recommended
base values for the regularization coefficients and other parameters
are given in Table~\ref{tab:Recommended-base-values}. We also find
that even though scaling the regularization coefficients by $\left\Vert e_{h}\left(y,\sigma,u\right)\right\Vert $
is often beneficial, sometimes, at low values of $\left\Vert e_{h}\left(y,\sigma,u\right)\right\Vert $,
the grid has already repositioned itself enough to resolve the anisotropic
structures in the flow, at which point anything more than marginal
adjustments to the grid can slow down residual convergence. To minimize
grid motion in this situation, $\lambda_{u}$ and $\lambda_{\Delta u}$
are automatically scaled in a global fashion by factors of 100 for
a prescribed number of iterations (e.g., 30) when repeatedly low values
of the increment factor (e.g., $\omega<0.02)$ are accompanied by
at least one rollback, which serves as a reliable indicator of said
scenario. 

Note that modifying the hard-coded values in Algorithm~\ref{alg:Line-search-algorithm}
(e.g., setting $N_{\mathrm{search}}$ to five and scaling $\lambda_{\Delta u}^{\kappa}$
by a factor of ten) can potentially yield better performance; nevertheless,
the chosen values work sufficiently well for the problems considered
in this study, and we remark that the primary concern of this work
is robustness in terms of maintaining grid validity. Future work will
explore strategies to accelerate iterative convergence, as well as
well-established methods to automatically adjust the global values
of $\lambda_{u}$ and/or $\lambda_{\Delta u}$ at each step~\citep{Tra12,Zah20}.
Furthermore, similar results can likely be obtained with locally adaptive
barrier functions (as opposed to penalty functions), although as previously
mentioned, a naive strategy may lead to frequent solver stagnation.
In addition, though not employed here, it may be useful to activate
penalty regularization over a localized region of elements, instead
of just one element, since doing so for only one element can lead
to movement of grid points that then causes degeneration in neighboring
cells. 

Finally, we remark that the full system can also be explicitly formulated
as a constrained-optimization problem with inequality constraints
$\det\left(\nabla u\right)>0$, where Lagrange multipliers corresponding
to those constraints are introduced as solution variables. Although
future work may explore a Lagrange-multiplier approach, the proposed
unconstrained-optimization formulation is appealing due to its simplicity
and smaller system size.

\begin{algorithm}
\caption{Line search algorithm for iteration $i+1$. $\omega\protect\leq1$
is a positive factor that scales the increment. The $\kappa$ superscripts
denote element-local quantities. $\mathcal{J}_{\min}^{\kappa}$ is
the minimum Jacobian determinant for cell $\kappa$. The compressible
Navier-Stokes system is assumed (Section~\ref{subsec:CNS}). \label{alg:Line-search-algorithm}}

\begin{algorithmic}[H]
\Require{$\left(y,\sigma,u\right)_{i}$}
%\Ensure{$X_k$} 
\State{converged $\gets$ False}
\While{not converged}
	\State{Compute $\Delta\left(y,\sigma,u\right)_{i}$}
	\State{$\omega \gets 1,\: \omega_a \gets 0, \: \omega_b \gets 1$}
	%\State{$R_h \gets \left\Vert e_{h}\left(y,\sigma,u\right)_i\right\Vert$}
	\State{$\left(y,\sigma,u\right)^*\gets\left(y,\sigma,u\right)_{i}$}
	\State{$N_\mathrm{search} \gets 5$}
	\For{$k \gets 0$ to $N_\mathrm{search}$}   
		\If{$k > 0$}
			\State{$\omega \gets (\omega_a + \omega_b)/2$}
		\EndIf
		\State{$\left(y,\sigma,u\right)_{i+1}\gets\left(y,\sigma,u\right)_{i}+\omega\Delta\left(y,\sigma,u\right)_{i}$}
		\State{$\mathsf{F}_\rho \gets \max_\Omega\left\{\left|\rho_{i+1} - \rho_i \right|/\rho_i\right\}$}
		\State{$\mathsf{F}_P \gets \max_\Omega\left\{\left|P\left(y_{i+1}\right) - P\left(y_{i}\right) \right|/P\left(y_{i}\right)\right\}$}
		%\State{$R_h \gets \left\Vert e_{h}\left(y,\sigma,u\right)_{i+1}\right\Vert$}
		\If{$\left\Vert e_{h}\left(y,\sigma,u\right)_{i+1}\right\Vert/\left\Vert e_{h}\left(y,\sigma,u\right)^*\right\Vert<1$ and $\mathsf{F}_\rho \leq \mathsf{f}$ and $\mathsf{F}_P \leq \mathsf{f}$}
			%\State{$\left(y,\sigma,u\right)_{i+1}\gets\left(y,\sigma,u\right)^*$}
			\If{$\omega = 1$}
				\State{\textbf{break}}
			\EndIf
			\State{$\left(y,\sigma,u\right)^*\gets\left(y,\sigma,u\right)_{i+1}$}
			\State{$\omega_a \gets \omega$}
		\Else
			\State{$\omega_b \gets \omega$}
		\EndIf
		\If{$\omega_a = 0$ and $N_\mathrm{search} = 5$}
			\State{$N_\mathrm{search} \gets N_\mathrm{search} + 10$ \Comment{Could not reduce $\left\Vert e_{h}\left(y,\sigma,u\right)\right\Vert$; increase $N_\mathrm{search}$}}
		\EndIf
	\EndFor
	\State{$\left(y,\sigma,u\right)_{i+1}\gets\left(y,\sigma,u\right)^*$}
	\State{converged $\gets$ True}
	\For{each element $\kappa$}
		\If{$\mathcal{J}_{\min}^{\kappa,i+1} < \mathcal{J}_b^{\kappa}$}
			\State{converged $\gets$ False}
			\State{$\lambda_{\Delta u}^{\kappa}\gets10\lambda_{\Delta u}^{\kappa}$, $\lambda_b^{\kappa}\gets10\lambda_b^{\kappa}$, $\lambda_1^{\kappa}\gets100\lambda_1^{\kappa}$}
			\State{$\mathcal{J}_b^{\kappa} \gets 10\mathcal{J}_{\min}^{\kappa,i}$}
		\EndIf
	\EndFor
\EndWhile
\end{algorithmic}
\end{algorithm}

\begin{table}

\caption{Recommended base values for regularization coefficients and other
parameters\label{tab:Recommended-base-values}}

\centering{}%
\begin{tabular}{ccccccc}
\toprule 
$\lambda_{u}$ & $\lambda_{\Delta u}$ & $\lambda_{1}$ & $\lambda_{b}$ & $\mathcal{J}_{b}$ & $\alpha$ & $\beta$\tabularnewline
\midrule
\midrule 
\addlinespace[0.2cm]
$10^{-7}$ & $10^{-5}$ & $10^{-3}$ & $10^{2}$ & $10^{-10}$ & $-1$ & $3$\tabularnewline
\bottomrule
\end{tabular}
\end{table}

\section{Results\label{sec:results}}

We apply MDG-ICE, equipped with the enhanced nonlinear solver described
in Section~\ref{sec:nonlinear-solver}, to compute viscous flows
with high-gradient features. Unsteady solutions are obtained using
a space-time discretization. All grids are generated using Gmsh~\citep{Geu09}.
The simulations in this study are performed using the JENRE® Multiphysics
Framework employed in previous work~\citep{Cor18,Ker20,Ker20_LS}
with the modifications and extensions described here. The three-dimensional
solutions are computed using hybrid shared- and distributed-memory
parallelism, the latter of which was not implemented in previous work.
In the present study, we employ a sparse, direct LDLT solver provided
by the MUMPS (MUltifrontal Massively Parallel Solver) package~\citep{Ame01,Ame19}
via the PETSc (Portable, Extensible Toolkit for Scientific Computation)
software library~\citep{Bal97,Bal23}. Future work will improve the
efficiency of solving the linear system by leveraging techniques such
as domain decomposition and multigrid algorithms.

\subsection{Space-time Burgers viscous shock formation\label{subsec:burgers}}

This section presents results for space-time Burgers viscous shock
formation, previously computed using MDG-ICE in~\citep{Ker20}. The
initial condition is given by

\begin{equation}
y(x,t=0)=\frac{1}{2\pi t_{s}}\sin\left(2\pi x\right)+y_{\infty},\label{eq:burgers_initial_condition}
\end{equation}
where $t_{s}=0.5$ is the time of shock formation and $y_{\infty}=0.2$
is the freestream velocity. The computational domain is $\Omega=\left[0,1\right]\times\left[0,1\right]$.
Inflow and outflow boundary conditions are applied at the left and
right boundaries, respectively. The solution is initialized by extruding
the initial condition~(\ref{eq:burgers_initial_condition}) in the
temporal direction. We employ continuation in $\mu$, i.e., a series
of problems is solved in which $\mu$ is successively decreased, with
the final solution of a given problem used as the initial condition
of the subsequent one. In previous work~\citep{Ker20}, subparametric
MDG-ICE($\mathcal{P}_{5}/\mathcal{P}_{1}$) solutions were computed,
also with $\mu$ continuation. Specifically, viscosities of $\mu=10^{-3}$,
$\mu=5\times10^{-4}$, and $\mu=10^{-4}$ were considered. The initial
grid consisted of 200 linear triangular elements obtained by splitting
each element of a uniform $10\times10$ quadrilateral mesh into two
triangles, resulting in a regular topology. Here, we compute subparametric
MDG-ICE($\mathcal{P}_{5}/\mathcal{P}_{2}$) solutions for a more challenging
series of viscosities: $\mu=10^{-3}$, $\mu=10^{-4}$, and $\mu=10^{-5}$.
We employ two different starting grids: the first is the same as
in~\citep{Ker20} (i.e., a regular grid with 200 triangular elements),
while the second is an irregular grid with 242 triangular elements.
A key difference with~\citep{Ker20} is that the elements are of
quadratic geometric order, significantly increasing the difficulty
of achieving solution convergence while maintaining a valid grid.

\subsubsection{Initial mesh topology: Regular}

\begin{figure}[H]
\begin{centering}
\subfloat[\label{fig:burgers_structured_planes_mu_1e-3_initial}Initial condition.]{\centering{}\includegraphics[width=0.48\columnwidth]{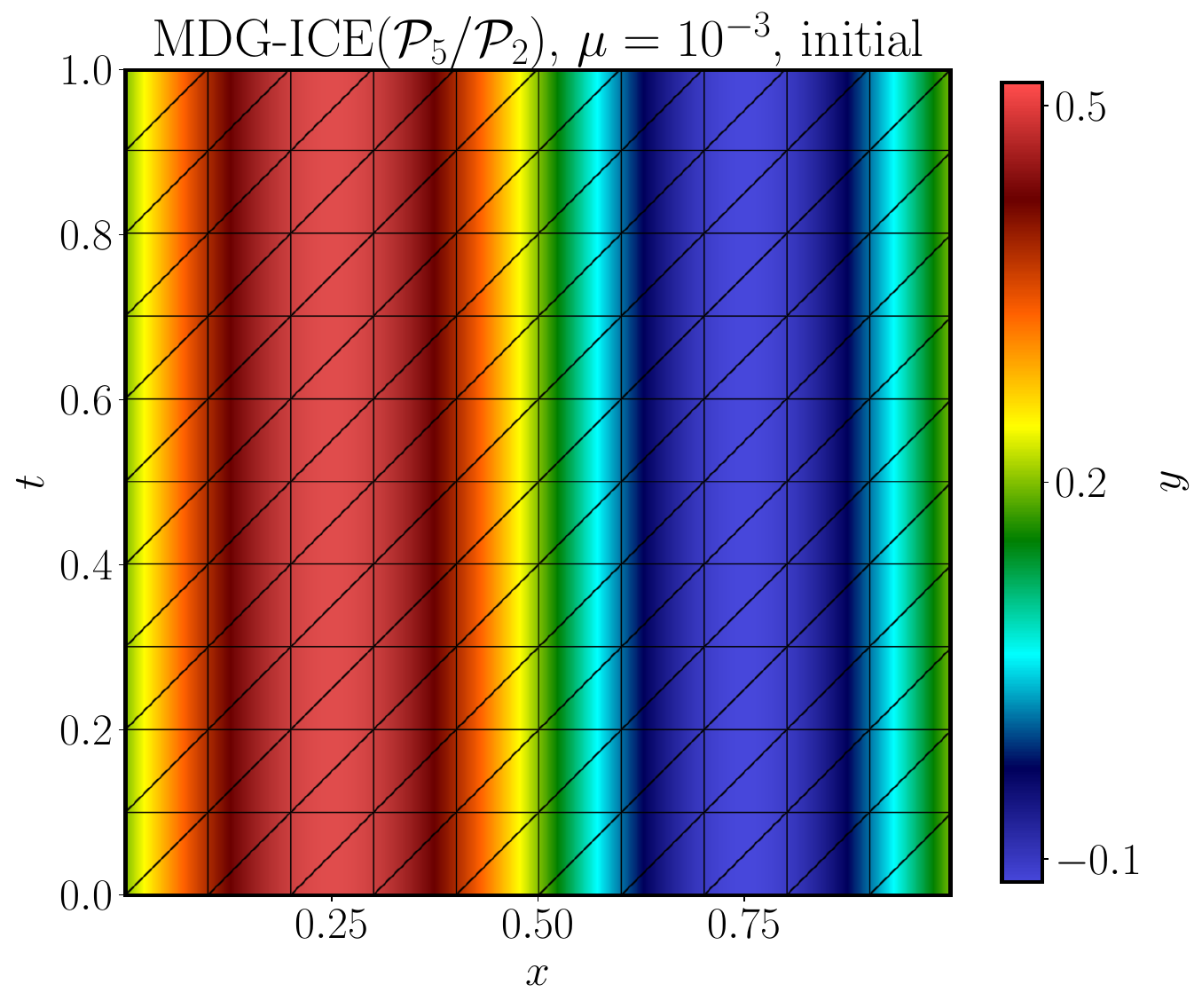}}\hfill{}\subfloat[\label{fig:burgers_structured_planes_mu_1e-3}$\mu=10^{-3}$ solution.]{\centering{}\includegraphics[width=0.48\columnwidth]{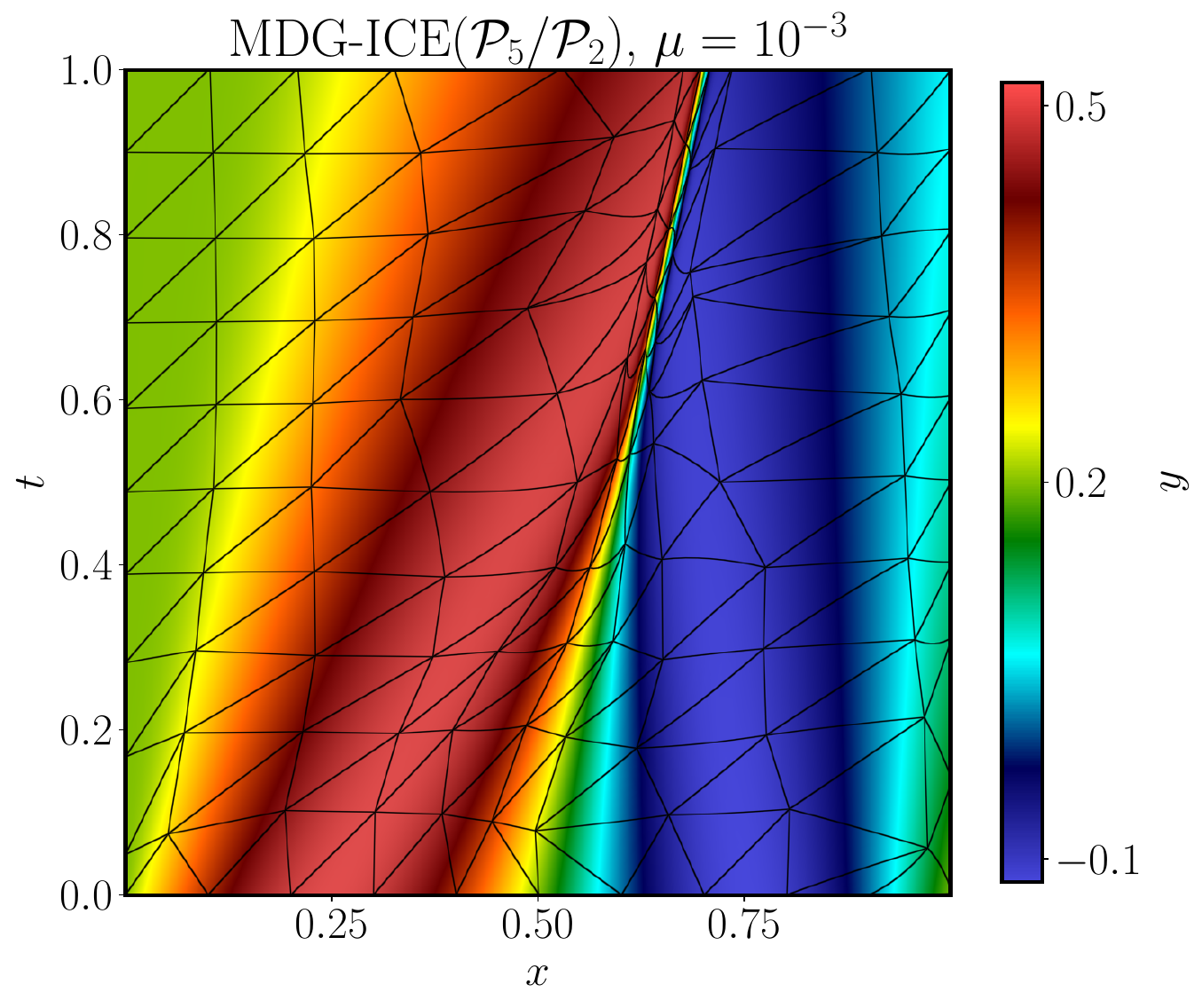}}\hfill{}\subfloat[\label{fig:burgers_structured_planes_mu_1e-4}$\mu=10^{-4}$ solution.]{\centering{}\includegraphics[width=0.48\columnwidth]{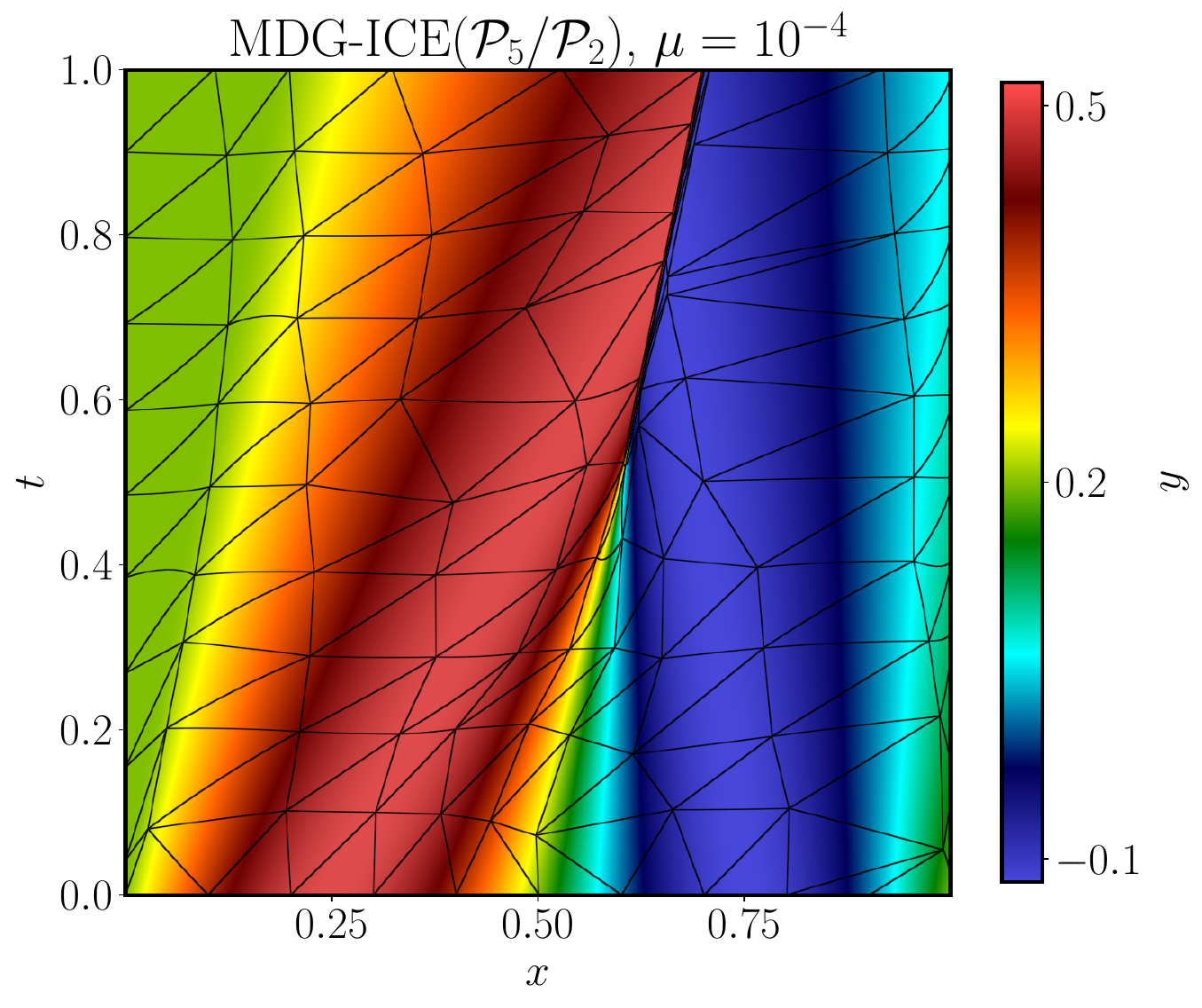}}\hfill{}\subfloat[\label{fig:burgers_structured_planes_mu_1e-5}$\mu=10^{-5}$ solution.]{\centering{}\includegraphics[width=0.48\columnwidth]{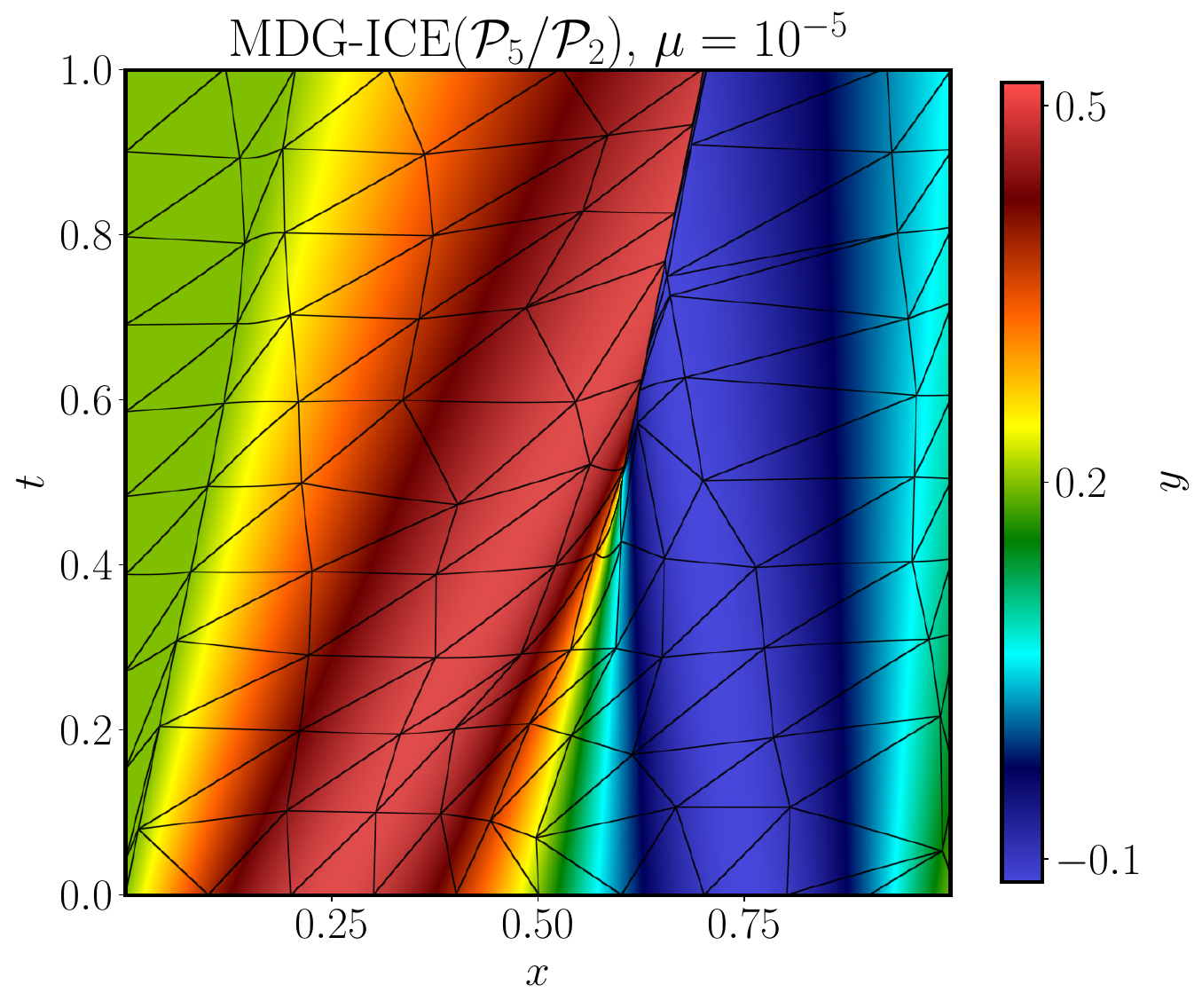}}
\par\end{centering}
\caption{\label{fig:burgers_structured_planes} The space-time Burgers initial
condition and final solutions for $\mu=10^{-3}$, $\mu=10^{-4}$,
and $\mu=10^{-5}$, with the corresponding grids superimposed. The
initial grid consists of 200 triangular elements of quadratic geometric
order with a regular topology. The initial condition is given in Equation~(\ref{eq:burgers_initial_condition}).}
\end{figure}

\begin{figure}[H]
\begin{centering}
\subfloat[\label{fig:burgers_structured_lines_mu_1e-3}$\mu=10^{-3}$ solution.]{\centering{}\includegraphics[width=0.48\columnwidth]{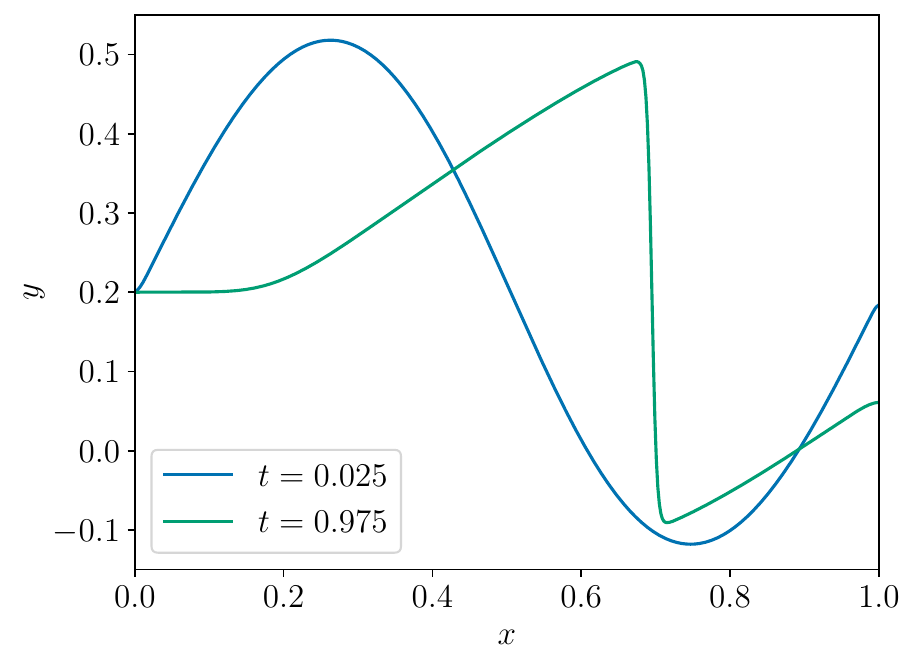}}\hfill{}\subfloat[\label{fig:burgers_structured_lines_mu_1e-4}$\mu=10^{-4}$ solution.]{\centering{}\includegraphics[width=0.48\columnwidth]{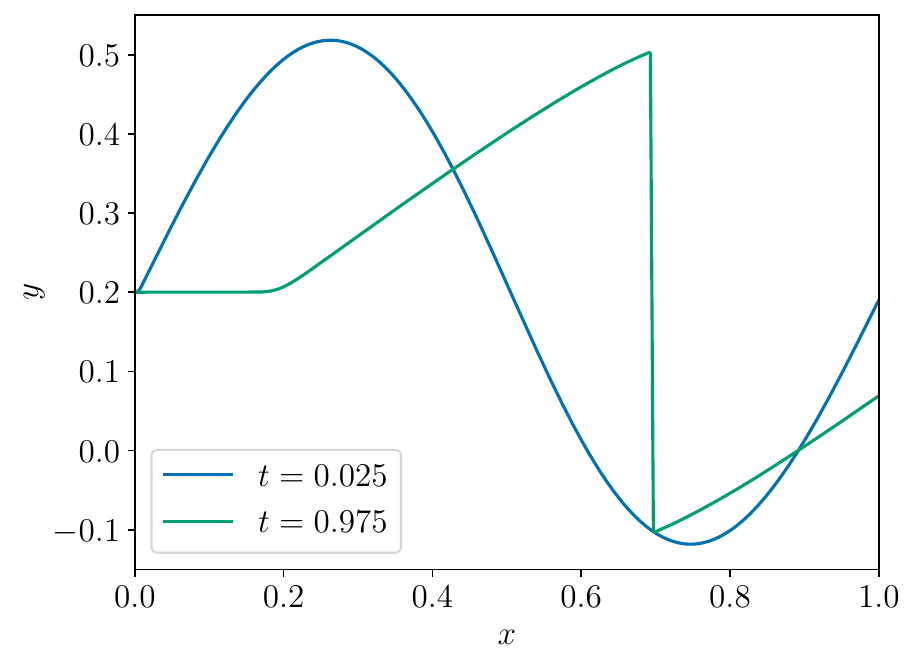}}\hfill{}\subfloat[\label{fig:burgers_structured_lines_mu_1e-5}$\mu=10^{-5}$ solution.]{\centering{}\includegraphics[width=0.48\columnwidth]{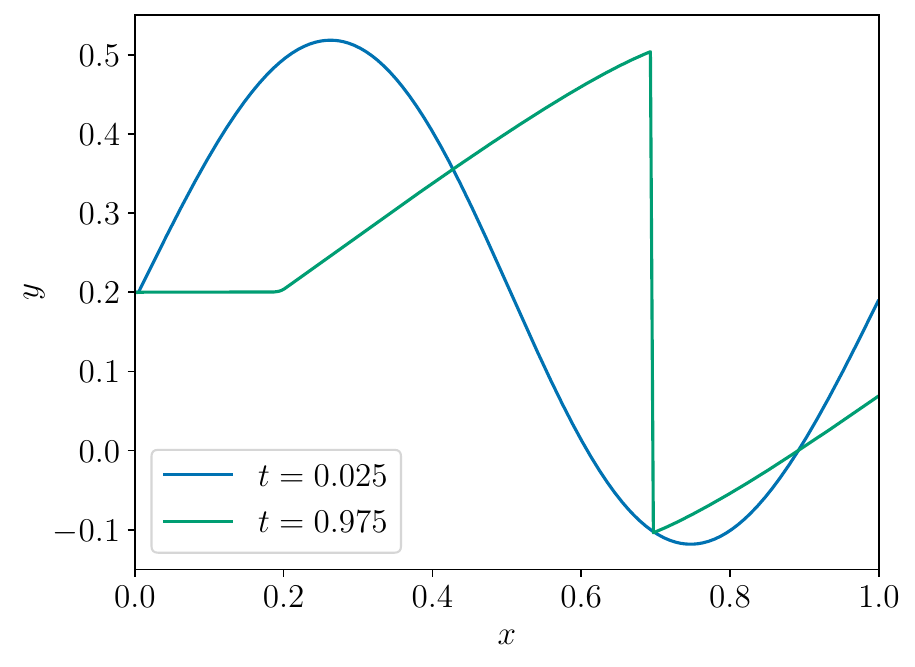}}
\par\end{centering}
\caption{\label{fig:burgers_structured_lines} One-dimensional profiles at
$t=0.025$ and $t=0.975$ for $\mu=10^{-3}$, $\mu=10^{-4}$, and
$\mu=10^{-5}$ for space-time Burgers viscous shock formation. The
initial grid consists of 200 triangular elements of quadratic geometric
order with a regular topology. The initial condition is given in Equation~(\ref{eq:burgers_initial_condition}).}
\end{figure}

Figure~\ref{fig:burgers_structured_planes_mu_1e-3_initial} presents
the space-time initial condition and final solutions for $\mu=10^{-3}$,
$\mu=10^{-4}$, and $\mu=10^{-5}$, with the corresponding grids superimposed.
Figure~\ref{fig:burgers_structured_lines} shows the corresponding
one-dimensional profiles at $t=0.025$ and $t=0.975$. MDG-ICE automatically
adjusts the grid geometry to resolve the viscous shock as a sharp
yet smooth profile without relying on artificial stabilization or
mesh-topology modification. As the viscosity is decreased, the viscous
shock becomes thinner and the aspect ratios of the nearby elements
are increased via anisotropic space-time $r$-adaptivity. The solutions
are free from spurious oscillations. The anisotropic, locally adaptive
penalty technique described in Section~(\ref{sec:nonlinear-solver})
enable excellent resolution of the thin viscous shock while maintaining
grid validity. 

\begin{figure}[ht]
\centering{}\includegraphics[clip,width=0.48\textwidth]{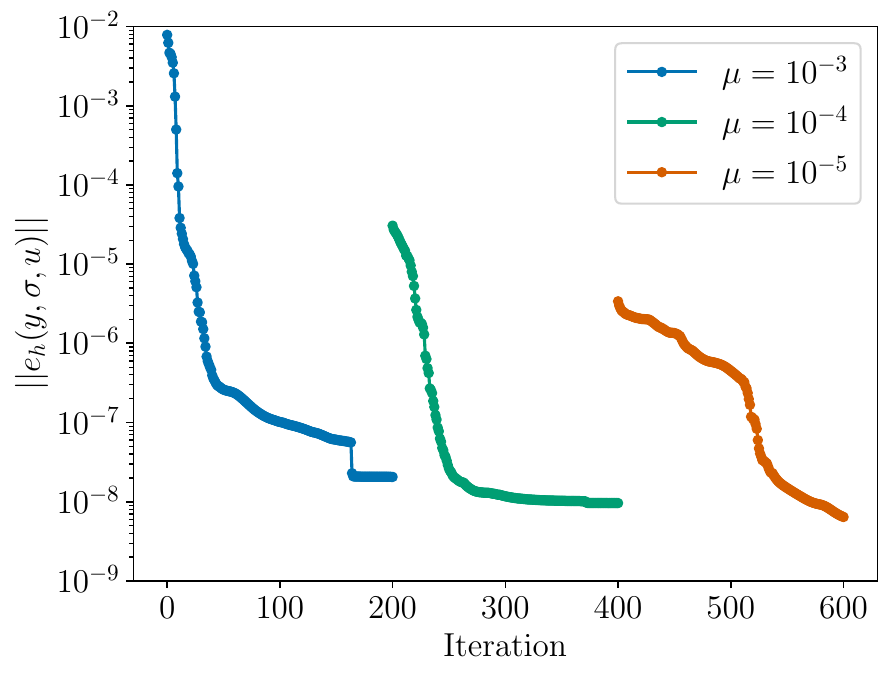}
\caption{Nonlinear convergence history for space-time Burgers viscous shock
formation. The initial grid consists of 200 triangular elements of
quadratic geometric order with a regular topology. The initial condition
is given in Equation~(\ref{eq:burgers_initial_condition}).}
\label{fig:burgers_structured_residual}
\end{figure}

The nonlinear convergence history is given in Figure~\ref{fig:burgers_structured_residual}.
The initial residual magnitudes for the $\mu=10^{-4}$ and $\mu=10^{-5}$
simulations are relatively small since the solver is restarted from
the corresponding higher-viscosity solution. 

\subsubsection{Initial mesh topology: Irregular}

\begin{figure}[H]
\begin{centering}
\subfloat[\label{fig:burgers_unstructured_planes_mu_1e-3_initial}Initial condition.]{\centering{}\includegraphics[width=0.48\columnwidth]{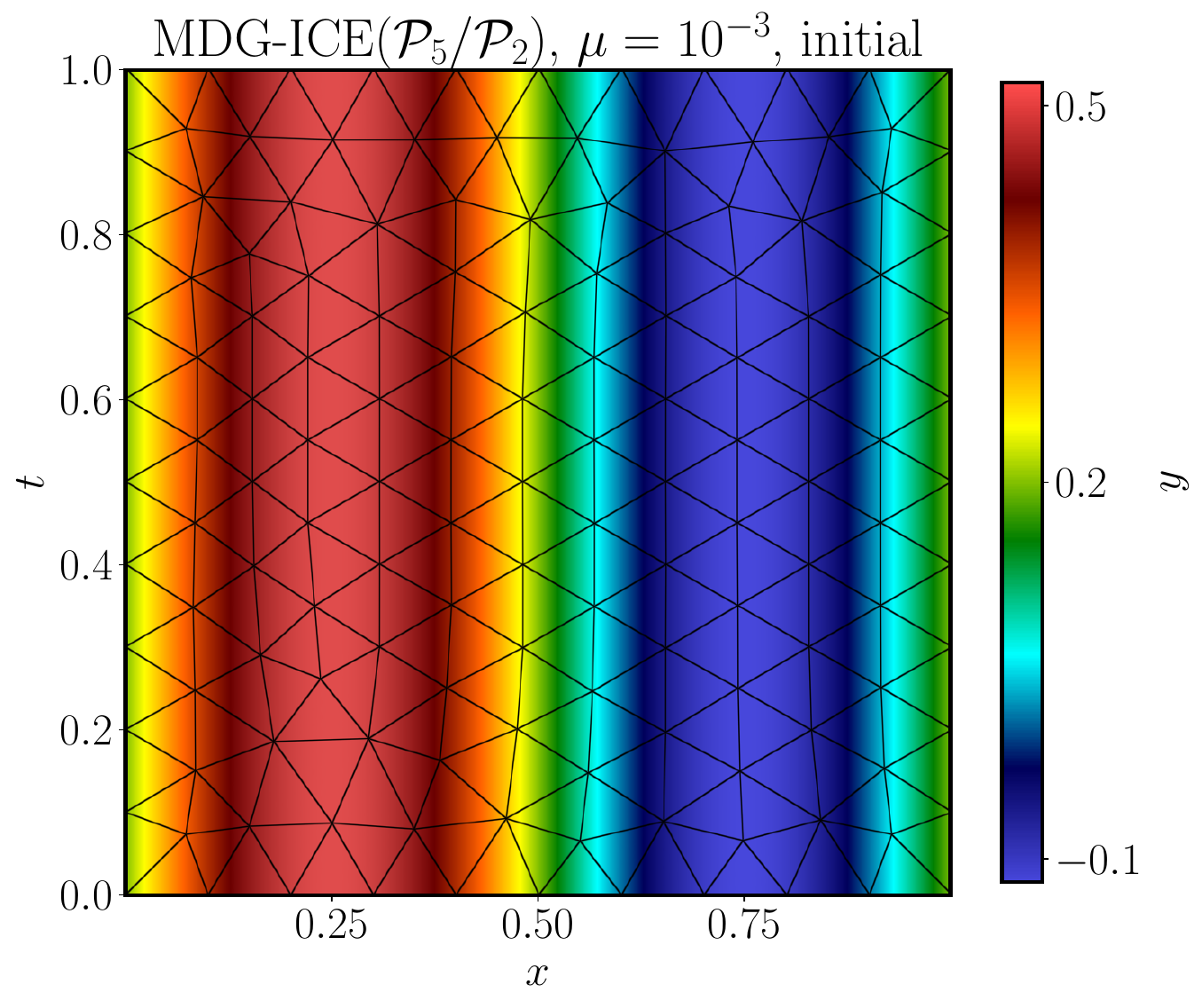}}\hfill{}\subfloat[\label{fig:burgers_unstructured_planes_mu_1e-3}$\mu=10^{-3}$ solution.]{\centering{}\includegraphics[width=0.48\columnwidth]{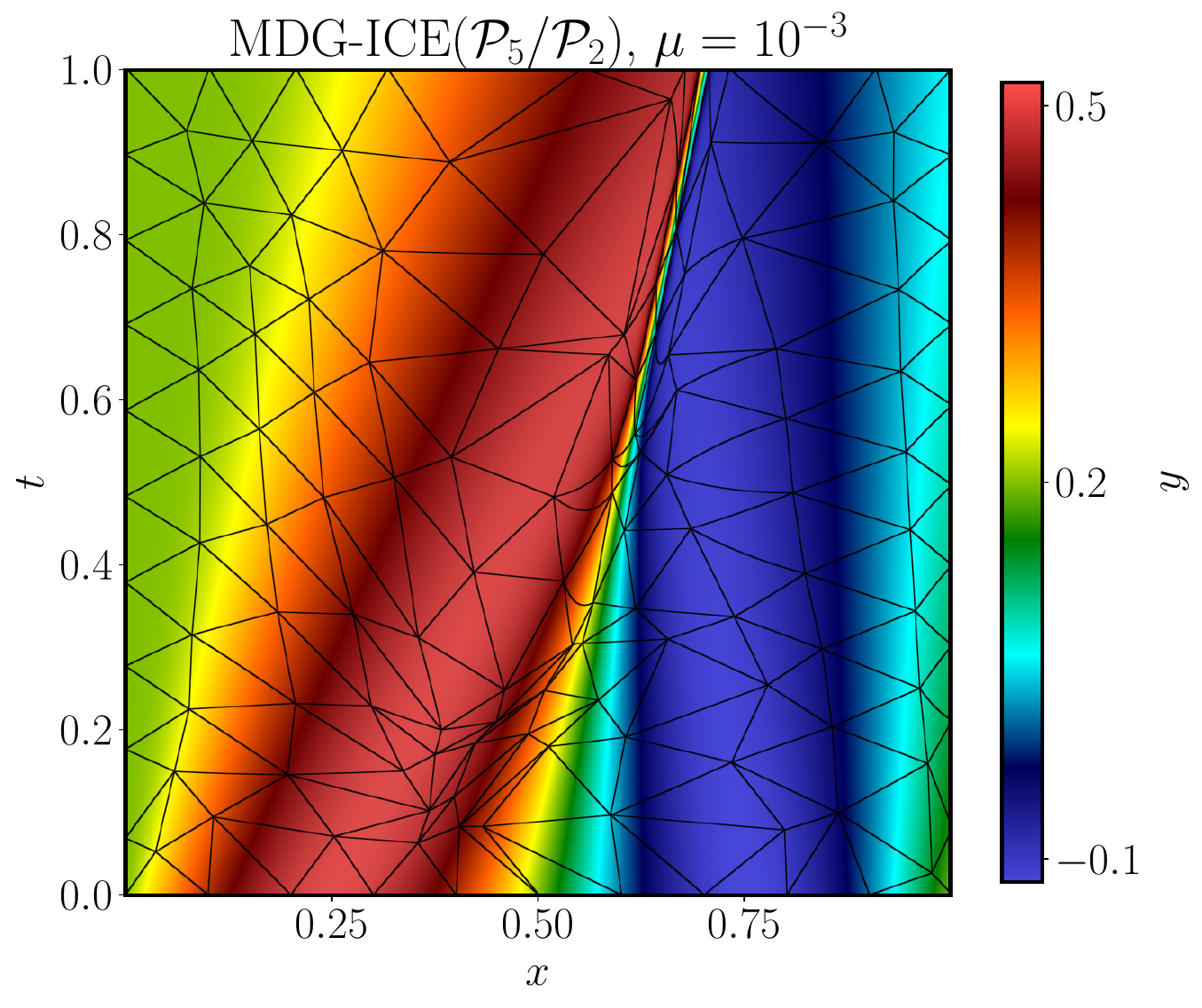}}\hfill{}\subfloat[\label{fig:burgers_unstructured_planes_mu_1e-4}$\mu=10^{-4}$ solution.]{\centering{}\includegraphics[width=0.48\columnwidth]{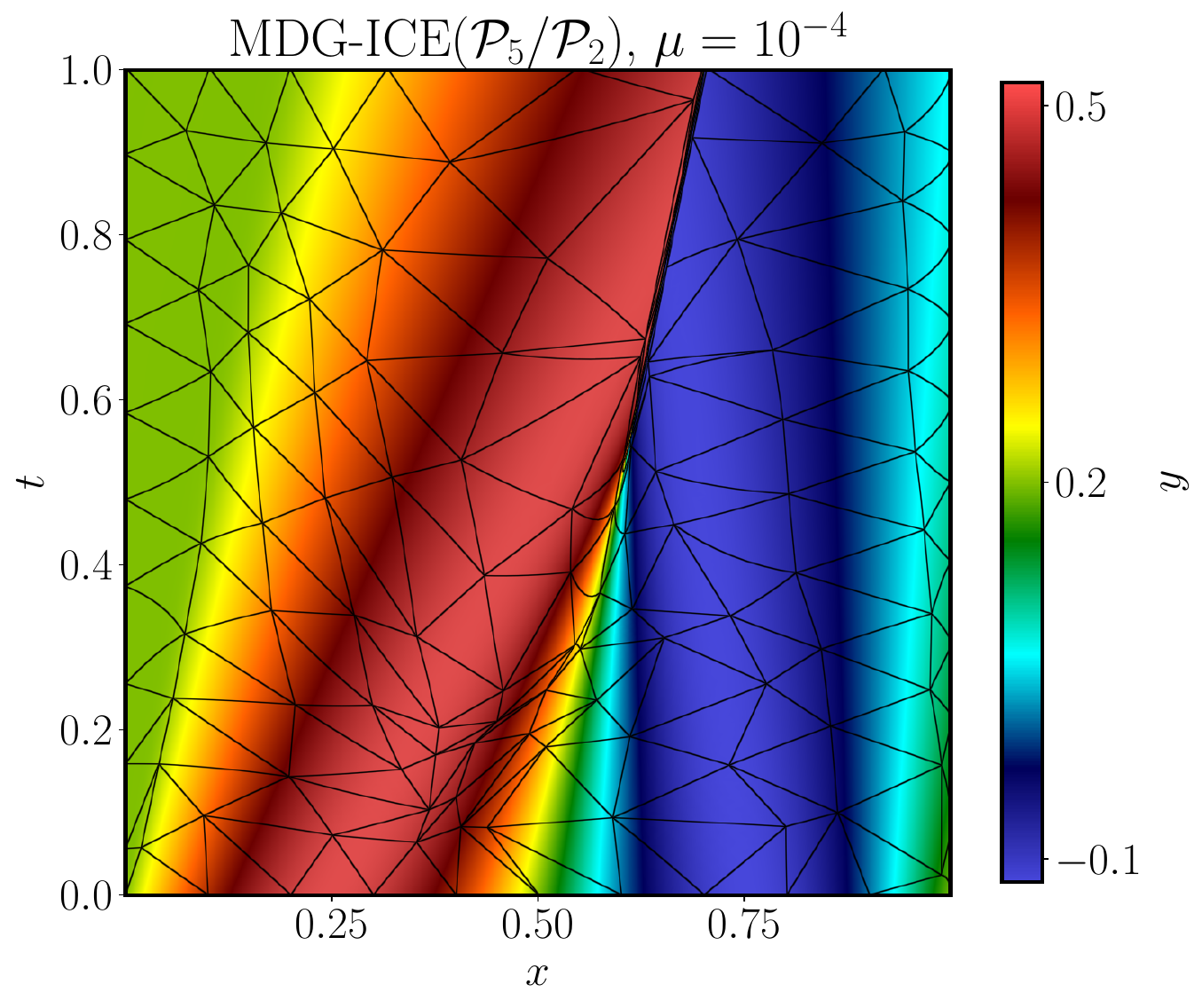}}\hfill{}\subfloat[\label{fig:burgers_unstructured_planes_mu_1e-5}$\mu=10^{-5}$ solution.]{\centering{}\includegraphics[width=0.48\columnwidth]{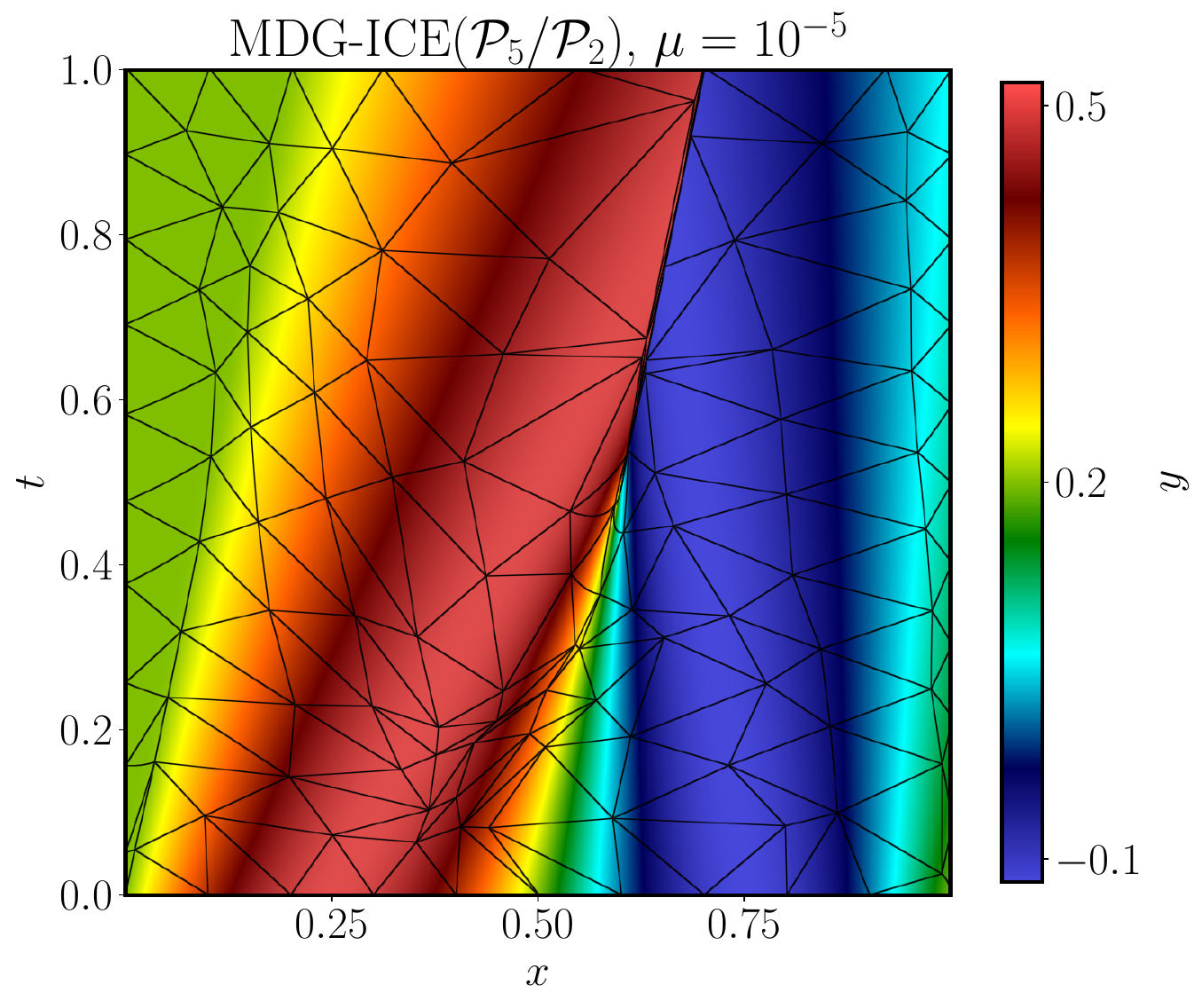}}
\par\end{centering}
\caption{\label{fig:burgers_unstructured_planes} The space-time Burgers initial
condition and final solutions for $\mu=10^{-3}$, $\mu=10^{-4}$,
and $\mu=10^{-5}$, with the corresponding grids superimposed. The
initial grid consists of 242 triangular elements of quadratic geometric
order with an irregular topology. The initial condition is given in
Equation~(\ref{eq:burgers_initial_condition}).}
\end{figure}

\begin{figure}[H]
\begin{centering}
\subfloat[\label{fig:burgers_unstructured_lines_mu_1e-3}$\mu=10^{-3}$ solution.]{\centering{}\includegraphics[width=0.48\columnwidth]{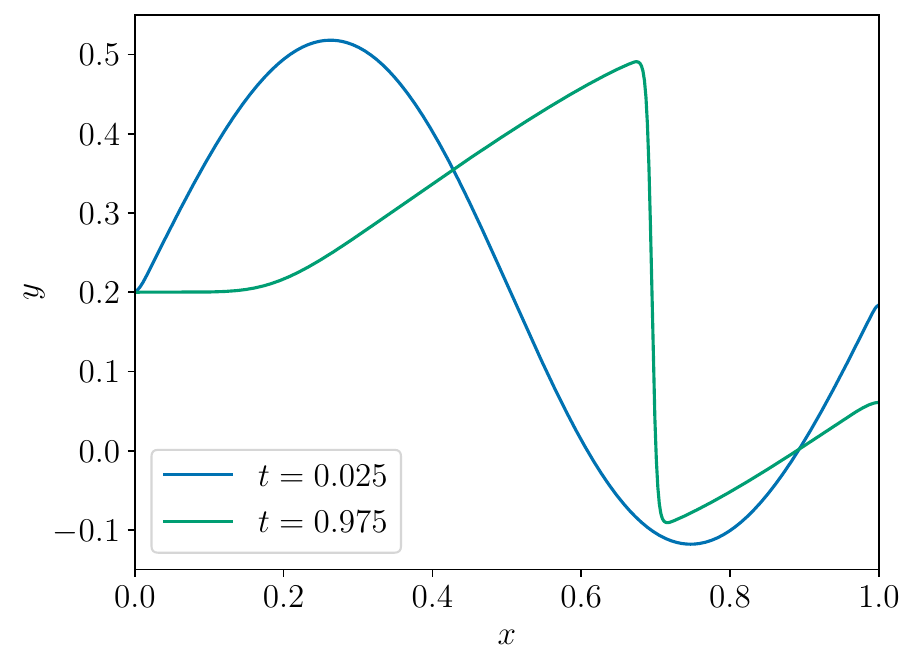}}\hfill{}\subfloat[\label{fig:burgers_unstructured_lines_mu_1e-4}$\mu=10^{-4}$ solution.]{\centering{}\includegraphics[width=0.48\columnwidth]{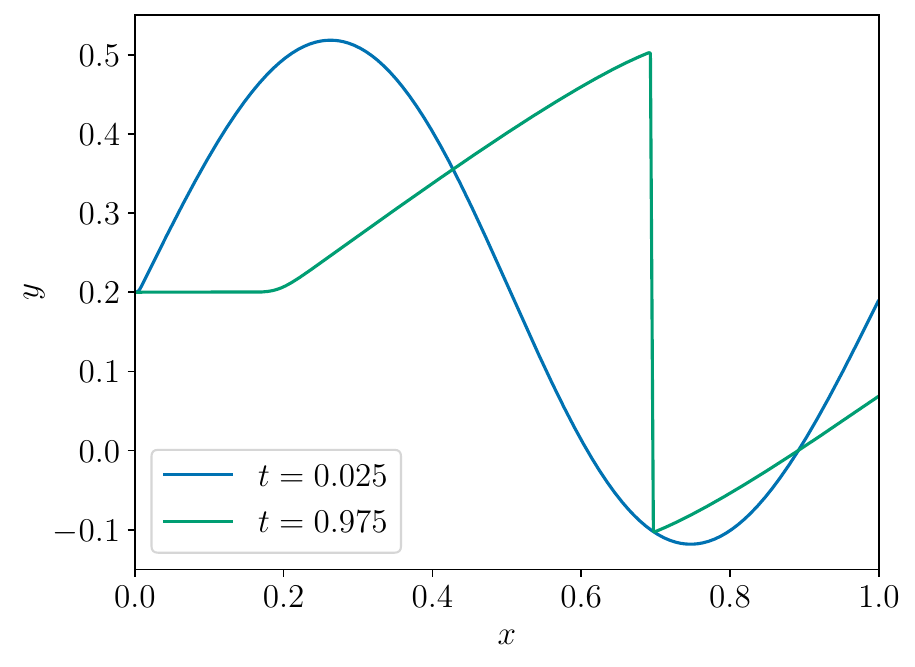}}\hfill{}\subfloat[\label{fig:burgers_unstructured_lines_mu_1e-5}$\mu=10^{-5}$ solution.]{\centering{}\includegraphics[width=0.48\columnwidth]{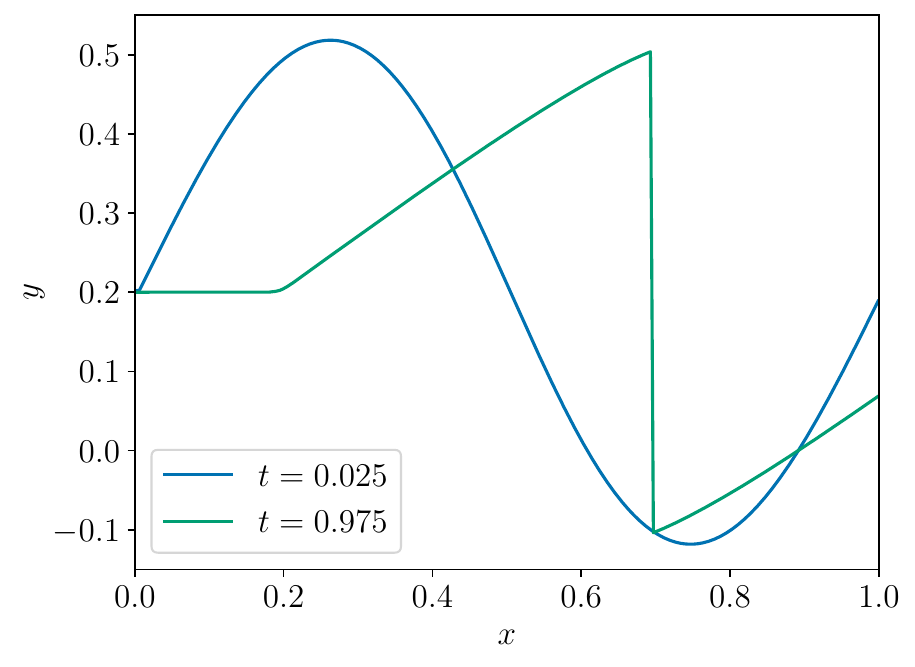}}
\par\end{centering}
\caption{\label{fig:burgers_unstructured_lines} One-dimensional profiles at
$t=0.025$ and $t=0.975$ for $\mu=10^{-3}$, $\mu=10^{-4}$, and
$\mu=10^{-5}$ for space-time Burgers viscous shock formation. The
initial grid consists of 242 triangular elements of quadratic geometric
order with an irregular topology. The initial condition is given in
Equation~(\ref{eq:burgers_initial_condition}).}
\end{figure}

\begin{figure}[ht]
\centering{}\includegraphics[clip,width=0.48\textwidth]{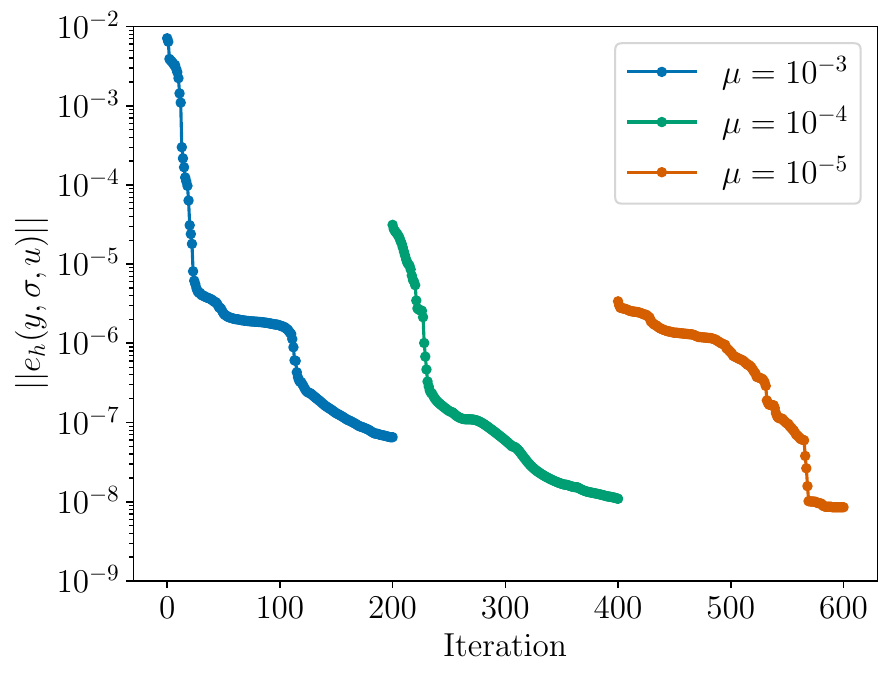}
\caption{Nonlinear convergence history for space-time Burgers viscous shock
formation. The initial grid consists of 242 triangular elements of
quadratic geometric order with an irregular topology. The initial
condition is given in Equation~(\ref{eq:burgers_initial_condition}).}
\label{fig:burgers_unstructured_residual}
\end{figure}

The initial 242-element irregular grid, along with the initial condition,
is displayed in Figure~\ref{fig:burgers_unstructured_planes_mu_1e-3_initial}.
The final solutions and grids for $\mu=10^{-3}$, $\mu=10^{-4}$,
and $\mu=10^{-5}$ are given in Figures~\ref{fig:burgers_unstructured_planes_mu_1e-3},~\ref{fig:burgers_unstructured_planes_mu_1e-4},
and~\ref{fig:burgers_unstructured_planes_mu_1e-5}, respectively.
Figure~\ref{fig:burgers_unstructured_lines} presents the corresponding
one-dimensional profiles at $t=0.025$ and $t=0.975$. Similar to
the regular-topology case, MDG-ICE automatically modifies the location,
size, and orientation of elements to resolve the viscous shock via
anisotropic space-time $r$-adaptivity. Additional grid adjustments
are introduced accordingly as the viscosity is decreased and the shock
becomes more difficult to resolve. The final solutions are well-resolved
and free from spurious oscillations. The nonlinear convergence history
is displayed in Figure~\ref{fig:burgers_unstructured_residual}.


\subsection{Mach 17.6 flow over two-dimensional cylinder\label{subsec:Viscous-Bow-Shock-Mach-17.6-2d}}

Next, we compute steady hypersonic viscous flow over a circular half-cylinder
in two spatial dimensions. The freestream Mach number and Reynolds
number (based on the cylinder radius) are 17.6 and $376,930$, respectively.
This problem is a common benchmark case for evaluating the ability
of numerical techniques to predict hypersonic flows. With conventional
finite volume techniques~\citep{Nom04,Gno04}, significant asymmetries
in the surface heat-flux profile were observed on both regular and
irregular simplicial grids. Note that fairly symmetric heating results
have been obtained on simplicial grids with DG schemes equipped with
smooth artificial viscosity~\citep{Bar10,Chi18}. In this work, we
aim to use the proposed MDG-ICE formulation to achieve symmetric solutions
without artificial dissipation. We employ two different starting grids:
the first is a regular grid with 392 elements, while the second is
an irregular grid with 526 elements. Freestream conditions are imposed
at the inflow boundary, defined as the ellipse $\left(\nicefrac{x}{6}\right)^{2}+\left(\nicefrac{y}{3}\right)^{2}=1$.
Extrapolation is employed at the outflow boundary, and the cylinder
boundary, a half-circle of unit radius, is an isothermal no-slip wall
with temperature $T_{\mathrm{wall}}=2.5T_{\infty}$.

\subsubsection{Initial mesh topology: Regular}

We employ continuation in both the Mach number and the Reynolds number,
starting with an isoparametric MDG-ICE($\mathcal{P}_{4}$) solution
corresponding to $\mathrm{Ma}=5,\mathrm{Re}=500$. This MDG-ICE solution
is initialized with a conventional DG solution stabilized with elementwise-constant
artificial viscosity of the form described in~\citep{Joh20}. Keeping
Re fixed at 500, the Mach number is consecutively increased by increments
of one until the target value of 17.6 is reached. These intermediate
solutions are not fully converged to a stationary point. Note that
the $\mathrm{Ma}=17.6,\mathrm{Re}=500$ solution was computed in previous
work~\citep{Chi22_ICCFD} prior to development of the adaptive, elementwise
regularization strategy (Section~\ref{subsec:adaptive-elementwise-regularization})
and the increment-limiting procedure (Section~\ref{subsec:increment-limiting});
cell degeneration was treated then using a standard cell-collapse
algorithm~\citep{Loh08}, resulting in a total of 388 elements. The
388-element mesh and temperature field for this $\mathrm{Ma}=17.6,\mathrm{Re}=500$
solution are presented in Figure~\ref{fig:Viscous-Bow-Shock-Mach-17.6-Re-500}.
The elements in the vicinity of the shock are anisotropic yet still
valid. The MDG-ICE solution is free from oscillations even in the
absence of artificial dissipation.
\begin{center}
\begin{figure}[H]
\begin{centering}
\subfloat[\label{fig:Viscous-Bow-Shock-Mach-17.6-Re-500-Mesh}Mesh]{\centering{}\includegraphics[clip,width=0.75\textwidth,keepaspectratio]{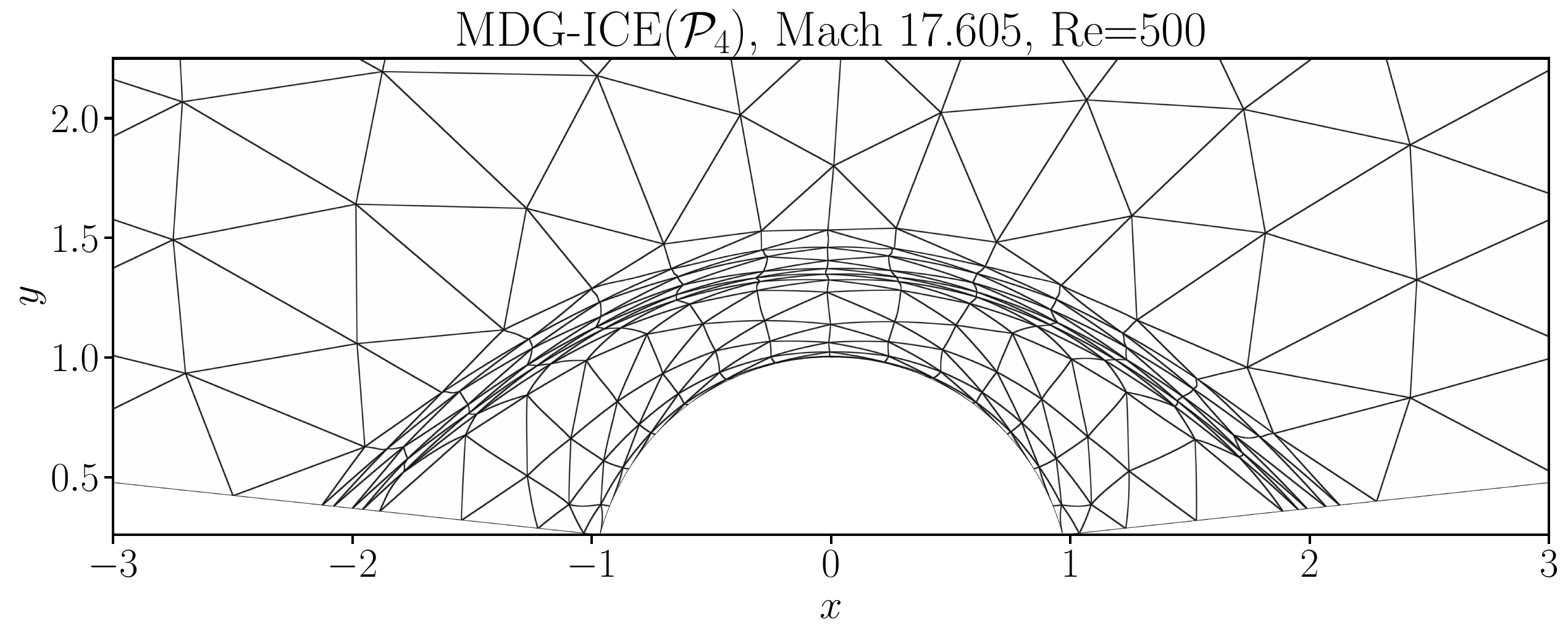}}\hfill{}\subfloat[\label{fig:Viscous-Bow-Shock-Mach-17.6-Re-500-Temperature}Temperature
field]{\centering{}\includegraphics[clip,width=0.75\textwidth,keepaspectratio]{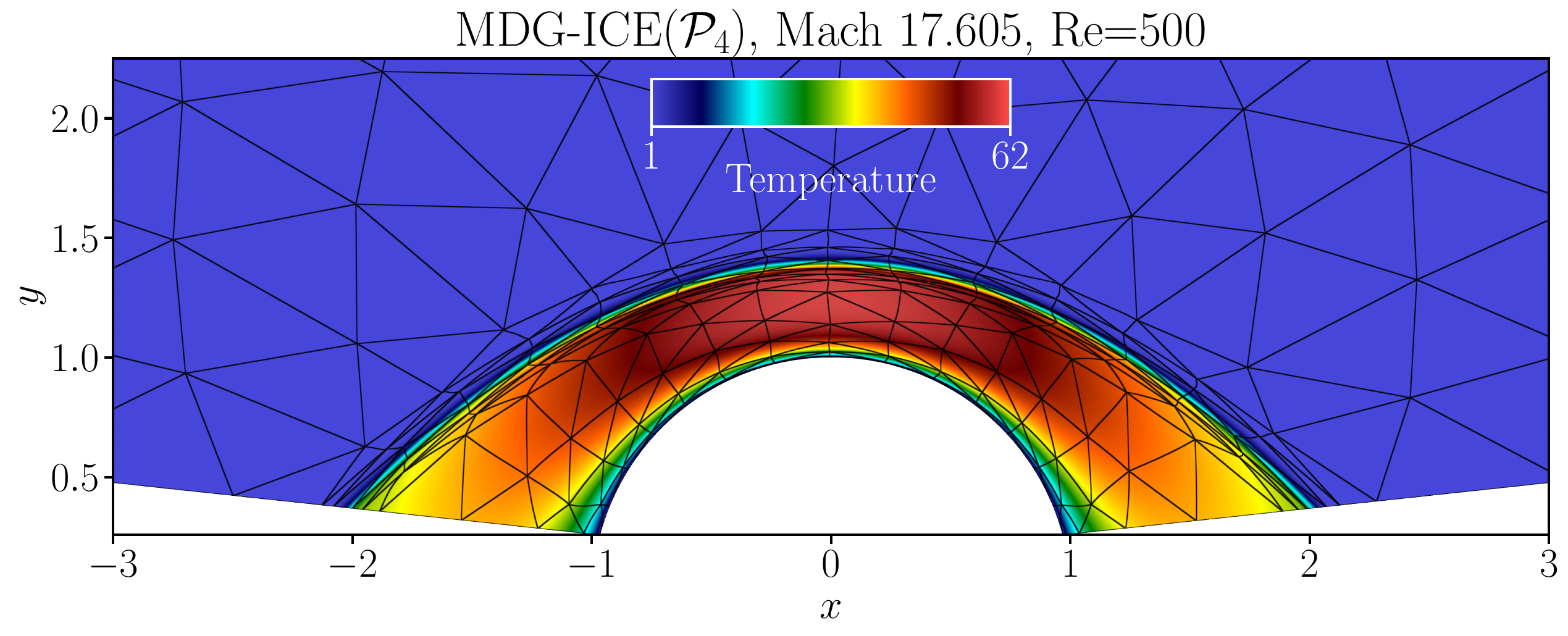}}
\par\end{centering}
\caption{\label{fig:Viscous-Bow-Shock-Mach-17.6-Re-500}The MDG-ICE solution
computed using 388 $\mathcal{P}_{4}$ isoparametric triangle elements
for two-dimensional Mach 17.6 flow over a cylinder at $\mathrm{Re}=500$.
The initial grid has a regular topology.}
\end{figure}
\par\end{center}

\begin{figure}[ht]
\centering{}\includegraphics[clip,width=0.48\textwidth]{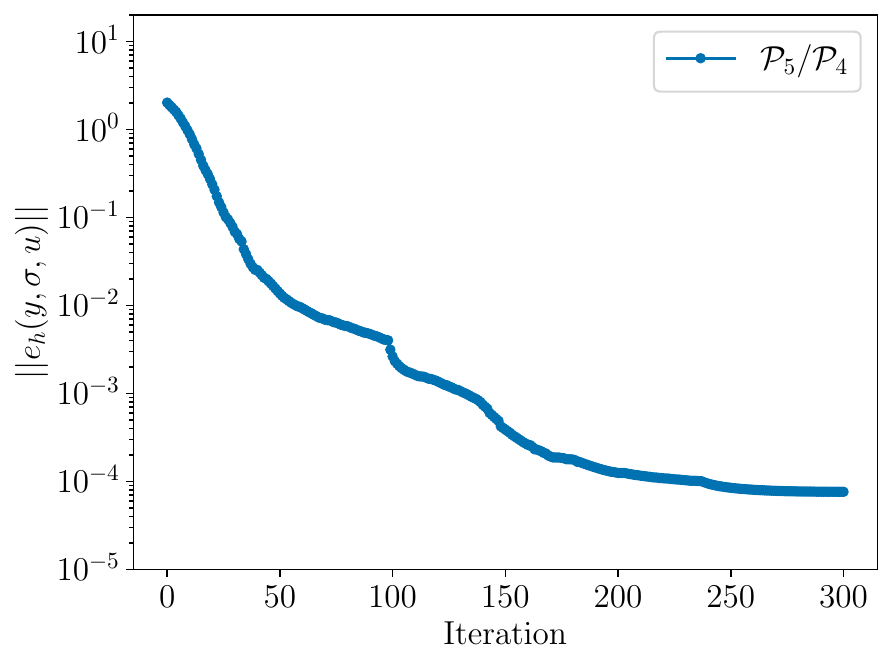}
\caption{Nonlinear convergence history for the subparametric MDG-ICE($\mathcal{P}_{5}/\mathcal{P}_{4}$)
solution to Mach 17.6 flow over a two-dimensional cylinder. The initial
grid has a regular topology.}
\label{fig:cylinder-2d-structured-residual}
\end{figure}

The Reynolds number is then consecutively increased by factors of
approximately two until the target value of $376,930$ is reached.
At this Reynolds number, we increase the polynomial degrees of $Y_{h}$
and $\Sigma_{h}$ from four to five in order to eliminate slight asymmetries
in the surface heat-flux profile, as will be shown later in this section.
The final mesh and temperature, Mach, and pressure fields are given
in Figure~\ref{fig:Viscous-Bow-Shock-Mach-17.6-Re-3.77e5}. Considerable
jumps in temperature and pressure are observed. MDG-ICE automatically
adjusts the grid geometry in order to resolve the viscous shock and
boundary layer, which become sharper as the Reynolds number is increased.
Highly anisotropic elements are observed at the shock and in the boundary
layer, yet grid validity is maintained as a result of the anisotropic
Laplacian-type regularization and the adaptive, elementwise regularization
strategy. The solution is well-resolved and free from spurious artifacts.
Despite the initially regular nature of the mesh, there is undoubtedly
very strong misalignment between the grid and the shock and boundary
layer. Figure~\ref{fig:cylinder-2d-structured-residual} displays
the nonlinear convergence history for the subparametric MDG-ICE($\mathcal{P}_{5}/\mathcal{P}_{4}$)
solution. The residual starts at a relatively small value since the
simulation is restarted from an isoparametric MDG-ICE($\mathcal{P}_{4}$)
solution. Future work will focus on incorporating stabilization mechanisms
(at least during early/intermediate iterations) and/or space-time
marching in order to accelerate convergence and circumvent the need
for continuation in Mach number and Reynolds number, which itself
can be considered akin to (global) artificial dissipation. Nevertheless,
this demonstrates how MDG-ICE can be naturally employed for parametric
studies in which the flow conditions are varied; other grid adaptation
strategies may need to incorporate coarsening techniques to maintain
efficiently refined grids.
\begin{center}
\begin{figure}[H]
\begin{centering}
\subfloat[\label{fig:Viscous-Bow-Shock-Mach-17.6-Re-3.77e5-Mesh}Mesh]{\centering{}\includegraphics[clip,width=0.75\textwidth]{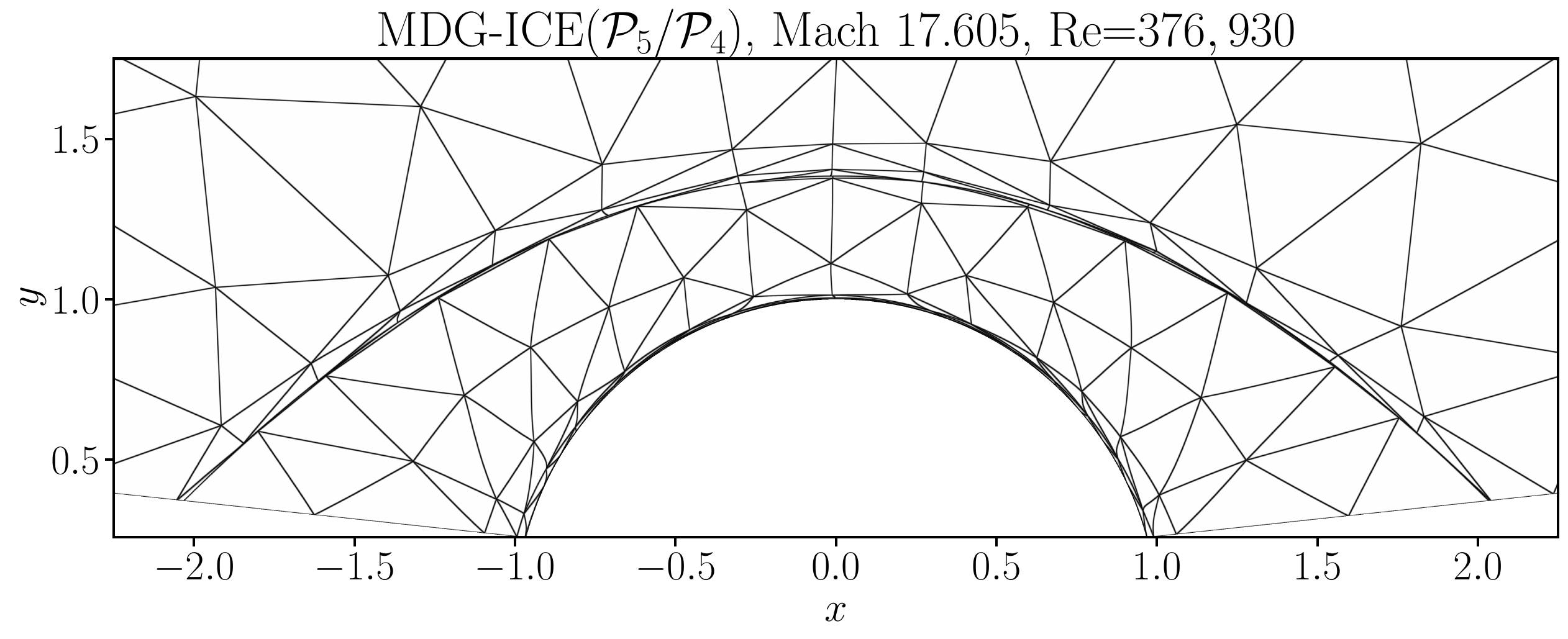}}\hfill{}\subfloat[\label{fig:Viscous-Bow-Shock-Mach-17.6-Re-3.77e5-Temperature}Temperature
field]{\centering{}\includegraphics[clip,width=0.75\textwidth]{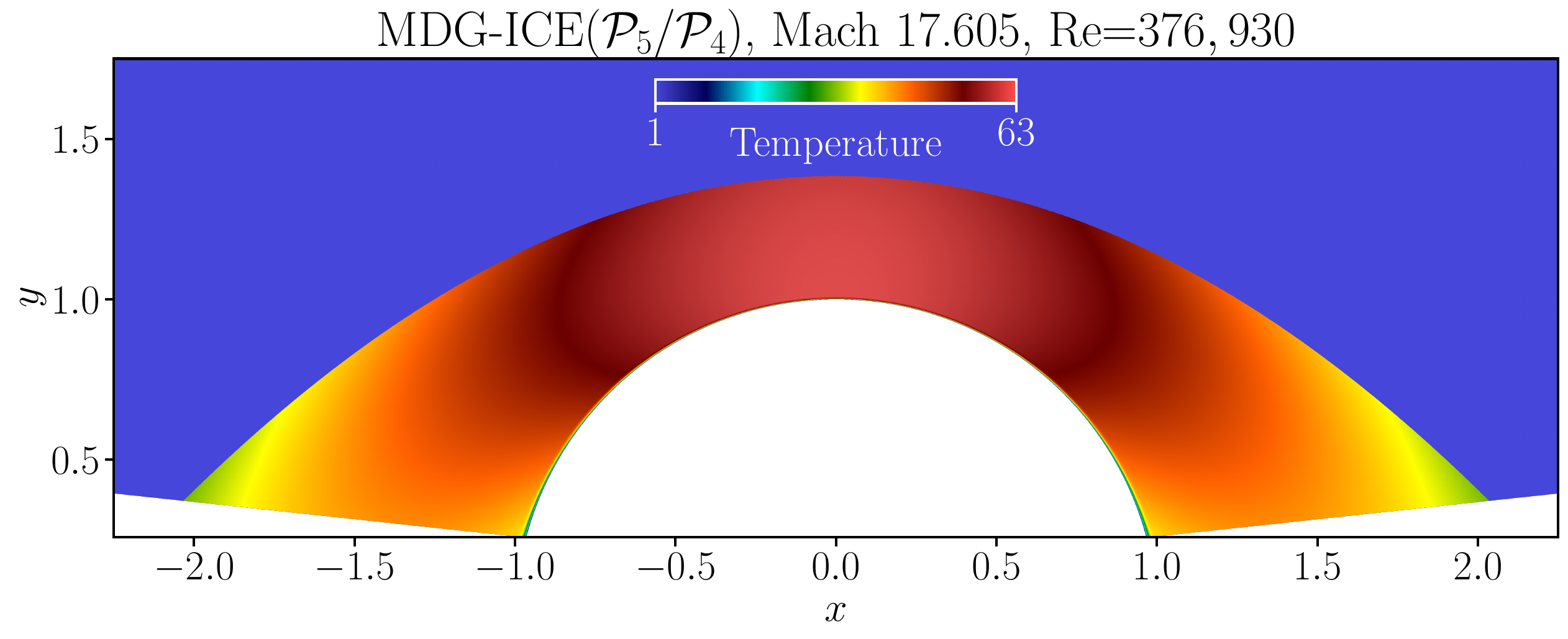}
}\hfill{}\subfloat[\label{fig:Viscous-Bow-Shock-Mach-17.6-Re-3.77e5-Pressure}Pressure
field]{\centering{}\includegraphics[clip,width=0.75\textwidth]{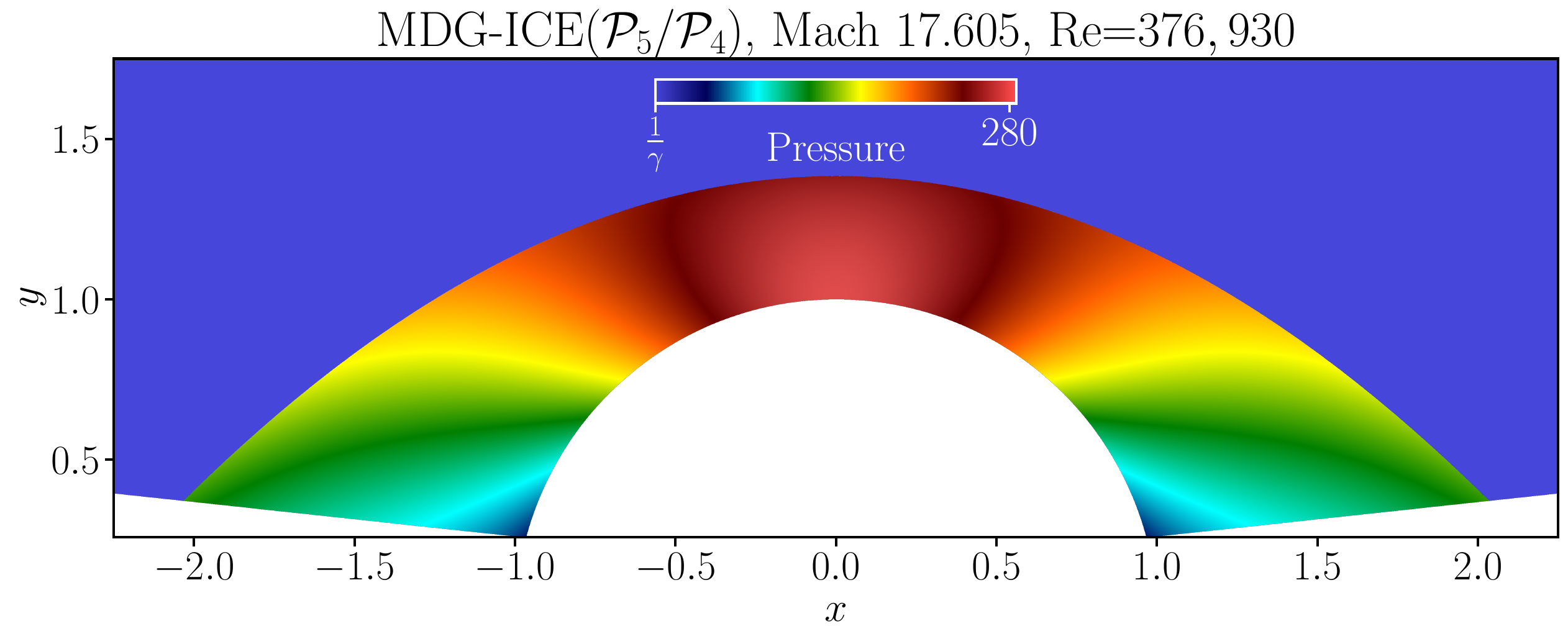}}
\par\end{centering}
\caption{\label{fig:Viscous-Bow-Shock-Mach-17.6-Re-3.77e5} The MDG-ICE solution
computed using 388 subparametric $\mathcal{P}_{5}/\mathcal{P}_{4}$
triangle elements for two-dimensional Mach 17.6 flow over a cylinder
at $\mathrm{Re}=376,930$. The initial grid has a regular topology.}
\end{figure}
\par\end{center}

\begin{figure}[H]
\subfloat[\label{fig:Bow-Shock-Mach-17.6-PressureCoefficient}Pressure coefficient
profile.]{\centering{}\includegraphics[clip,width=0.48\textwidth]{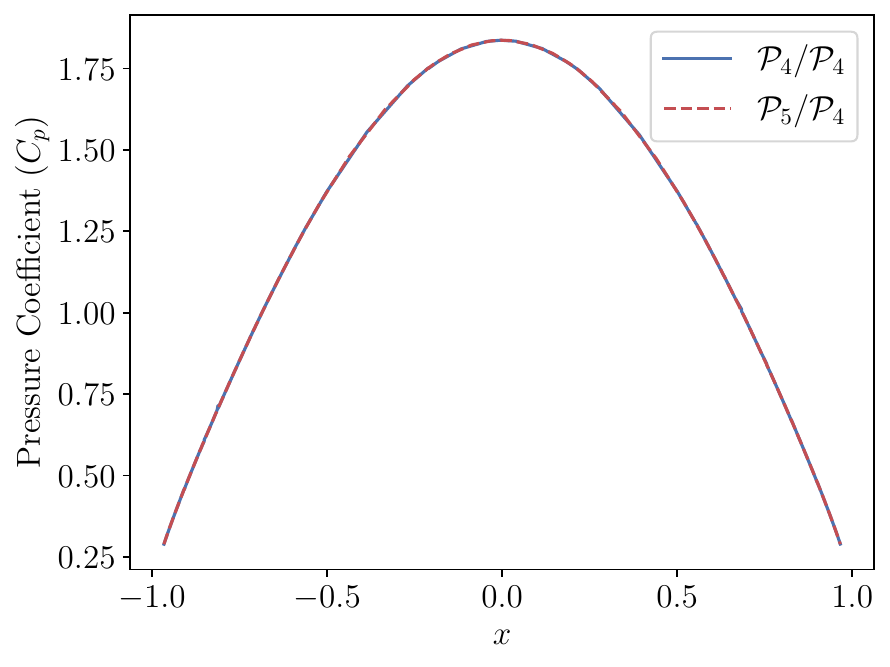}
}\hfill{}\subfloat[\label{fig:Bow-Shock-Mach-17.6-StantonNumber}Stanton number profile.]{\centering{}\includegraphics[clip,width=0.48\textwidth]{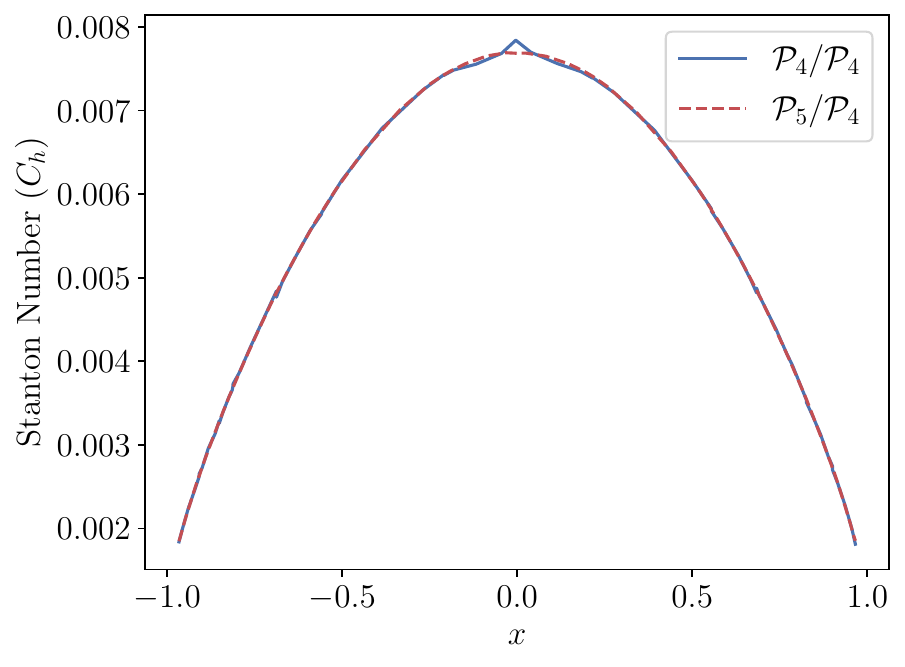}
}

\caption{\label{fig:Bow-Shock-Mach-17.6-Surface-Quantities} Surface profiles
of pressure coefficient and Stanton number for the MDG-ICE solution
computed using 388 $\mathcal{P}_{5}/\mathcal{P}_{4}$ triangle elements
for for two-dimensional Mach 17.6 flow over a cylinder at $\mathrm{Re}=376,930$.
The initial grid has a regular topology.}
\end{figure}

Figure~\ref{fig:Bow-Shock-Mach-17.6-Surface-Quantities} presents
the isoparametric MDG-ICE($\mathcal{P}_{4}$) and subparametric MDG-ICE($\mathcal{P}_{5}/\mathcal{P}_{4}$)
predictions of the pressure coefficient and Stanton number, defined
as
\[
C_{p}=\frac{P-P_{\infty}}{\frac{1}{2}\rho_{\infty}v_{\infty}^{2}}
\]
and
\[
C_{h}=\frac{q_{n}}{c_{p}\rho_{\infty}v_{\infty}\left(T_{t,\infty}-T_{\mathrm{wall}}\right)},
\]
respectively, where $q_{n}$ is the normal heat flux and $T_{t}$
is the stagnation temperature. The pressure profiles for both MDG-ICE
solutions are highly symmetric, but the heat-flux profile for the
isoparametric MDG-ICE($\mathcal{P}_{4}$) solution exhibits a slight
nonphysical cusp at the stagnation point. Nevertheless, this cusp
is eliminated with $p$-refinement. The stagnation-point Stanton number
in the MDG-ICE($\mathcal{P}_{5}/\mathcal{P}_{4}$) solution is approximately
0.0077. Note that there exists some variation in the stagnation-point
Stanton number reported in the literature; for example, 0.0085 in~\citep{Gno04},
0.0076 in~\citep{Kit13_2}, and 0.0082 in~\citep{Bar10}.

\subsubsection{Initial mesh topology: Irregular}

Continuation in the Mach and Reynolds numbers is once again employed,
starting with $\mathrm{Ma}=5,\mathrm{Re}=100$. The 526-element mesh
and temperature field for an intermediate MDG-ICE solution at $\mathrm{Ma}=5,\mathrm{Re}=100$
are presented in Figure~\ref{fig:Viscous-Bow-Shock-Mach-17.6-Re-100-unstructured}.
The temperature field is smooth. At this low Reynolds number, only
slight grid repositioning is required to resolve the flow.
\begin{center}
\begin{figure}[H]
\begin{centering}
\subfloat[\label{fig:Viscous-Bow-Shock-Mach-17.6-Re-100-Mesh-unstructured}Mesh]{\centering{}\includegraphics[clip,width=0.75\textwidth]{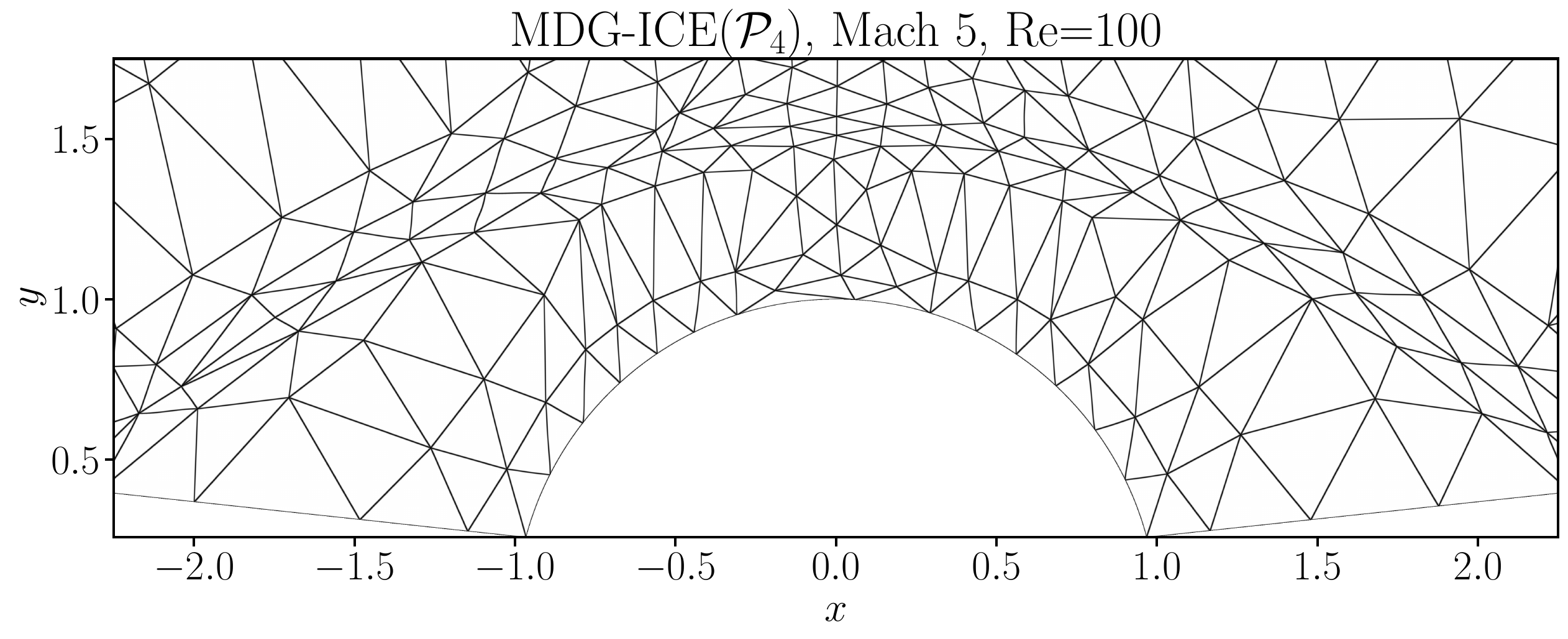}}\hfill{}\subfloat[\label{fig:Viscous-Bow-Shock-Mach-17.6-Re-100-Temperature-unstructured}Temperature
field]{\centering{}\includegraphics[clip,width=0.75\textwidth]{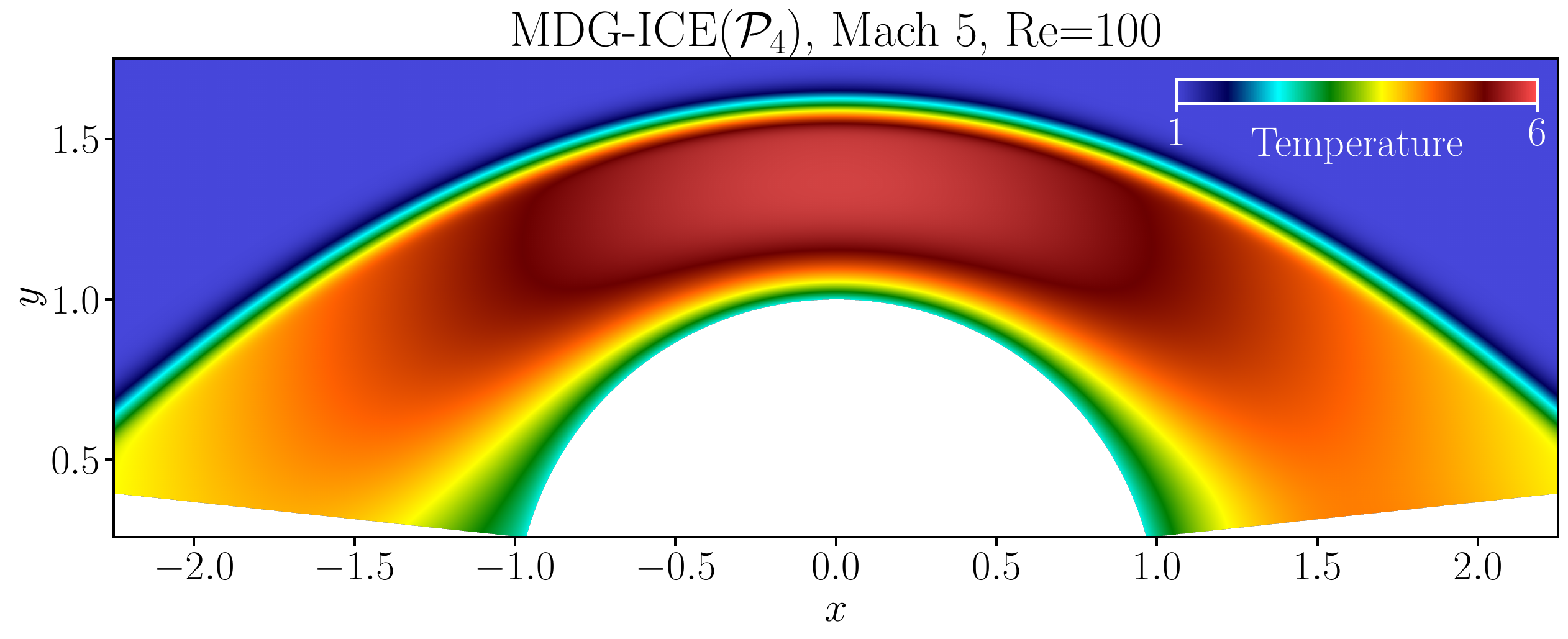}}
\par\end{centering}
\caption{\label{fig:Viscous-Bow-Shock-Mach-17.6-Re-100-unstructured}The MDG-ICE
solution computed using 526 $\mathcal{P}_{4}$ isoparametric triangle
elements for for two-dimensional Mach 17.6 flow over a cylinder at
$\mathrm{Re}=100$. The initial grid has an irregular topology.}
\end{figure}
\par\end{center}

Global $p$-refinement of the state and auxiliary-variable approximations
is again employed. The final subparametric MDG-ICE($\mathcal{P}_{5}/\mathcal{P}_{4}$)
solution at $\mathrm{Ma}=17.6,\mathrm{Re}=376,930$ is given in Figure~\ref{fig:Viscous-Bow-Shock-Mach-17.6-Re-3.77e5-unstructured}.
The long, thin elements oriented orthogonal to the shock are a direct
consequence of two main factors associated with the initially irregular
nature of the grid: (a) moreso than in the regular case, incrementing
the Mach and Reynolds numbers (in the absence of additional stabilization)
causes transient artifacts to appear upstream of the shock, which
then induce appreciable grid motion; (b) the compression of the grid
at the shock and boundary layer is less likely to pull in outlying
elements accordingly than in the regular case. Note that starting
with intermediate Reynolds numbers, $\alpha$ is set to zero in order
to alleviate excessive stiffening of the anisotropic, shock-orthogonal
elements due to the Laplacian regularization~(\ref{eq:regularization_laplacian-new}).
Although it is not clear how detrimental this type of element may
be in more complex configurations (if at all), the formation of such
elements can be mitigated by, as previously discussed, the use of
artificial dissipation during intermediate iterations and/or metric-based
remeshing. Nevertheless, the thin viscous structures remain sharp
and free from spurious oscillations.
\begin{center}
\begin{figure}[H]
\begin{centering}
\subfloat[\label{fig:Viscous-Bow-Shock-Mach-17.6-Re-3.77e5-Mesh-unstructured}Mesh]{\centering{}\includegraphics[clip,width=0.75\textwidth]{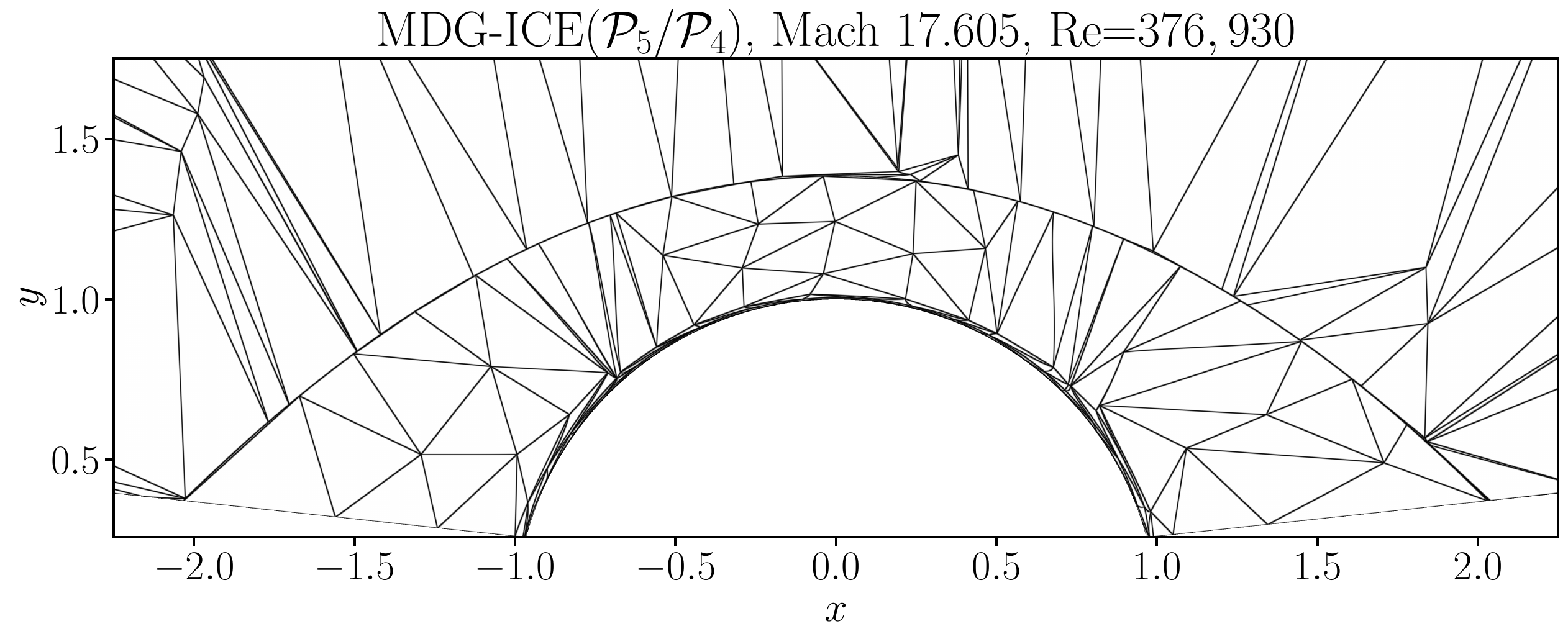}}\hfill{}\subfloat[\label{fig:Viscous-Bow-Shock-Mach-17.6-Re-3.77e5-Temperature-unstructured}Temperature
field]{\centering{}\includegraphics[clip,width=0.75\textwidth]{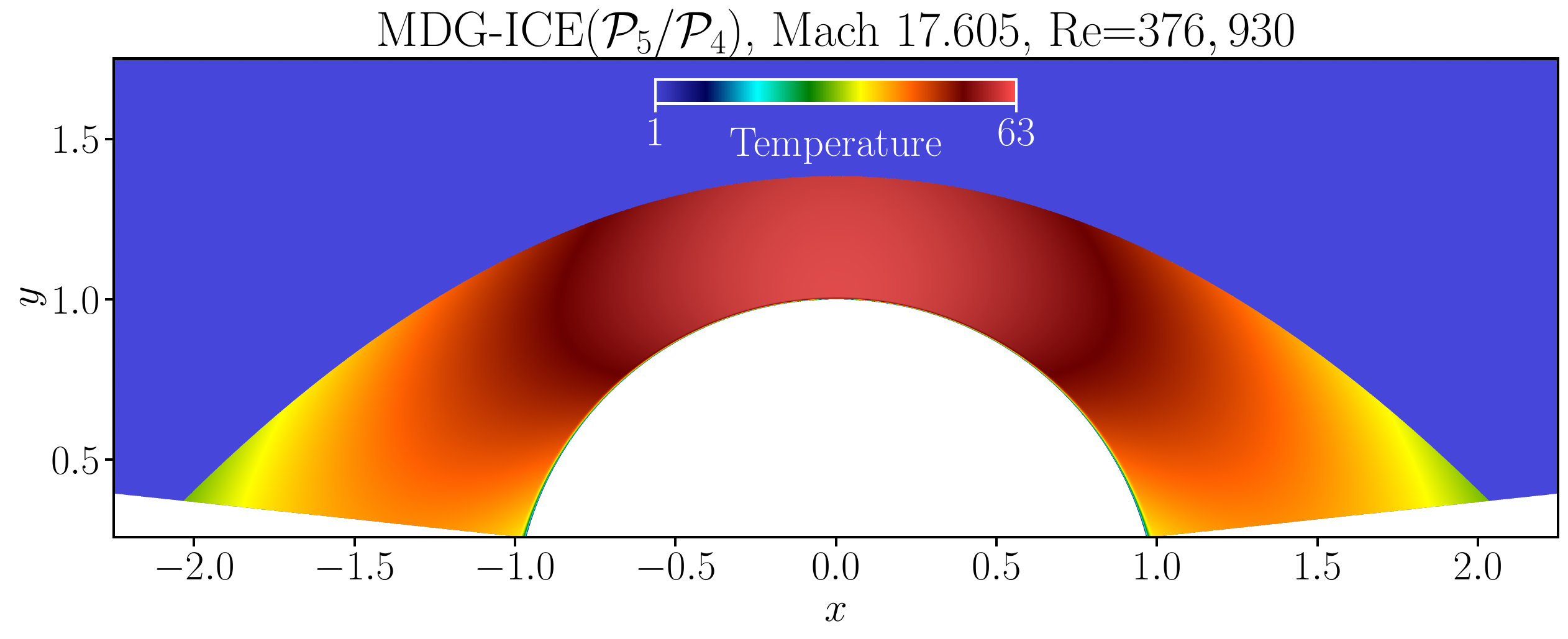}
}\hfill{}\subfloat[\label{fig:Viscous-Bow-Shock-Mach-17.6-Re-3.77e5-Pressure-unstructured}Pressure
field]{\centering{}\includegraphics[clip,width=0.75\textwidth]{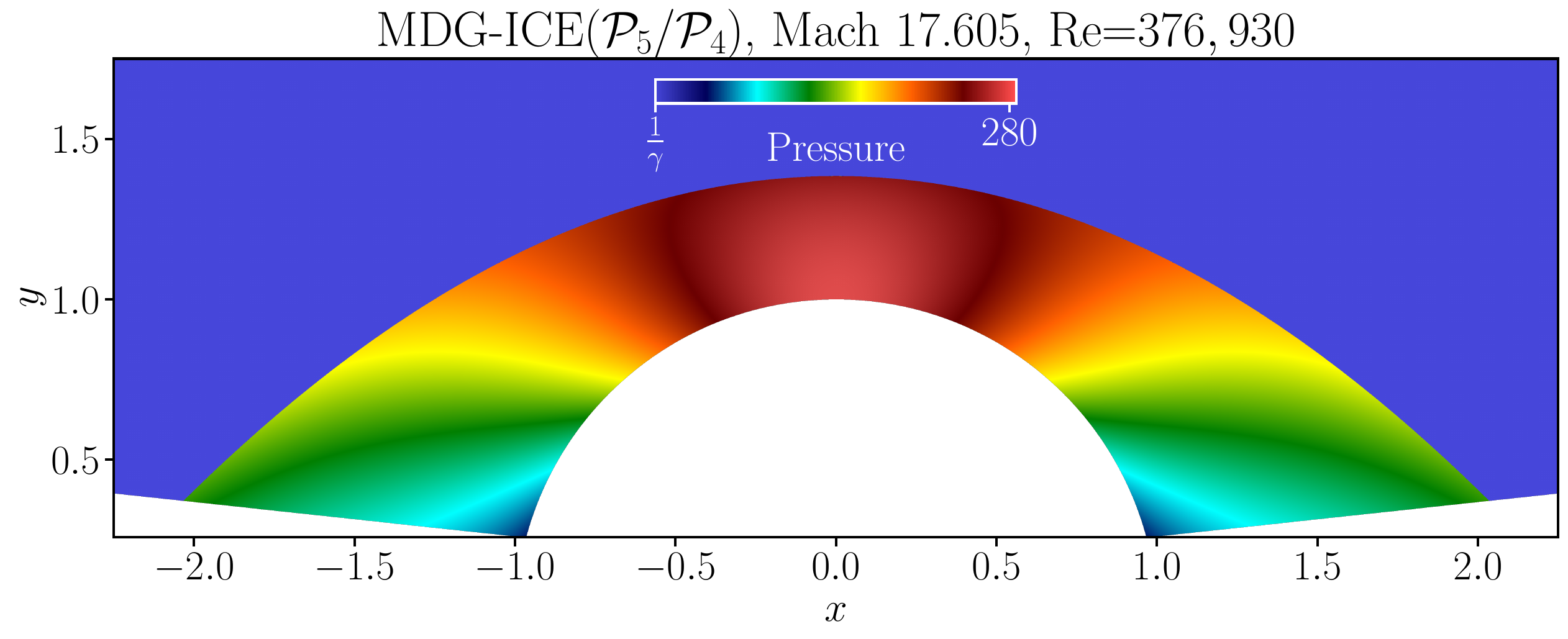}}
\par\end{centering}
\caption{\label{fig:Viscous-Bow-Shock-Mach-17.6-Re-3.77e5-unstructured} The
MDG-ICE solution computed using 526 $\mathcal{P}_{5}/\mathcal{P}_{4}$
subparametric triangle elements for two-dimensional Mach 17.6 flow
over a cylinder at $\mathrm{Re}=376,930$. The initial grid has an
irregular topology.}
\end{figure}
\par\end{center}

Figure~\ref{fig:cylinder-2d-structured-residual} displays the nonlinear
convergence history for the subparametric MDG-ICE($\mathcal{P}_{5}/\mathcal{P}_{4}$)
solution. The initial residual is already low since the simulation
is restarted from an isoparametric MDG-ICE($\mathcal{P}_{4}$) solution.

\begin{figure}[ht]
\centering{}\includegraphics[clip,width=0.48\textwidth]{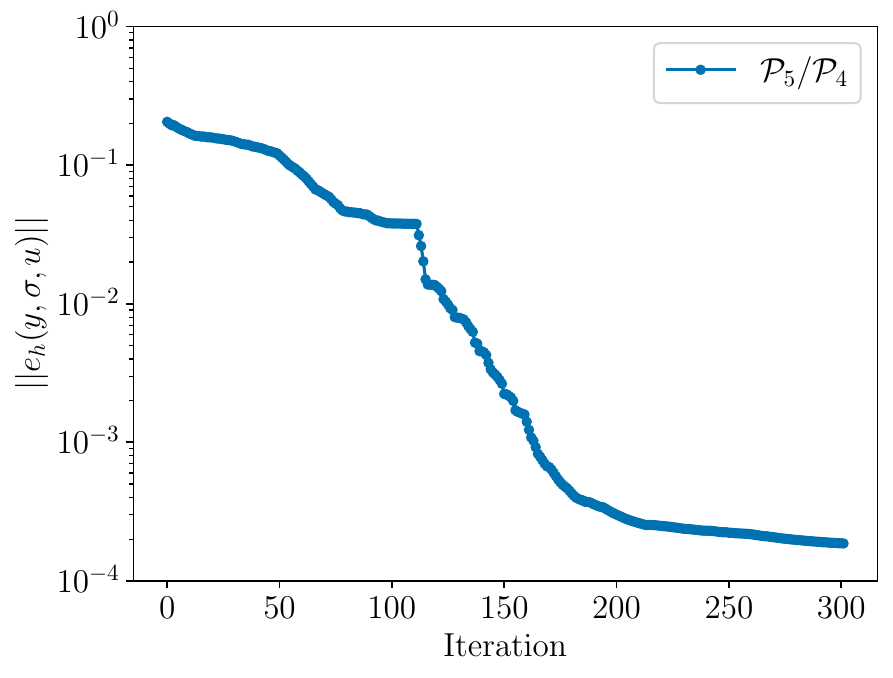}
\caption{Nonlinear convergence history for the subparametric MDG-ICE($\mathcal{P}_{5}/\mathcal{P}_{4}$)
solution to Mach 17.6 flow over a two-dimensional cylinder. The initial
grid has an irregular topology.}
\label{fig:cylinder-2d-unstructured-residual}
\end{figure}

\begin{figure}[H]
\subfloat[\label{fig:Bow-Shock-Mach-17.6-PressureCoefficient-unstructured}Pressure
coefficient profile.]{\centering{}\includegraphics[clip,width=0.48\textwidth]{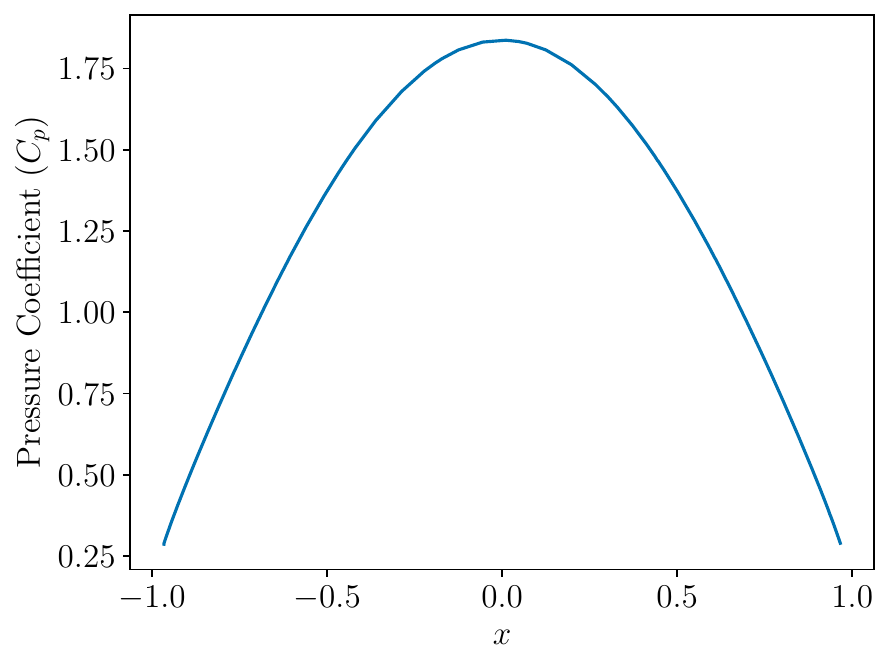}
}\hfill{}\subfloat[\label{fig:Bow-Shock-Mach-17.6-StantonNumber-unstructured}Stanton
number profile.]{\centering{}\includegraphics[clip,width=0.48\textwidth]{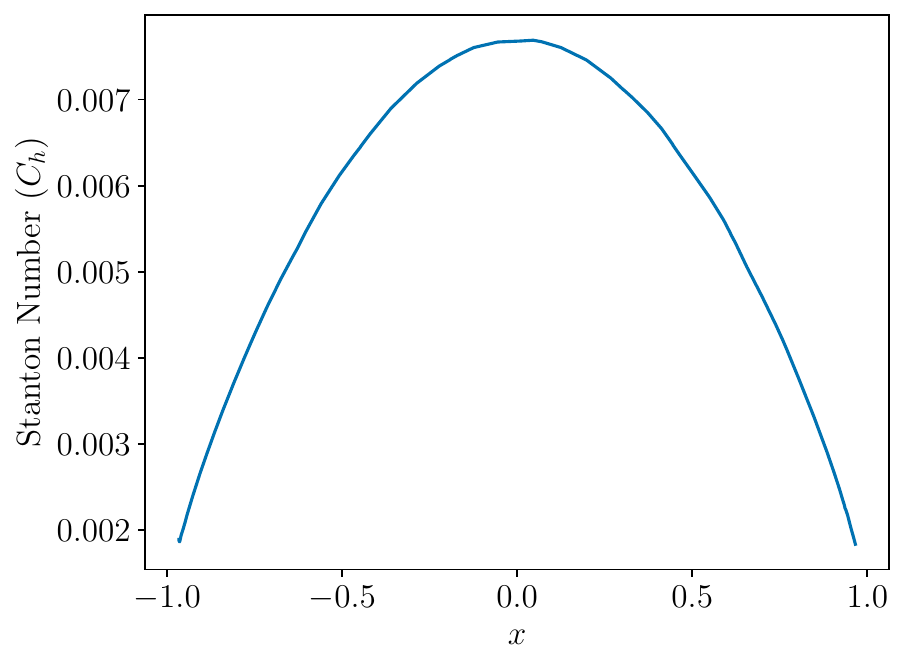}
}

\caption{\label{fig:Bow-Shock-Mach-17.6-Surface-Quantities-unstructured} Surface
profiles of pressure coefficient and Stanton number obtained with
the MDG-ICE solution computed using 526 $\mathcal{P}_{5}/\mathcal{P}_{4}$
subparametric triangle elements for two-dimensional Mach 17.6 flow
over a cylinder at $\mathrm{Re}=376,930$. The initial grid has an
irregular topology.}
\end{figure}

Figure~\ref{fig:Bow-Shock-Mach-17.6-Surface-Quantities-unstructured}
presents the surface profiles of pressure coefficient and Stanton
number for the subparametric MDG-ICE($\mathcal{P}_{5}/\mathcal{P}_{4}$)
solution. The stagnation-point Stanton number is approximately 0.0077.
Despite the evidently asymmetric grid and extremely strong grid-shock
misalignment, the surface profiles are highly symmetric and agree
well with those in Figure~(\ref{fig:Bow-Shock-Mach-17.6-Surface-Quantities}).

\subsection{Mach 17.6 flow over three-dimensional cylinder}

This test case is the three-dimensional version of the previous configuration.
The consideration of three spatial dimensions allows for the emergence
of instabilities and asymmetries that are naturally suppressed in
the two-dimensional case. In particular, it is with a three-dimensional
domain that nonphysical artifacts in the heat-flux profiles obtained
with finite volume schemes are most prominent~\citep{Gno04,Nom04}.
The two-dimensional domain is extruded one layer of cells in the homogeneous
direction by two (nondimensional) units. Slip conditions are applied
at the symmetry boundaries. Unlike in the previous subsection, the
inflow boundary is defined as the circle $x^{2}+y^{2}=6.5^{2}$. The
reason for this modification is that as the Mach number is increased
during the continuation strategy, spurious transients appear upstream
of the shock and interact with the boundary, causing noticeable instabilities
that can be dampened by positioning the inflow boundary far away from
the shock. Although also present in the two-dimensional simulations,
these instabilities are exacerbated in the three-dimensional setting.
Again, the use of artificial dissipation would likely resolve this
issue. As in the two-dimensional case, we employ two different starting
grids: the first is a regular grid with 3024 elements, while the second
is an irregular grid with 3183 elements. 

\subsubsection{Initial mesh topology: Regular}

\begin{figure}[ht]
\centering{}\includegraphics[clip,width=0.75\textwidth]{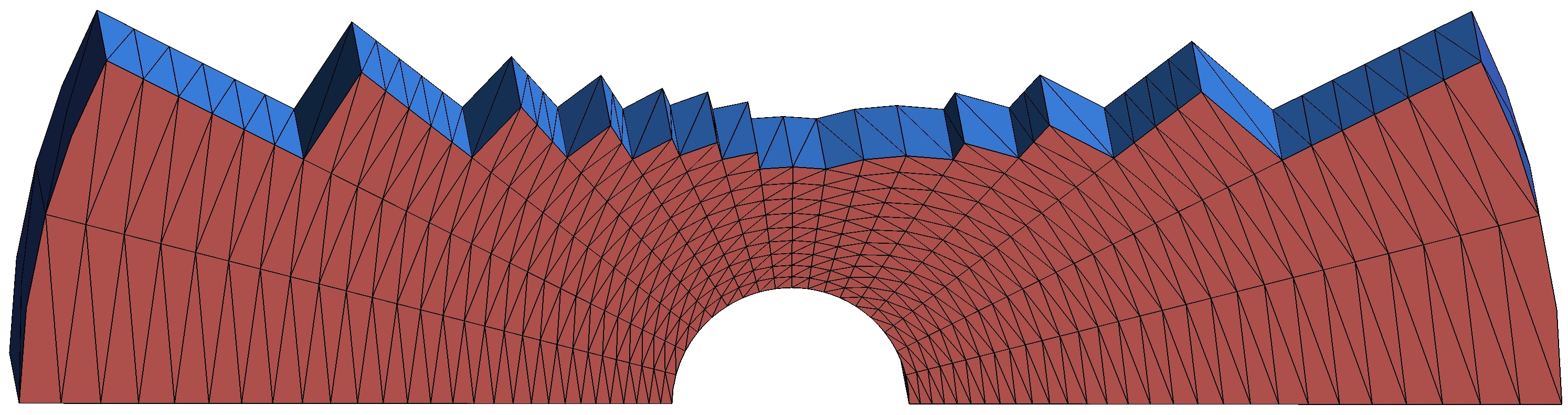}
\caption{Clipped, three-dimensional perspective of the initial grid with a
regular topology.}
\label{fig:cylinder-3d-DG-initial-grid-structured}
\end{figure}

\begin{figure}[ht]
\centering{}\includegraphics[clip,width=0.75\textwidth]{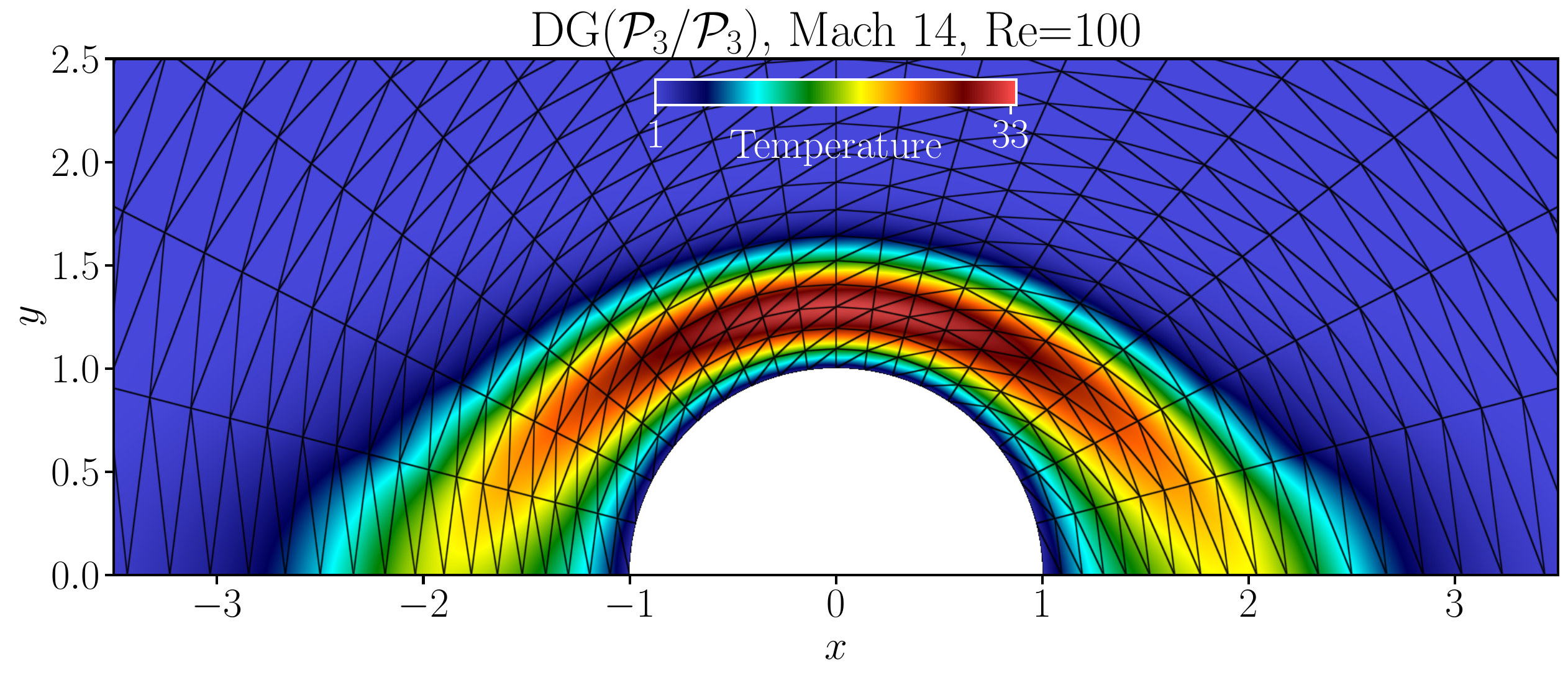}
\caption{Isoparametric DG($\mathcal{P}_{3}$) solution at $\mathrm{Ma}=14,\mathrm{Re}=100$
on 3024 tetrahedral cells along the $z=2$ symmetry plane. This solution
is used as the initial condition for the MDG-ICE continuation strategy.
The initial grid has a regular topology.}
\label{fig:cylinder-3d-DG-initialization-structured}
\end{figure}

Figure~\ref{fig:cylinder-3d-DG-initial-grid-structured} provides
a clipped, three-dimensional perspective of the initial grid. An isoparametric
DG($\mathcal{P}_{3}$) solution at $\mathrm{Ma}=14,\mathrm{Re}=100$
is used as the initial condition for the MDG-ICE continuation with
superparametric $\mathcal{P}_{3}/\mathcal{P}_{4}$ cells. A positivity-preserving
and entropy-based linear-scaling limiter~\citep{Lv15,Jia18,Chi22}
is employed to help maintain stability in the DG solution. The temperature
field and initial grid for the DG solution along the $z=2$ symmetry
plane are presented in Figure~\ref{fig:cylinder-3d-DG-initialization-structured}.
Note the coarseness of the grid with respect to the expected high-gradient
features at the target conditions. After reaching $\mathrm{Ma}=17.6,\mathrm{Re}=376,930$,
two global $p$-refinements of the state and auxiliary-variable approximations
are performed. The final subparametric MDG-ICE($\mathcal{P}_{5}/\mathcal{P}_{4}$)
solution is displayed in Figure~\ref{fig:cylinder-3d-DG-initialization-structured}.
Just as in the two-dimensional case, the grid is automatically adapted
to resolve the viscous shock and boundary layer while maintaining
grid validity. Figure~\ref{fig:Viscous-Bow-Shock-Mach-17.6-Re-3.77e5-3D-zoom-structured}
zooms in on the shock layer along the stagnation line. Extremely high-aspect-ratio
elements at the shock and boundary layer are observed. In particular,
the cells at the shock are almost visually indistinguishable. Furthermore,
in the temperature field, although the smoothness of the boundary
layer can be discerned, the viscous shock resembles a truly discontinuous
feature.

\begin{center}
\begin{figure}[H]
\begin{centering}
\subfloat[\label{fig:Viscous-Bow-Shock-Mach-17.6-Re-3.77e5-Mesh-3D-structured}Mesh]{\centering{}\includegraphics[clip,width=0.75\textwidth]{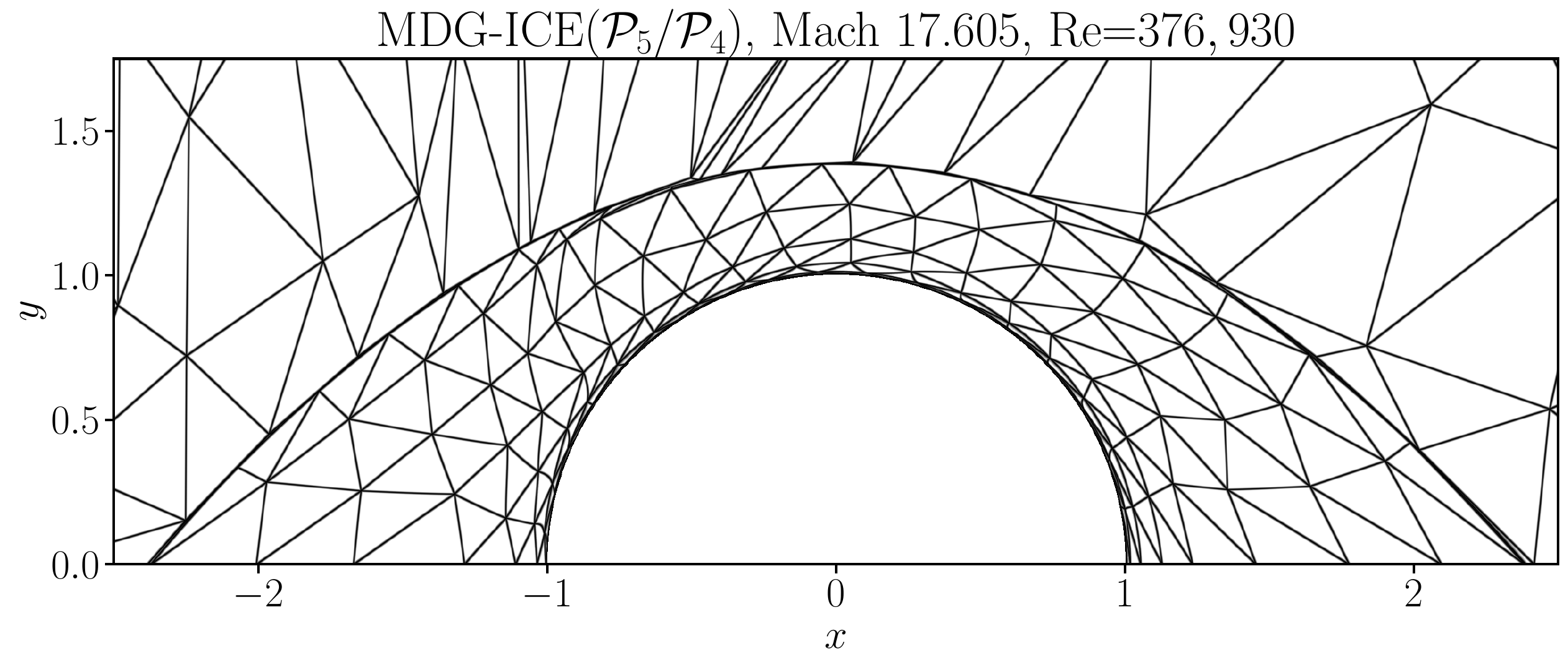}}\hfill{}\subfloat[\label{fig:Viscous-Bow-Shock-Mach-17.6-Re-3.77e5-Temperature-3D-structured}Temperature
field]{\centering{}\includegraphics[clip,width=0.75\textwidth]{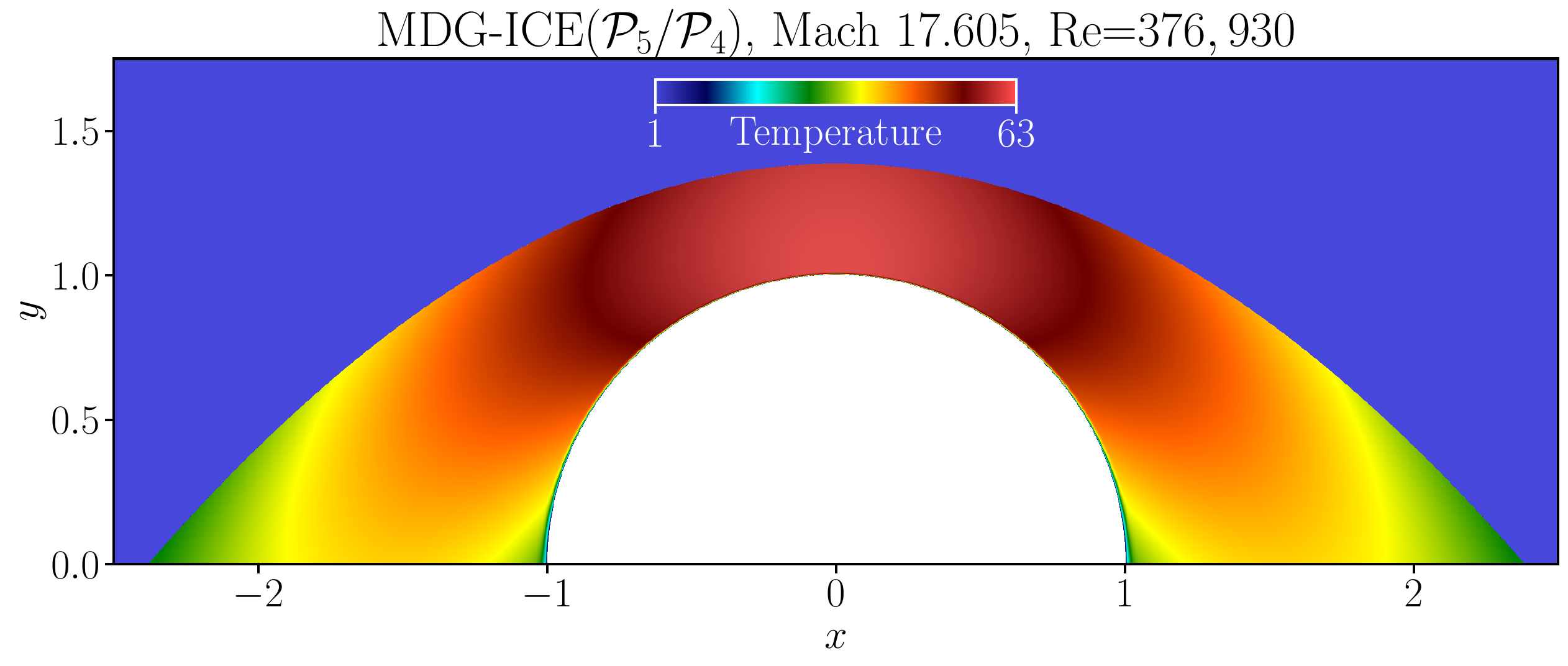}
}\hfill{}\subfloat[\label{fig:Viscous-Bow-Shock-Mach-17.6-Re-3.77e5-Pressure-3D-structured}Pressure
field]{\centering{}\includegraphics[clip,width=0.75\textwidth]{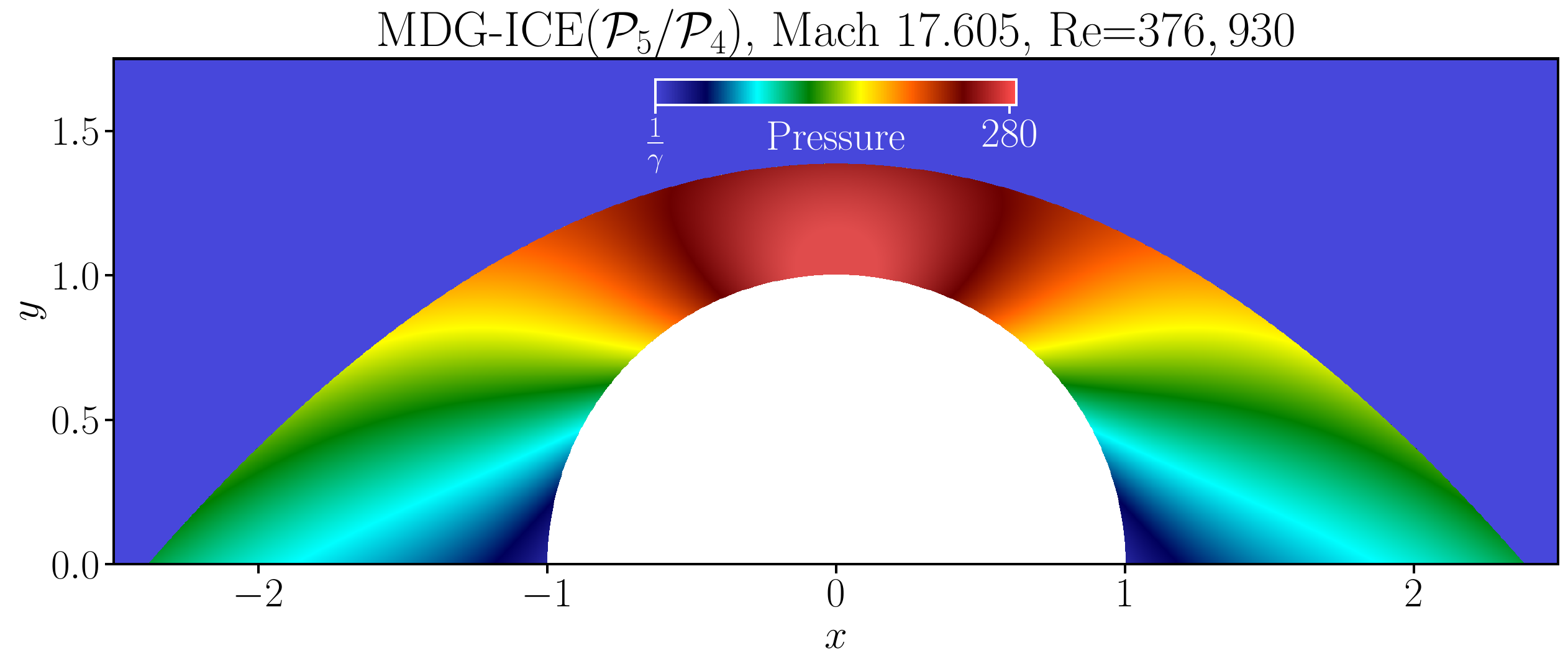}}
\par\end{centering}
\caption{\label{fig:Viscous-Bow-Shock-Mach-17.6-Re-3.77e5-3D} The MDG-ICE
solution (shown along the $z=2$ symmetry plane) computed using 3024
$\mathcal{P}_{5}/\mathcal{P}_{4}$ subparametric tetrahedral elements
for three-dimensional Mach 17.6 flow over a cylinder at $\mathrm{Re}=376,930$.
The initial grid has a regular topology.}
\end{figure}
\par\end{center}

\begin{center}
\begin{figure}[H]
\begin{centering}
\subfloat[\label{fig:Viscous-Bow-Shock-Mach-17.6-Re-3.77e5-Mesh-3D-zoom-structured}Mesh]{\centering{}\includegraphics[clip,width=0.48\textwidth]{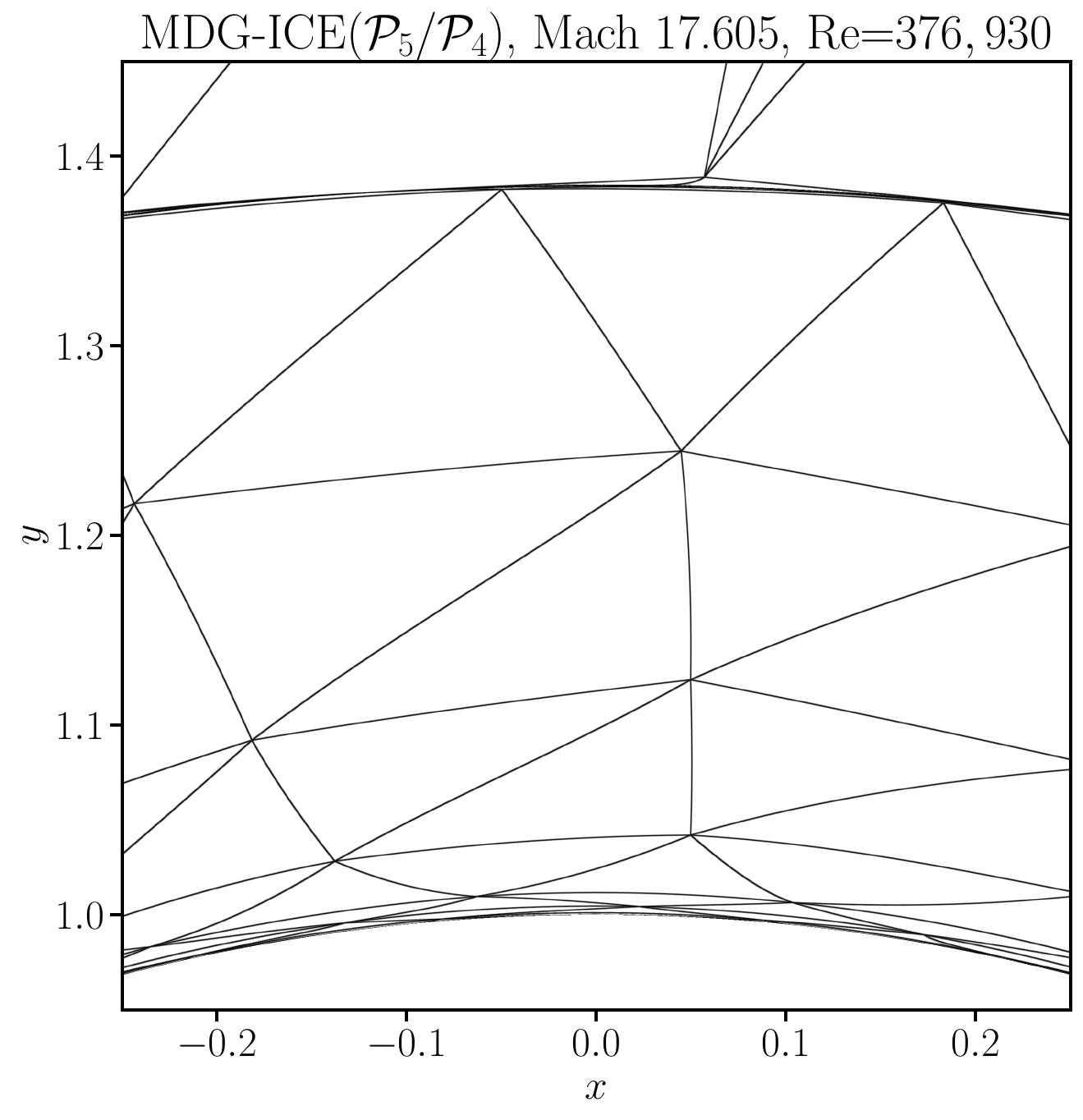}}\hfill{}\subfloat[\label{fig:Viscous-Bow-Shock-Mach-17.6-Re-3.77e5-Temperature-3D-zoom-structured}Temperature
field]{\centering{}\includegraphics[clip,width=0.48\textwidth]{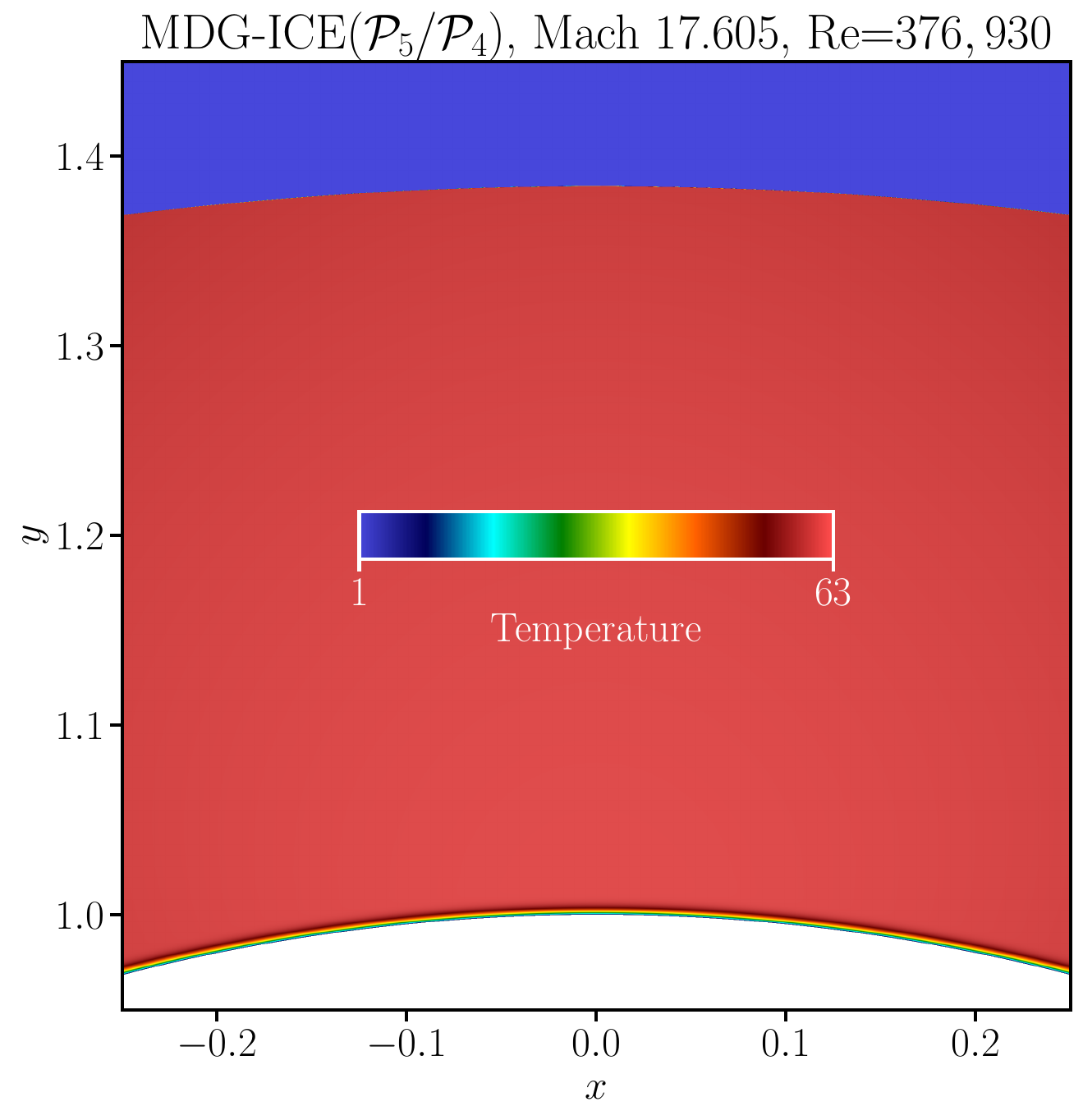}
}
\par\end{centering}
\caption{\label{fig:Viscous-Bow-Shock-Mach-17.6-Re-3.77e5-3D-zoom-structured}
Zoomed-in view of the subparametric MDG-ICE($\mathcal{P}_{5}/\mathcal{P}_{4}$)
solution (shown along the $z=2$ symmetry plane) to three-dimensional
Mach 17.6 flow over a cylinder at $\mathrm{Re}=376,930$. The initial
grid has a regular topology.}
\end{figure}
\par\end{center}

\begin{figure}[ht]
\centering{}\includegraphics[clip,width=0.48\textwidth]{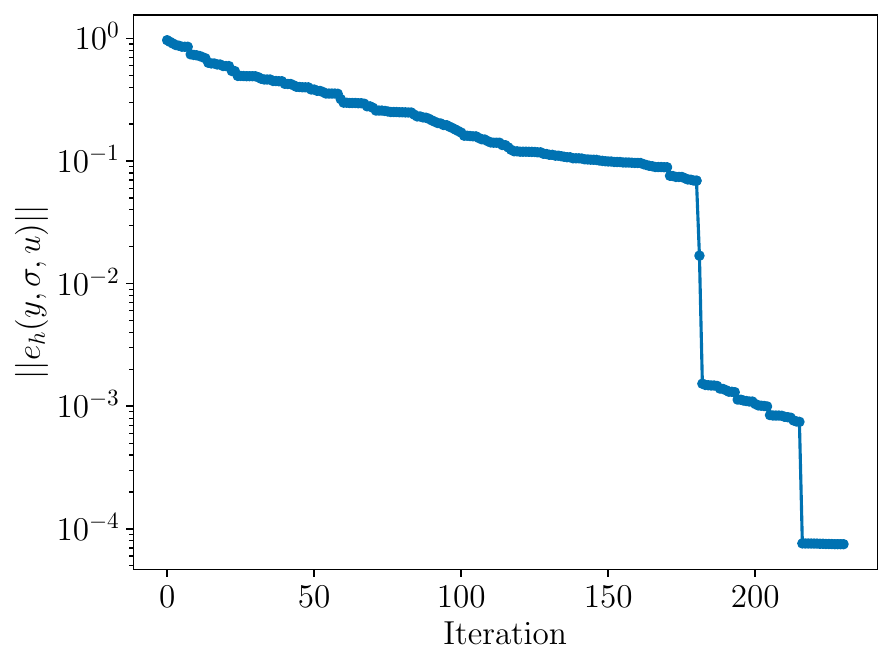}
\caption{Nonlinear convergence history for the subparametric MDG-ICE($\mathcal{P}_{5}/\mathcal{P}_{4}$)
solution to Mach 17.6 flow over a three-dimensional cylinder. The
initial grid has a regular topology.}
\label{fig:cylinder-3d-residual-structured}
\end{figure}

\begin{figure}[H]
\subfloat[\label{fig:Bow-Shock-Mach-17.6-stagnation-line-3D-temperature-structured}Temperature.]{\centering{}\includegraphics[clip,width=0.48\textwidth]{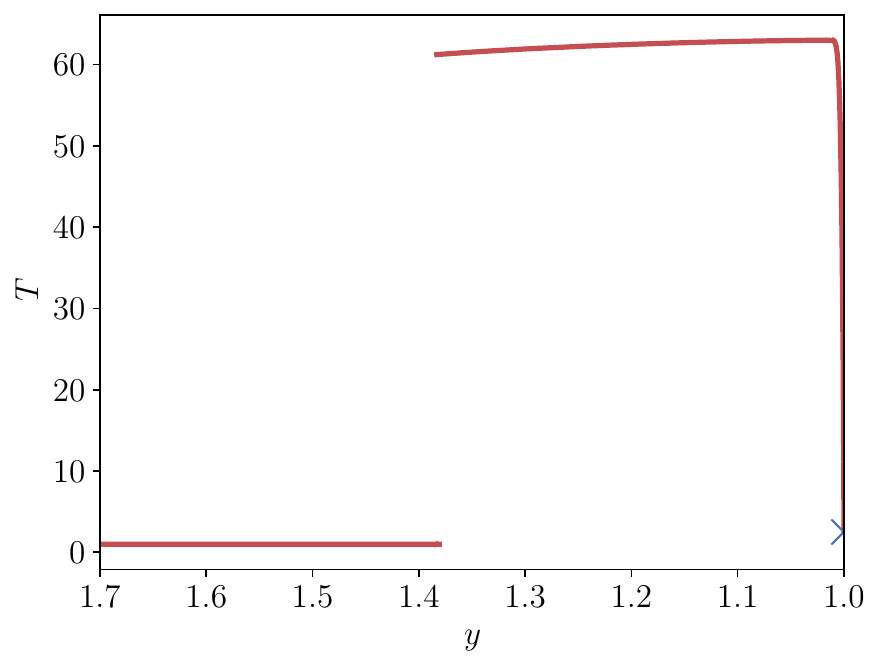}
}\hfill{}\subfloat[\label{fig:Bow-Shock-Mach-17.6-stagnation-line-3D-pressure-structured}Pressure.]{\centering{}\includegraphics[clip,width=0.48\textwidth]{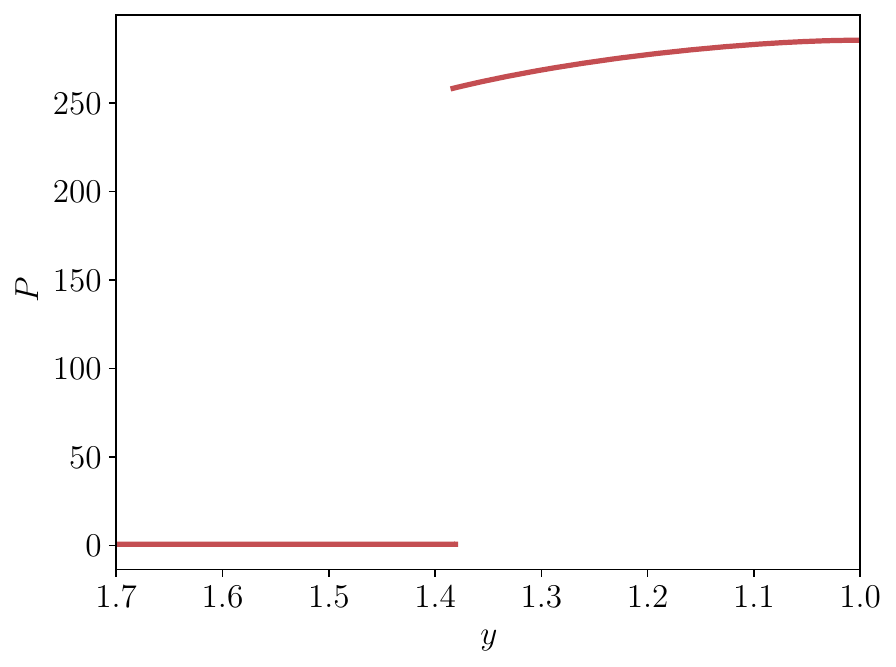}
}

\caption{\label{fig:Bow-Shock-Mach-17.6-stagnation-line-3D} Stagnation-line
profiles of temperature and pressure obtained with the MDG-ICE solution
computed using 3024 $\mathcal{P}_{5}/\mathcal{P}_{4}$ subparametric
tetrahedral elements for three-dimensional Mach 17.6 flow over a cylinder
at $\mathrm{Re}=376,930$. The stagnation point is located at $y=1$.
The stagnation-line quantities are plotted on a per-cell basis, i.e.,
only points within the same cell are connected. The exact stagnation-point
temperature, $T=2.5$, is marked with the symbol $\times$. The initial
grid has a regular topology.}
\end{figure}

The convergence history for the subparametric MDG-ICE($\mathcal{P}_{5}/\mathcal{P}_{4}$)
solution is given in Figure~\ref{fig:cylinder-3d-residual-structured}.
The residual magnitude starts at a relatively small value since the
solution is restarted from an isoparametric MDG-ICE($\mathcal{P}_{4}$)
calculation.  Figure~\ref{fig:Bow-Shock-Mach-17.6-stagnation-line-3D}
shows the stagnation-line profiles of temperature and pressure. The
stagnation point is located at $y=1$. The stagnation-line quantities
are plotted on a per-cell basis, i.e., only points within the same
cell are connected. Note that it is very difficult to capture the
interior of the extremely thin viscous shock using the employed line
sampler, which probes the solution at discrete points. This explains
why the shock is presented as a discontinuity in Figure~\ref{fig:Bow-Shock-Mach-17.6-stagnation-line-3D}.
These results further confirm that the solution is free from spurious
oscillations, despite the substantial gradient across the shock. 

\begin{figure}[H]
\subfloat[\label{fig:Bow-Shock-Mach-17.6-PressureCoefficient-3D-structured}Pressure
coefficient profile.]{\centering{}\includegraphics[clip,width=0.48\textwidth]{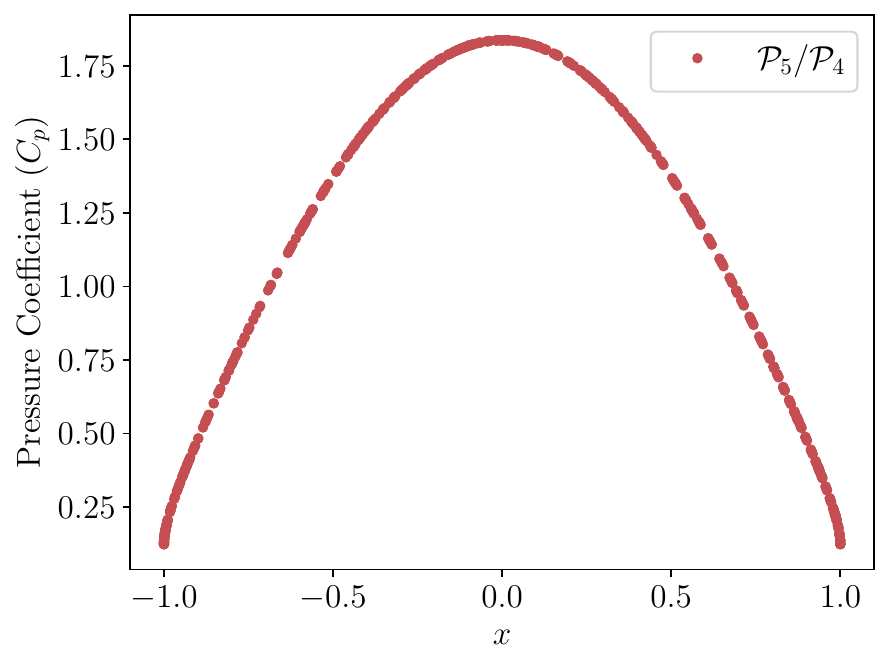}
}\hfill{}\subfloat[\label{fig:Bow-Shock-Mach-17.6-StantonNumber-3D-structured}Stanton
number profile.]{\centering{}\includegraphics[clip,width=0.48\textwidth]{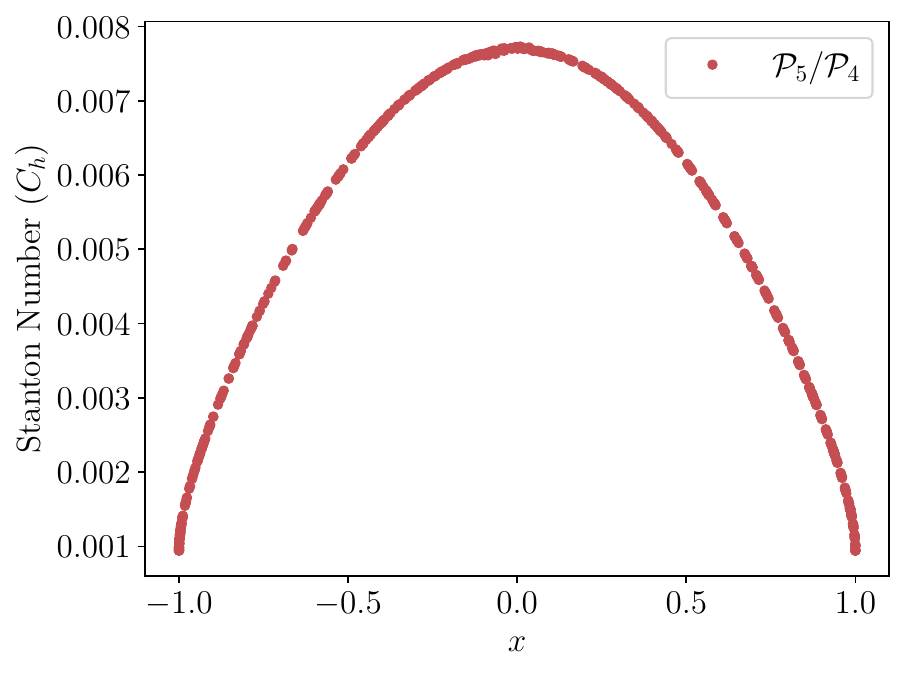}
}

\caption{\label{fig:Bow-Shock-Mach-17.6-Surface-Quantities-3D-structured}
Surface profiles of pressure coefficient and Stanton number for the
subparametric MDG-ICE($\mathcal{P}_{5}/\mathcal{P}_{4}$) solution
to three-dimensional Mach 17.6 flow over a cylinder at $\mathrm{Re}=376,930$.
Pressure and heat-flux values at all auxiliary-variable degrees of
freedom along the cylinder wall are shown. The initial grid has a
regular topology.}
\end{figure}

Figure~\ref{fig:Bow-Shock-Mach-17.6-Surface-Quantities-3D-structured}
gives the surface pressure and heat flux evaluated at all degrees
of freedom corresponding to $\Sigma_{h}$ for the subparametric MDG-ICE($\mathcal{P}_{5}/\mathcal{P}_{4}$)
solution. The pressure profile is essentially perfectly symmetric.
Though very slight asymmetries are still present in the heat-flux
profile, it is nevertheless highly symmetric. The stagnation-point
Stanton number in the MDG-ICE($\mathcal{P}_{5}/\mathcal{P}_{4}$)
solution is approximately 0.0077.

\subsubsection{Initial mesh topology: Irregular\label{subsec:cylinder-3d-irregular}}

\begin{figure}[ht]
\centering{}\includegraphics[clip,width=0.75\textwidth]{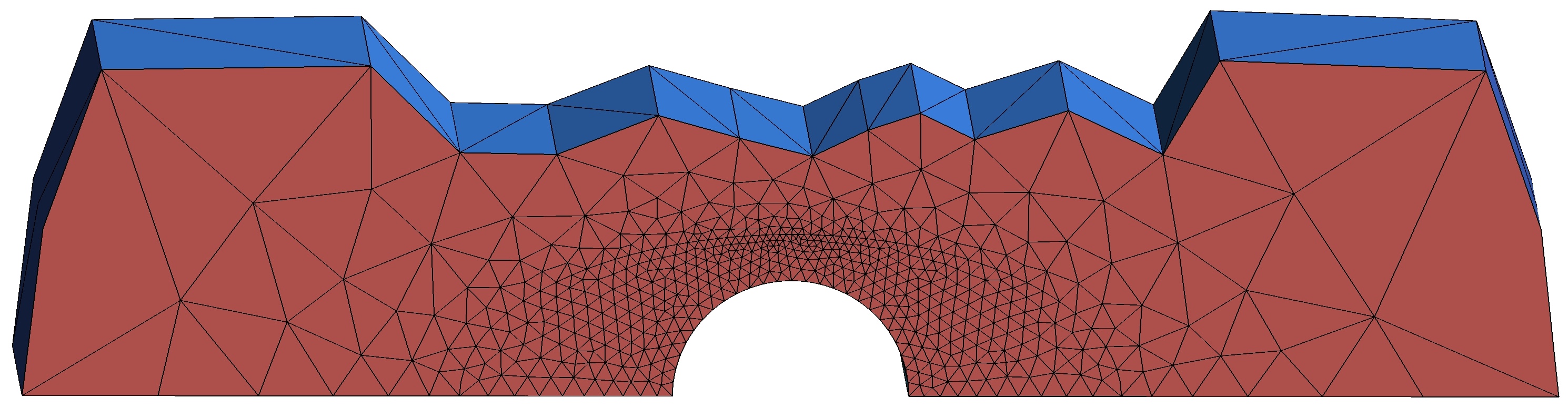}
\caption{Clipped, three-dimensional perspective of the initial grid with an
irregular topology.}
\label{fig:cylinder-3d-DG-initial-grid-unstructured}
\end{figure}

\begin{figure}[ht]
\centering{}\includegraphics[clip,width=0.75\textwidth]{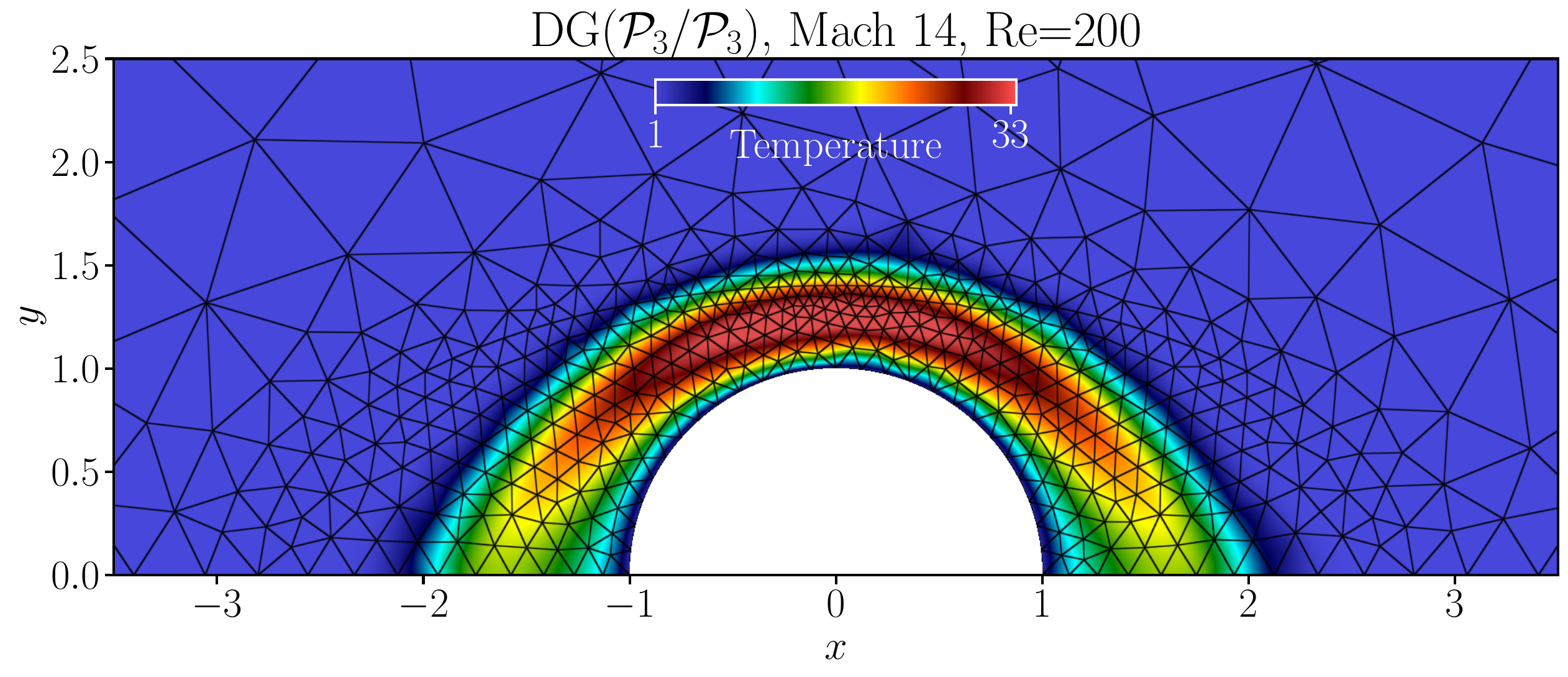}
\caption{Isoparametric DG($\mathcal{P}_{3}$) solution at $\mathrm{Ma}=14,\mathrm{Re}=200$
on 3183 tetrahedral cells along the $z=2$ symmetry plane. This solution
is used as the initial condition for the MDG-ICE continuation strategy.
The initial grid has an irregular topology.}
\label{fig:cylinder-3d-DG-initialization-unstructured}
\end{figure}

A clipped, three-dimensional view of the initial grid is presented
in Figure~\ref{fig:cylinder-3d-DG-initial-grid-unstructured}. In
order to reduce the computational cost and memory footprint of the
linear solver, the initial grid is made finer in the expected shock-layer
region than near the inflow boundary. An isoparametric DG($\mathcal{P}_{3}$)
solution at $\mathrm{Ma}=14,\mathrm{Re}=200$ is used as the initial
condition for the MDG-ICE continuation with superparametric $\mathcal{P}_{3}/\mathcal{P}_{4}$
cells. Figure~\ref{fig:cylinder-3d-DG-initialization-unstructured}
presents the temperature field and initial grid for the DG solution
along the $z=2$ symmetry plane. Given the irregular nature of the
grid, continuation in Reynolds number, especially at higher Reynolds
numbers, leads to the appearance of temperature undershoots that are
especially difficult to dampen in this problem in the absence of artificial
dissipation. To mitigate these instabilities, we employ the alternative
least-squares MDG-ICE formulation with optimal test functions~\citep{Ker20_LS}
based on the discontinuous Petrov-Galerkin methodology by Demkowicz
and Gopalakrishnan~\citep{Dem10,Dem11,Dem15_20}. This formulation
is found to be significantly more robust (in terms of preventing spurious
undershoots/overshoots in flow quantities) than the standard MDG-ICE
formulation. The anisotropic, locally adaptive penalty method described
in Section~\ref{sec:nonlinear-solver} is incorporated in an analogous
manner. Furthermore, the irregular nature of the grid induces larger
grid motion in the shock layer, especially in response to spurious
transients as the Reynolds number is increased, which can lead to
excessively anisotropic grid interfaces along the cylinder surface.
To prevent the appearance of such thin grid interfaces along the cylinder
surface, we freeze the cylinder surface grid in the geometric projection
operator, $b(u)$. Although it is not yet clear how detrimental the
presence of unnecessarily high-aspect-ratio grid interfaces along
a domain boundary may be in more complex configurations, the formation
of such grid interfaces can be mitigated by the use of artificial
dissipation (which can also alleviate the aforementioned instabilities)
during intermediate iterations and/or metric-based remeshing.

After reaching $\mathrm{Ma}=17.6,\mathrm{Re}=376,930$, two global
$p$-refinements of the state and auxiliary-variable approximations
are performed, and we revert to the standard MDG-ICE formulation since
the alternative least-squares MDG-ICE formulation with optimal test
functions~\citep{Ker20_LS} is no longer needed at this point. Figure~\ref{fig:cylinder-3d-DG-initialization-unstructured}
displays the final subparametric MDG-ICE($\mathcal{P}_{5}/\mathcal{P}_{4}$)
solution. The grid is ostensibly more irregular than the previous
grid and, as in the two-dimensional case, features long, thin elements
oriented orthogonal to the shock. Nevertheless, the shock and boundary
layer remain well-resolved. Figure~\ref{fig:Viscous-Bow-Shock-Mach-17.6-Re-3.77e5-3D-zoom-unstructured}
zooms in on the shock layer along the stagnation line. Again, the
cells at the shock are almost visually indistinguishable, and the
viscous shock resembles a discontinuous feature. The solution is free
from spurious oscillations.
\begin{center}
\begin{figure}[H]
\begin{centering}
\subfloat[\label{fig:Viscous-Bow-Shock-Mach-17.6-Re-3.77e5-Mesh-3D-unstructured}Mesh]{\centering{}\includegraphics[clip,width=0.75\textwidth]{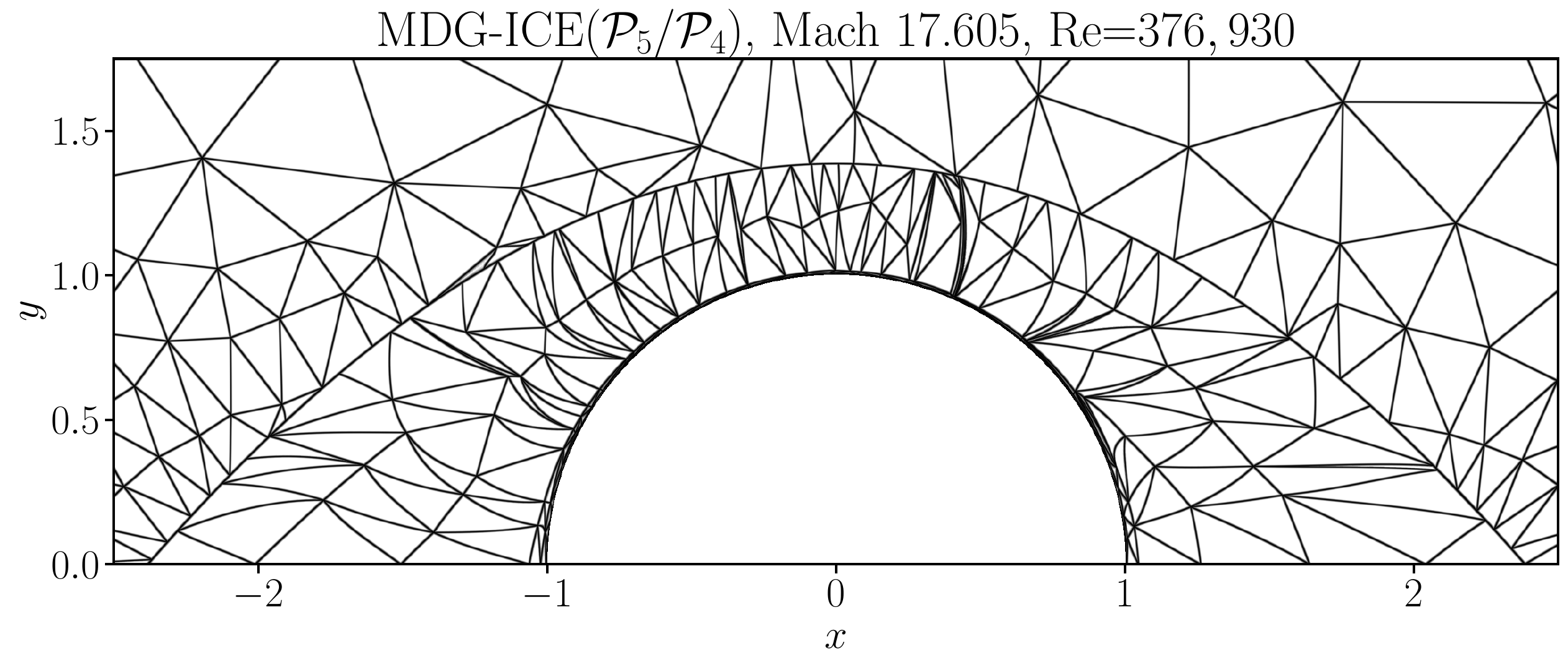}}\hfill{}\subfloat[\label{fig:Viscous-Bow-Shock-Mach-17.6-Re-3.77e5-Temperature-3D-unstructured}Temperature
field]{\centering{}\includegraphics[clip,width=0.75\textwidth]{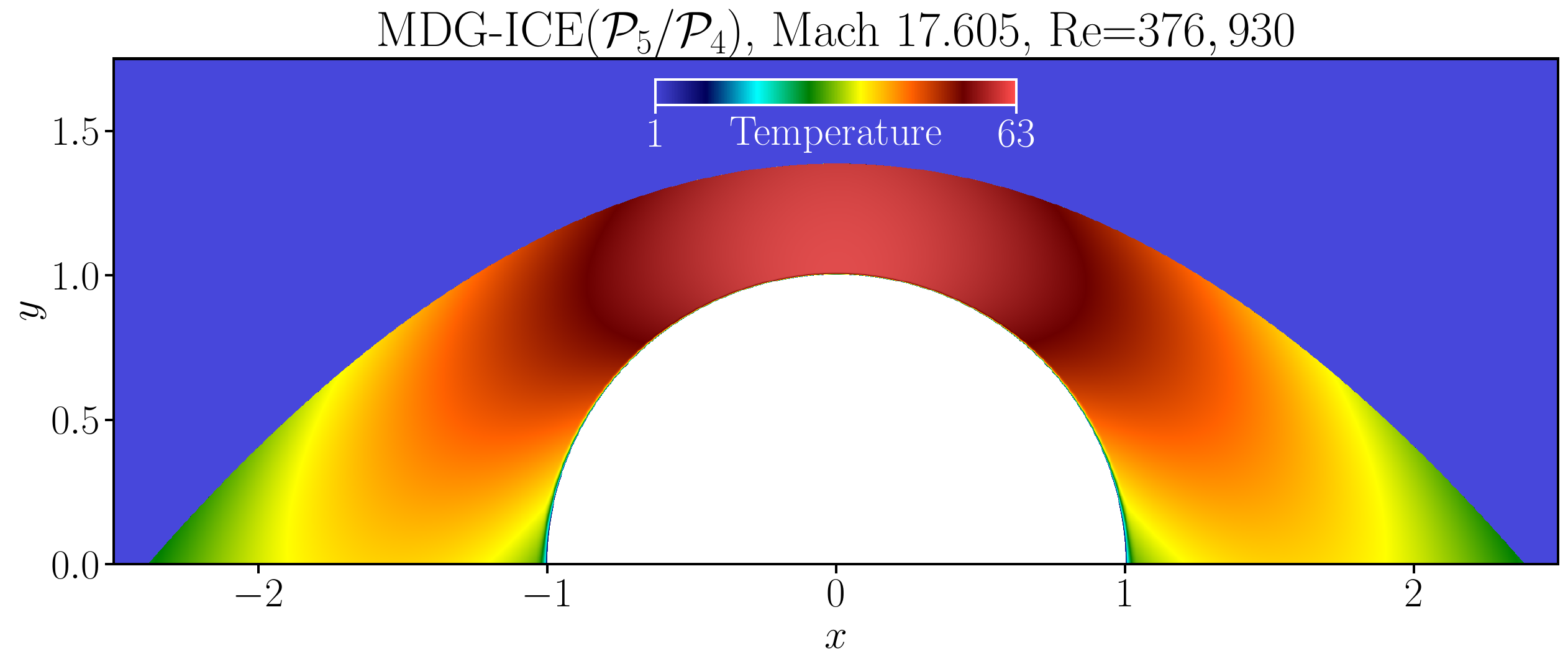}
}\hfill{}\subfloat[\label{fig:Viscous-Bow-Shock-Mach-17.6-Re-3.77e5-Pressure-3D-unstructured}Pressure
field]{\centering{}\includegraphics[clip,width=0.75\textwidth]{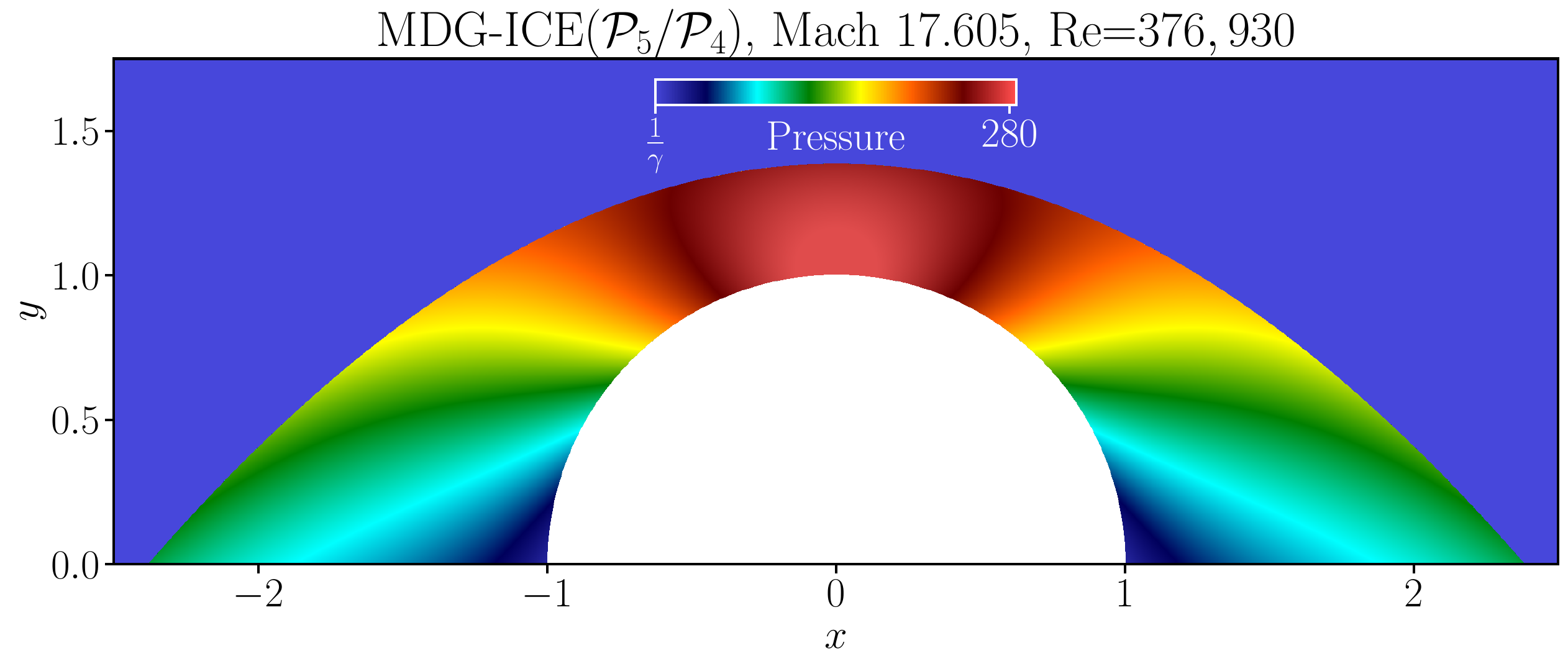}}
\par\end{centering}
\caption{\label{fig:Viscous-Bow-Shock-Mach-17.6-Re-3.77e5-3D-unstructured}
The MDG-ICE solution (shown along the $z=2$ symmetry plane) computed
using 3183 $\mathcal{P}_{5}/\mathcal{P}_{4}$ subparametric tetrahedral
elements for three-dimensional Mach 17.6 flow over a cylinder at $\mathrm{Re}=376,930$.
The initial grid has an irregular topology.}
\end{figure}
\par\end{center}

\begin{center}
\begin{figure}[H]
\begin{centering}
\subfloat[\label{fig:Viscous-Bow-Shock-Mach-17.6-Re-3.77e5-Mesh-3D-zoom-unstructured}Mesh]{\centering{}\includegraphics[clip,width=0.48\textwidth]{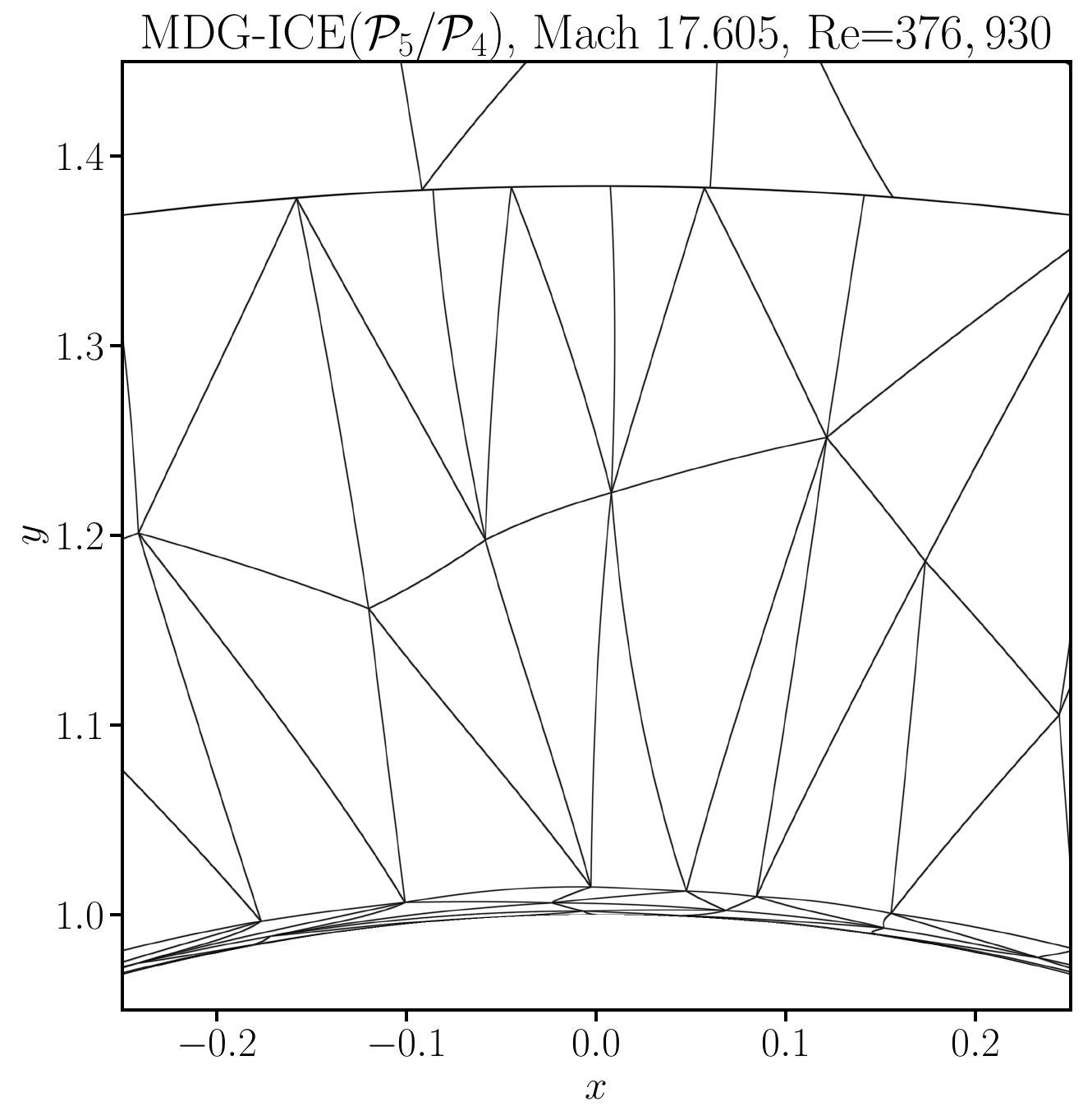}}\hfill{}\subfloat[\label{fig:Viscous-Bow-Shock-Mach-17.6-Re-3.77e5-Temperature-3D-zoom-unstructured}Temperature
field]{\centering{}\includegraphics[clip,width=0.48\textwidth]{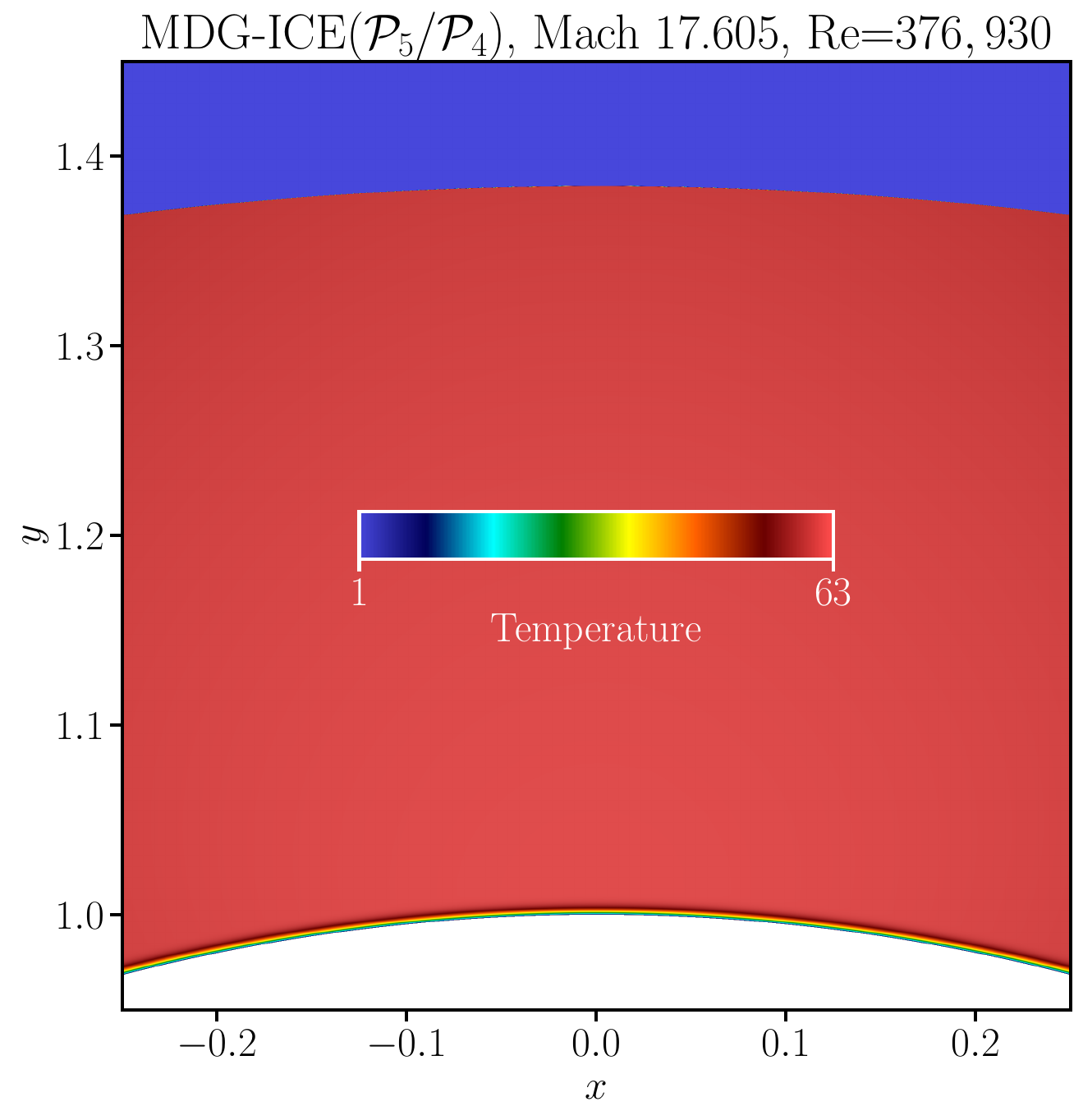}
}
\par\end{centering}
\caption{\label{fig:Viscous-Bow-Shock-Mach-17.6-Re-3.77e5-3D-zoom-unstructured}
Zoomed-in view of the subparametric MDG-ICE($\mathcal{P}_{5}/\mathcal{P}_{4}$)
solution (shown along the $z=2$ symmetry plane) to three-dimensional
Mach 17.6 flow over a cylinder at $\mathrm{Re}=376,930$. The initial
grid has an irregular topology.}
\end{figure}
\par\end{center}

\begin{figure}[ht]
\centering{}\includegraphics[clip,width=0.48\textwidth]{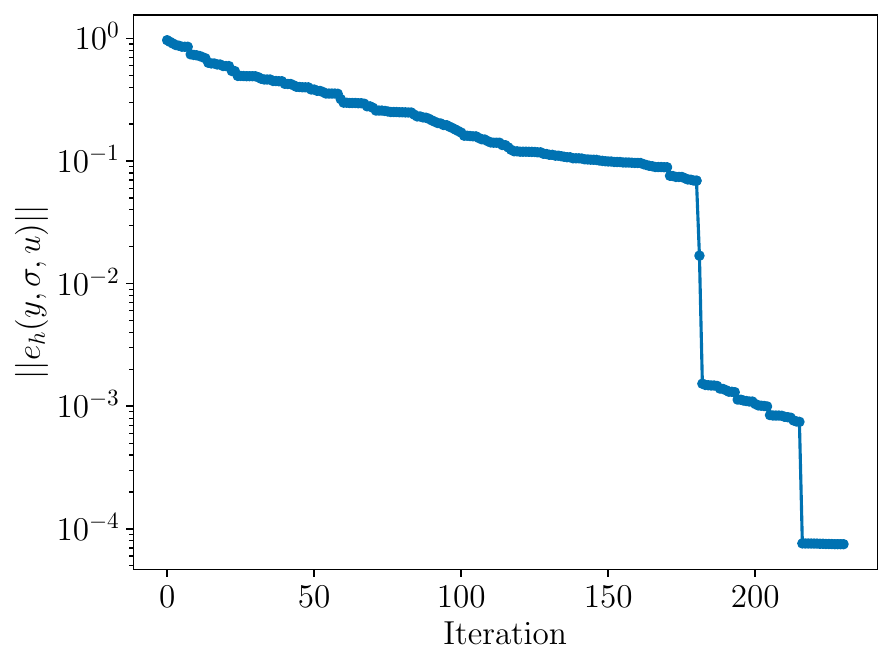}
\caption{Nonlinear convergence history for the subparametric MDG-ICE($\mathcal{P}_{5}/\mathcal{P}_{4}$)
solution to Mach 17.6 flow over a three-dimensional cylinder. The
initial grid has an irregular topology.}
\label{fig:cylinder-3d-residual-unstructured}
\end{figure}

The convergence history for the subparametric MDG-ICE($\mathcal{P}_{5}/\mathcal{P}_{4}$)
solution is given in Figure~\ref{fig:cylinder-3d-residual-unstructured}.
The residual magnitude starts at a relatively small value since the
solution is restarted from an isoparametric MDG-ICE($\mathcal{P}_{4}$)
calculation. 

\begin{figure}[H]
\subfloat[\label{fig:Bow-Shock-Mach-17.6-stagnation-line-3D-temperature-unstructured}Temperature.]{\centering{}\includegraphics[clip,width=0.48\textwidth]{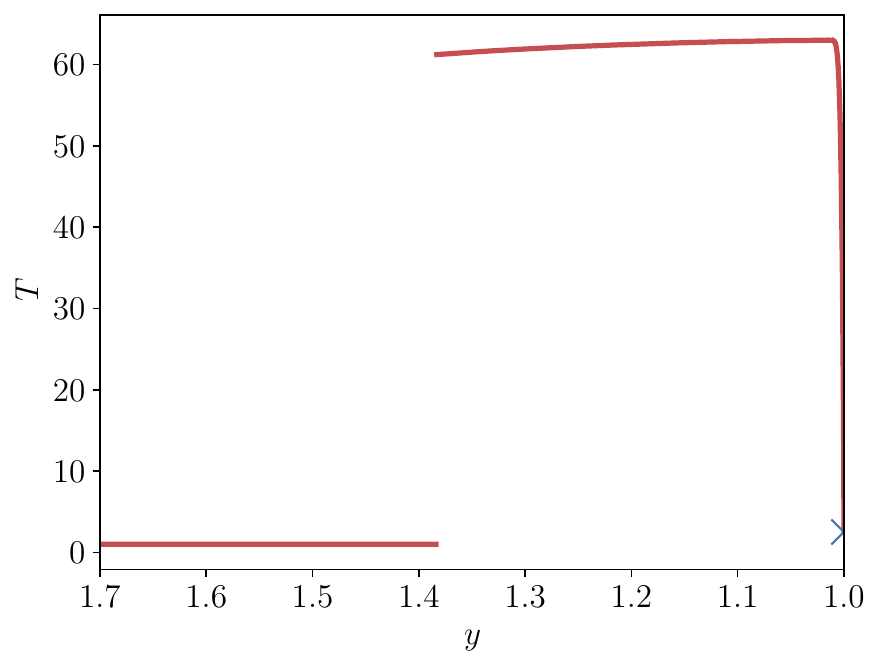}
}\hfill{}\subfloat[\label{fig:Bow-Shock-Mach-17.6-stagnation-line-3D-pressure-unstructured}Pressure.]{\centering{}\includegraphics[clip,width=0.48\textwidth]{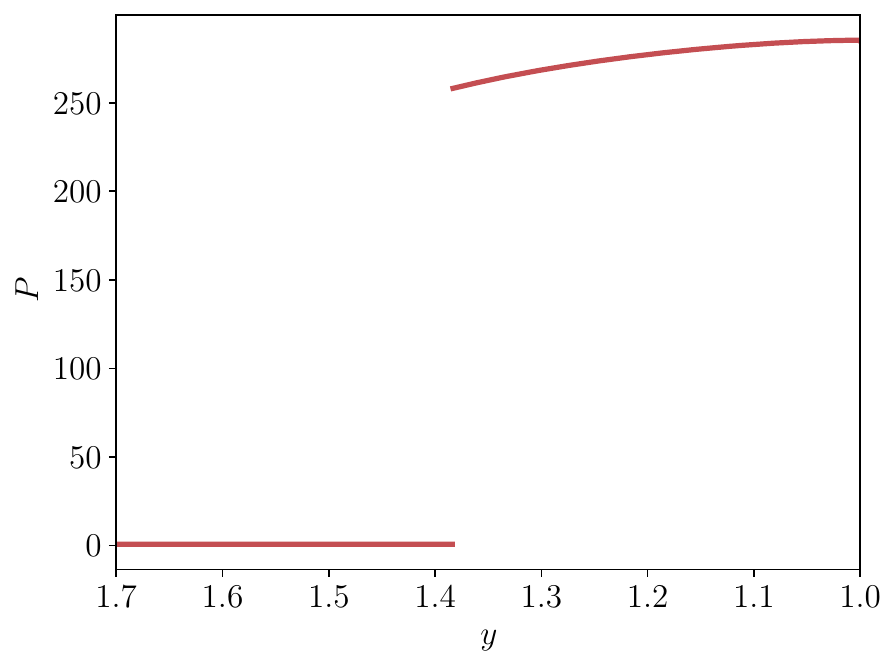}
}

\caption{\label{fig:Bow-Shock-Mach-17.6-stagnation-line-3D-unstructured} Stagnation-line
profiles of temperature and pressure obtained with the MDG-ICE solution
computed using 3183 $\mathcal{P}_{5}/\mathcal{P}_{4}$ subparametric
tetrahedral elements for three-dimensional Mach 17.6 flow over a cylinder
at $\mathrm{Re}=376,930$. The stagnation point is located at $y=1$.
The stagnation-line quantities are plotted on a per-cell basis, i.e.,
only points within the same cell are connected. The exact stagnation-point
temperature, $T=2.5$, is marked with the symbol $\times$. The initial
grid has an irregular topology.}
\end{figure}

The stagnation-line profiles of temperature and pressure are given
in Figure~\ref{fig:Bow-Shock-Mach-17.6-stagnation-line-3D-unstructured}.
The stagnation point is located at $y=1$. The stagnation-line quantities
are plotted on a per-cell basis, which means only points within the
same cell are connected. Again, it is very difficult to capture the
interior of the highly anisotropic viscous shock using the employed
line sampler, which probes the solution at discrete points. This explains
why the shock resembles a true discontinuity in Figure~\ref{fig:Bow-Shock-Mach-17.6-stagnation-line-3D-unstructured}.
The shock and boundary layer are well-resolved. 

\begin{figure}[H]
\subfloat[\label{fig:Bow-Shock-Mach-17.6-PressureCoefficient-3D-unstructured}Pressure
coefficient profile.]{\centering{}\includegraphics[clip,width=0.48\textwidth]{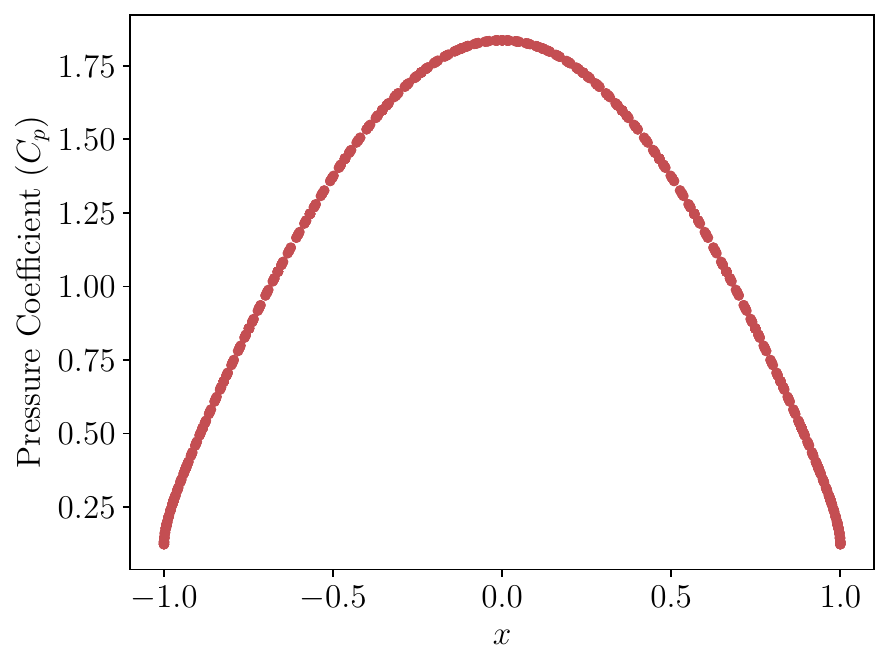}
}\hfill{}\subfloat[\label{fig:Bow-Shock-Mach-17.6-StantonNumber-3D-unstructured}Stanton
number profile.]{\centering{}\includegraphics[clip,width=0.48\textwidth]{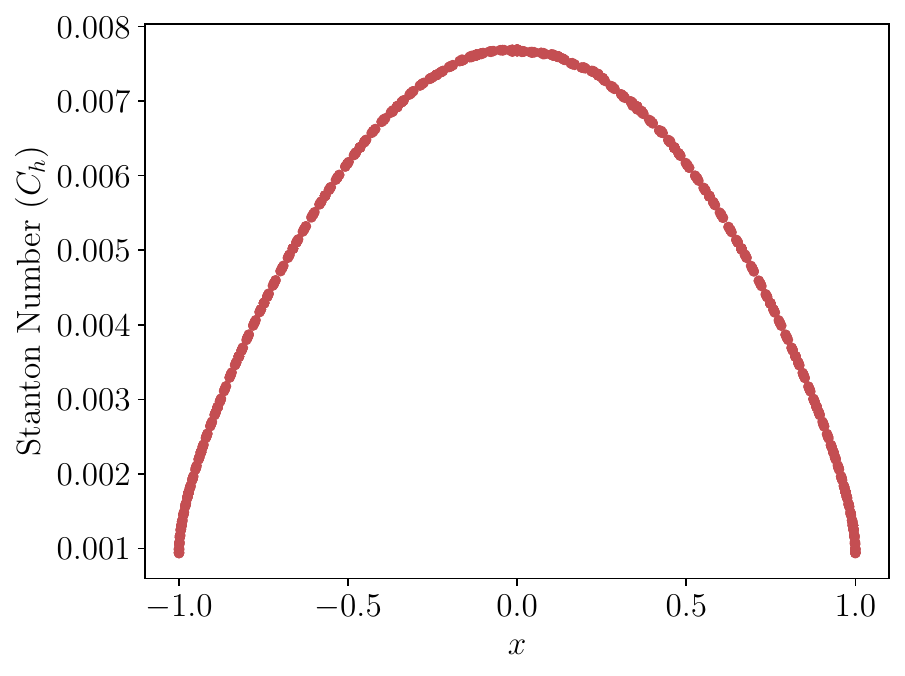}
}

\caption{\label{fig:Bow-Shock-Mach-17.6-Surface-Quantities-3D-unstructured}
Surface profiles of pressure coefficient and Stanton number for the
isoparametric MDG-ICE($\mathcal{P}_{4}$) and subparametric MDG-ICE($\mathcal{P}_{5}/\mathcal{P}_{4}$)
solutions to three-dimensional Mach 17.6 flow over a cylinder at $\mathrm{Re}=376,930$.
Pressure and heat-flux values at all auxiliary-variable degrees of
freedom along the cylinder wall are shown. The initial grid has an
irregular topology.}
\end{figure}

Figure~\ref{fig:Bow-Shock-Mach-17.6-Surface-Quantities-3D-unstructured}
gives the surface pressure and heat flux evaluated at all degrees
of freedom corresponding to $\Sigma_{h}$ for the subparametric MDG-ICE($\mathcal{P}_{5}/\mathcal{P}_{4}$)
solution. Excellent symmetry is observed in the pressure and heat-flux
profiles. Though very slight asymmetries can be discerned in the latter,
these asymmetries are considerably smaller than those observed in
finite-volume predictions~\citep{Gno04,Nom04}). The stagnation-point
Stanton number in the MDG-ICE($\mathcal{P}_{5}/\mathcal{P}_{4}$)
solution is approximately 0.0077.

\subsection{Mach 5 flow over three-dimensional sphere}

In our final test case, we compute steady Mach 5 laminar flow over
a hemisphere in three spatial dimensions. Only one quarter of the
hemisphere, which is of unit radius, is considered due to the memory
footprint and computational cost of the linear solver. The Reynolds
number (based on the sphere radius) is $3.775\times10^{6}$. The sphere
boundary is an isothermal no-slip wall with temperature $T_{\mathrm{wall}}=1.308T_{\infty}$.
These conditions are similar to those considered by Blottner~\citep{Blo90},
who employed an axisymmetric solver, and in the 2022 High-Fidelity
CFD Workshop~\citep{Fis21,Mur23}, except the Reynolds number here
is four times higher. At the freestream conditions by Blottner~\citep{Blo90},
large-scale asymmetries and mesh imprinting have been observed in
three-dimensional finite volume and finite element solutions~\citep{Mur23,Kir10,Cou18},
even with structured hexahedral grids. Freestream conditions are imposed
at the inflow boundary, defined as a quarter hemisphere with radius
2.3 times larger than that of the spherical body. Extrapolation is
applied at the outflow boundary. 

Continuation in Reynolds number is employed. Figure~\ref{fig:sphere-DG-initialization}
presents the initial 7650-cell tetrahedral grid and an isoparametric
DG($\mathcal{P}_{2}$) solution at $\mathrm{Ma}=5,\mathrm{Re}=100$,
which is used to initialize the MDG-ICE continuation with isoparametric
$\mathcal{P}_{3}$ elements. The sphere surface and the $y=0$ and
$z=0$ planes are displayed. The freestream flow is in the $-x$-direction.
Mesh imprinting along the sphere boundary and a highly diffused shock
are observed in the DG solution. Given the cost of the LDLT linear
solver, to maximize efficiency, the initial mesh is generated such
that the azimuthal resolution is higher near the stagnation line than
elsewhere. Furthermore, the wall-normal resolution is decreased near
the inflow boundary. We also find that as the Reynolds number is increased,
due to higher gradients at the shock than at the boundary layer, the
solver tends to move cells from the near-wall region to the vicinity
of the shock. To alleviate this loss of boundary-layer resolution,
a very thin layer of cells is constructed adjacent to the wall. This
issue can likely be resolved with localized artificial dissipation
(to reduce gradients at the shock), a robust remeshing strategy, and/or
local $p$-refinement, all of which will be pursued in the future.
Nevertheless, we note that building an initial grid with additional
resolution near the boundary surface is significantly simpler than
constructing a grid with the required near-wall resolution while simultaneously
aligning the grid interfaces with shocks, whose locations are generally
unknown a priori. With our planned advancements to the MDG-ICE formulation
(including a more efficient linear-solver strategy), we will consider
the full hemispherical domain, as well as an initially irregular topology,
in future work.
\begin{center}
\begin{figure}[H]
\begin{centering}
\subfloat[\label{fig:sphere-DG-initialization-mesh}Mesh]{\centering{}\includegraphics[clip,width=0.48\textwidth]{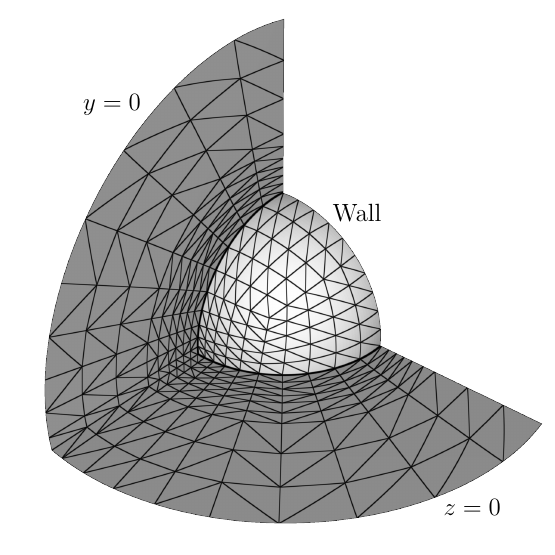}}\hfill{}\subfloat[\label{fig:sphere-DG-initialization-pressure}Pressure field]{\centering{}\includegraphics[clip,width=0.48\textwidth]{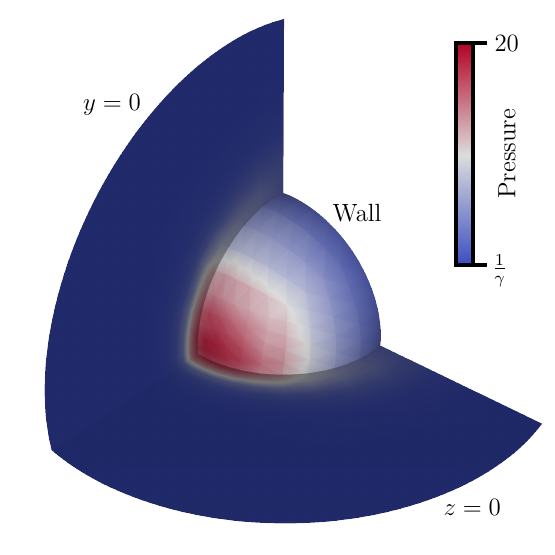}
}
\par\end{centering}
\caption{\label{fig:sphere-DG-initialization} Isoparametric DG($\mathcal{P}_{2}$)
solution at $\mathrm{Ma}=5,\mathrm{Re}=100$ on 7650 tetrahedral cells.
This solution is used as the initial condition for the MDG-ICE continuation
strategy. The sphere surface and the $y=0$ and $z=0$ planes are
displayed. The freestream flow is in the $-x$-direction.}
\end{figure}
\par\end{center}

\begin{figure}[ht]
\centering{}\subfloat[\label{fig:sphere-final-grid}Mesh.]{\centering{}\includegraphics[clip,width=0.48\textwidth]{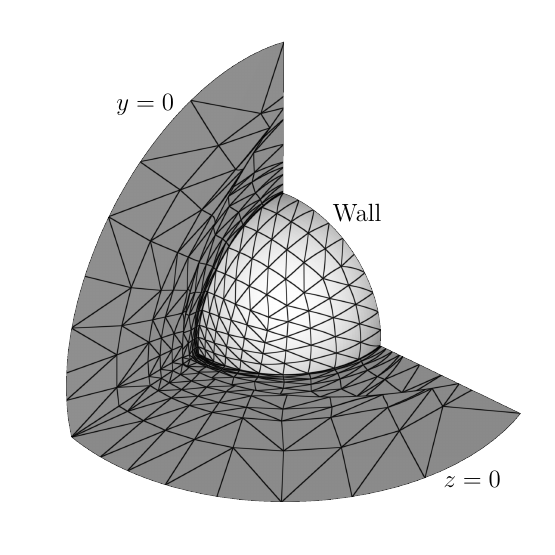}
}\hfill{}\subfloat[\label{fig:sphere-final-pressure}Pressure field.]{\centering{}\includegraphics[clip,width=0.48\textwidth]{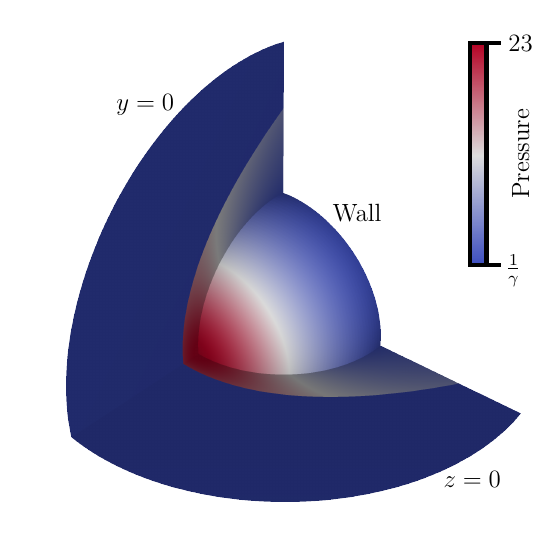}
}\hfill{}\subfloat[\label{fig:sphere-final-heat-flux}Surface heat flux.]{\centering{}\includegraphics[clip,width=0.48\textwidth]{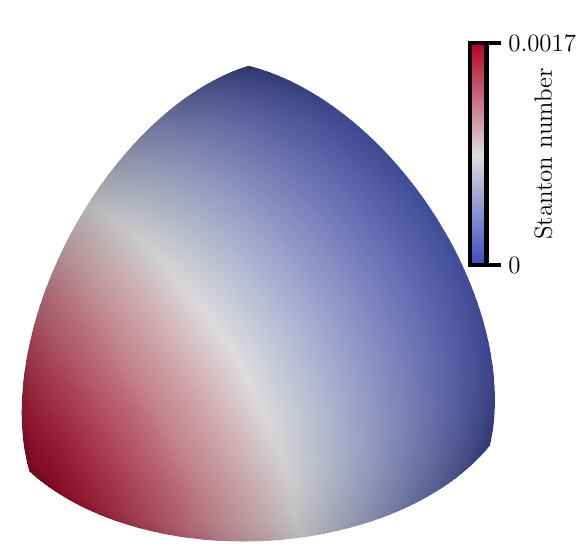}
}\hfill{}\subfloat[\label{fig:sphere-final-temperature-zoom}Zoomed-in view of temperature
profile along $z=0$ plane.]{\centering{}\includegraphics[clip,width=0.48\textwidth]{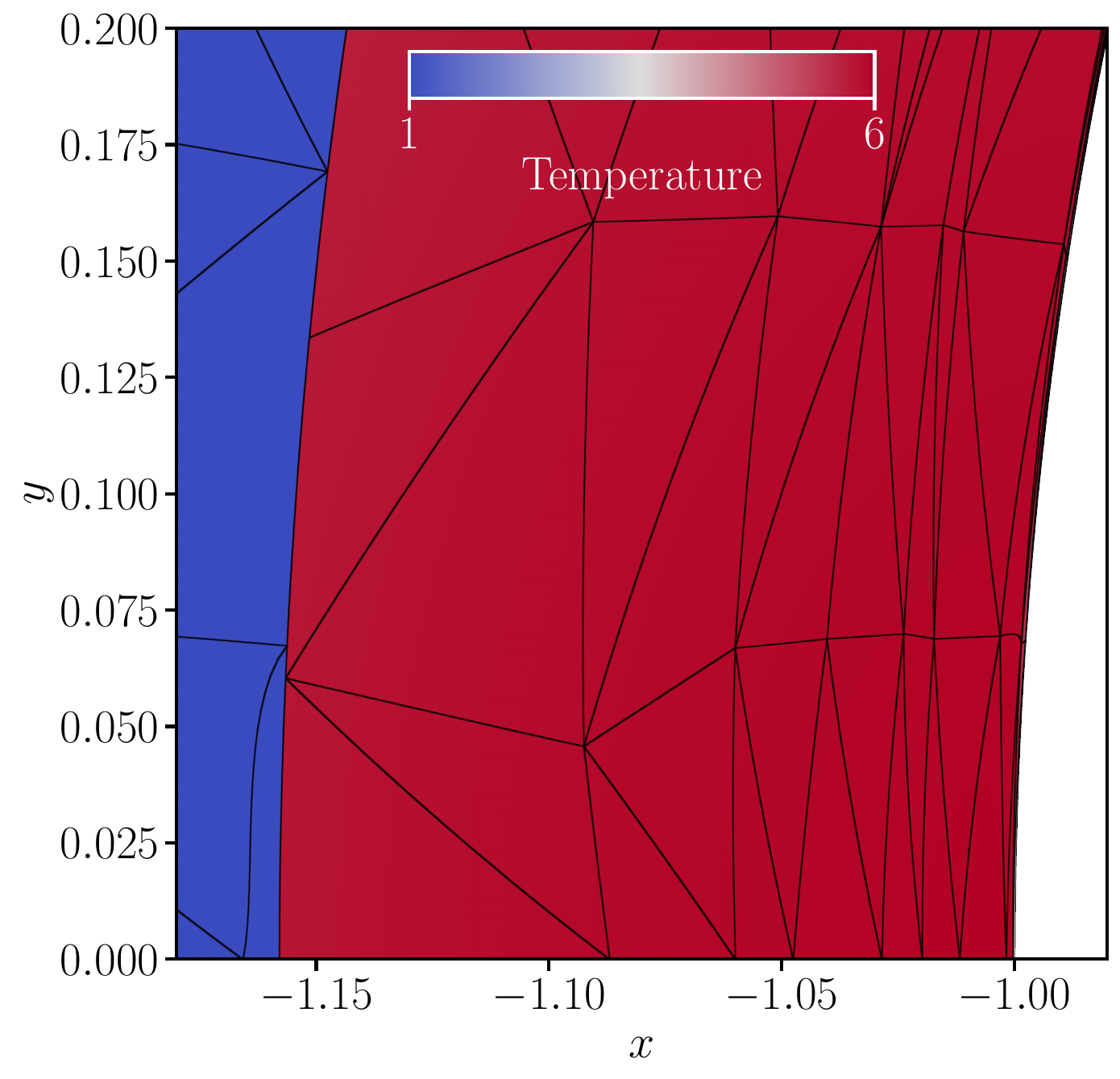}
}\caption{\label{fig:sphere-final}Final grid and pressure field for the subparametric
MDG-ICE($\mathcal{P}_{5}/\mathcal{P}_{3}$) solution to hypersonic
flow over a sphere at $\mathrm{Ma}=5,\mathrm{Re}=3.775\times10^{6}$.
In Figures~\ref{fig:sphere-final-grid} and~\ref{fig:sphere-final-pressure},
the $y=0$ and $z=0$ planes and the sphere wall are displayed. The
freestream flow is in the $-x$-direction.}
\end{figure}

Successive increases in the Reynolds number leads to the appearance
of temperature undershoots that are especially difficult to dampen
in this problem in the absence of artificial dissipation. As in Section~\ref{subsec:cylinder-3d-irregular},
to reduce these instabilities, we employ the alternative least-squares
MDG-ICE formulation with optimal test functions~\citep{Ker20_LS}.
After reaching the final Reynolds number, we perform two global $p$-refinements
of the state and auxiliary-variable approximations while retaining
the alternative least-squares MDG-ICE formulation. Figure~\ref{fig:sphere-final}
displays the final subparametric MDG-ICE($\mathcal{P}_{5}/\mathcal{P}_{3}$)
solution at $\mathrm{Ma}=5,\mathrm{Re}=3.775\times10^{6}$. To reduce
the number of unnecessary elements in the freestream region, we project
the inflow boundary to a quarter hemisphere of radius 2.15. Cell collapses~\citep{Loh08}
are then performed to remove invalid cells, resulting in 7161 elements.
The final grid is shown in Figure~\ref{fig:sphere-final-grid}. The
surface mesh is noticeably coarse. The pressure field and surface
heat-flux profile, presented in Figures~\ref{fig:sphere-final-pressure}
and~\ref{fig:sphere-final-heat-flux}, respectively, are free from
mesh imprinting and spurious oscillations. Figure~\ref{fig:sphere-final-temperature-zoom}
provides a zoomed-in view of the temperature field along the $z=0$
plane. The temperature gradient in the boundary layer can be observed.
The viscous shock resembles a true discontinuity, and the cells resolving
the shock cannot be individually discerned. Nevertheless, grid validity
is maintained. Note the closer proximity of the shock to the boundary
layer than in the cylinder problem, despite a lower Mach number, which
helps explain the aforementioned issue in which the solver moves grid
points from the boundary layer to the shock.

\begin{figure}[ht]
\centering{}\includegraphics[clip,width=0.48\textwidth]{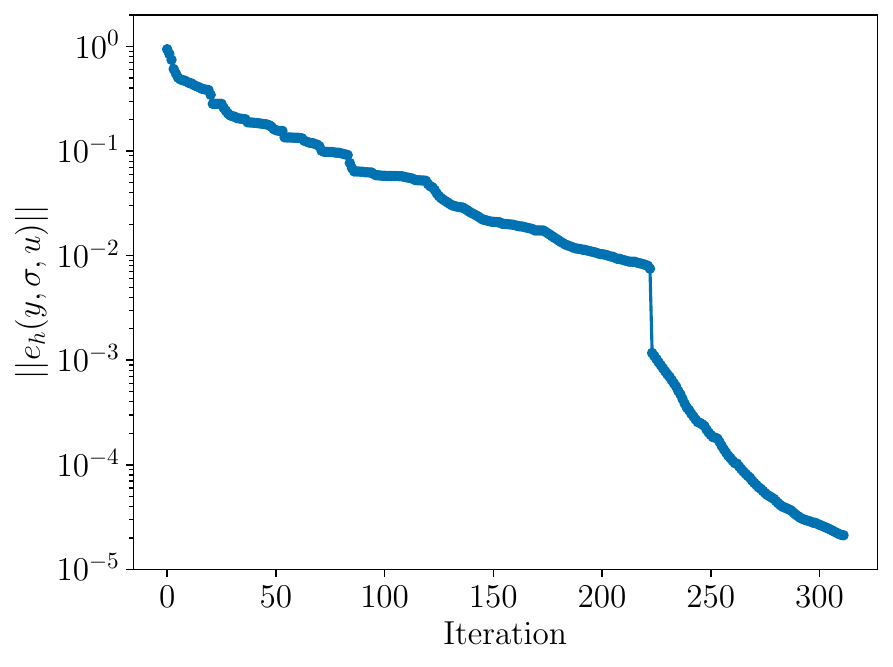}
\caption{Nonlinear convergence history for the subparametric MDG-ICE($\mathcal{P}_{5}/\mathcal{P}_{3}$)
solution to Mach 5 flow over a sphere.}
\label{fig:sphere-residual}
\end{figure}

\begin{figure}[H]
\subfloat[\label{fig:sphere-stagnation-line-temperature}Temperature.]{\centering{}\includegraphics[clip,width=0.48\textwidth]{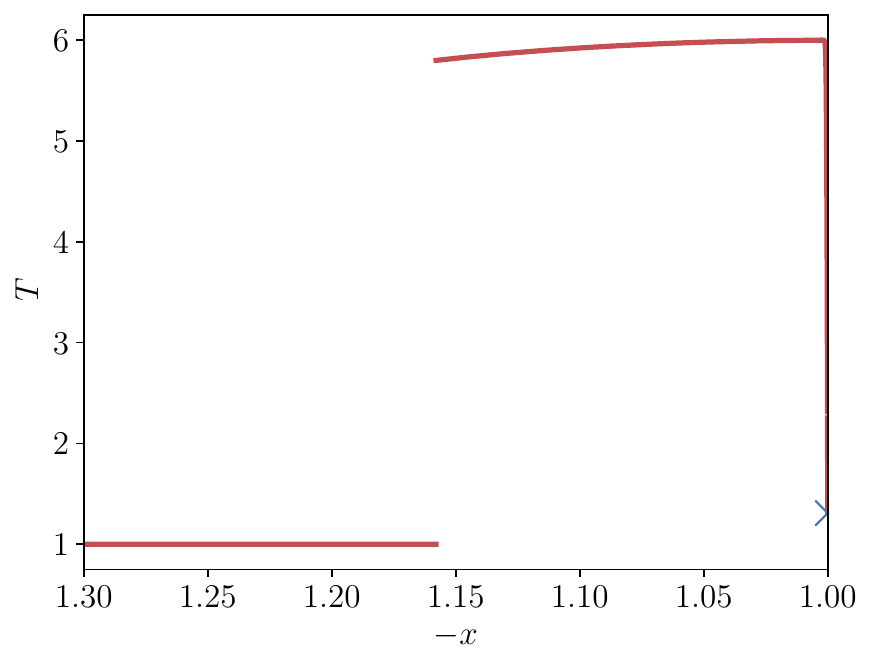}
}\hfill{}\subfloat[\label{fig:sphere-stagnation-line-pressure}Pressure.]{\centering{}\includegraphics[clip,width=0.48\textwidth]{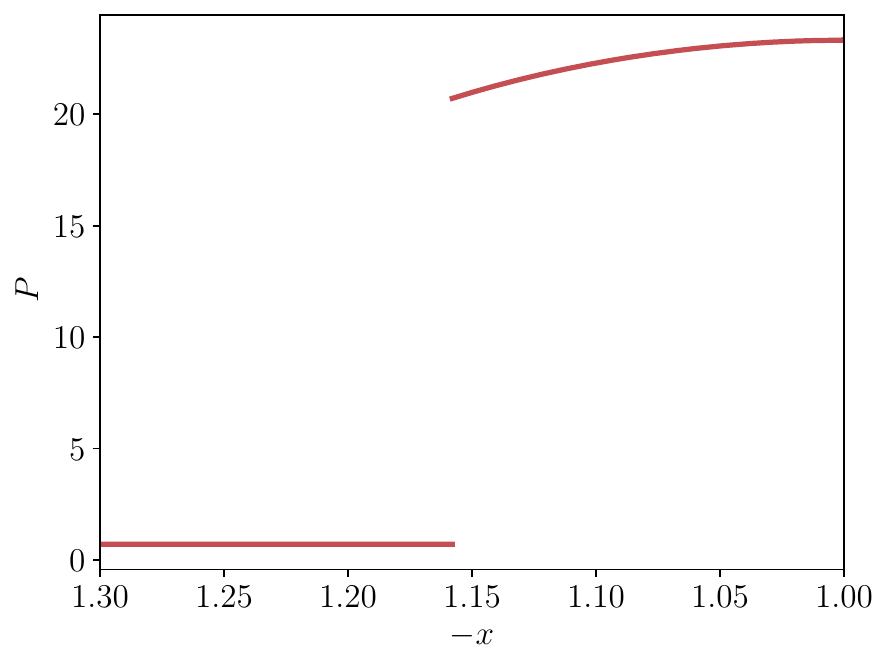}
}

\caption{\label{fig:sphere-stagnation-line} Stagnation-line profiles of temperature
and pressure obtained with the MDG-ICE solution computed using 7161
$\mathcal{P}_{5}/\mathcal{P}_{3}$ subparametric triangle elements
for three-dimensional Mach 5 flow over a sphere at $\mathrm{Re}=3.775\times10^{6}$.
The stagnation point is located at $x=-1$. The stagnation-line quantities
are plotted on a per-cell basis, such that only points within the
same cell are connected. The exact stagnation-point temperature, $T=1.308$,
is marked with the symbol $\times$.}
\end{figure}

The convergence history for the subparametric MDG-ICE($\mathcal{P}_{5}/\mathcal{P}_{3}$)
solution is presented in Figure~\ref{fig:sphere-residual}. The initial
residual is already low since we restart from a subparametric MDG-ICE($\mathcal{P}_{4}/\mathcal{P}_{3}$)
calculation. Figure~\ref{fig:sphere-stagnation-line} display profiles
of temperature and pressure along the stagnation line. The stagnation
point is located at $x=-1$. The stagnation-line quantities are plotted
on a per-cell basis, such that only points within the same cell are
connected. Again, it is very difficult to capture the interior of
the extremely thin viscous shock using the employed line sampler,
which explains why the shock is presented as a discontinuity. These
results further exemplify the sharpness of the shock profile and the
absence of spurious artifacts in the solution.

\begin{figure}[H]
\subfloat[\label{fig:sphere-pressure-coefficient}Pressure coefficient profile.]{\centering{}\includegraphics[clip,width=0.48\textwidth]{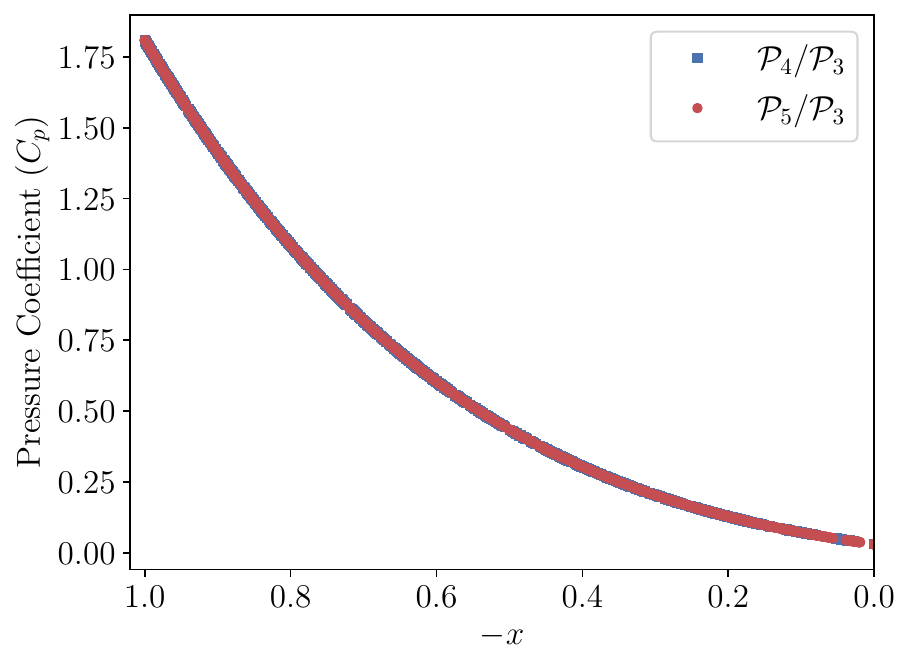}
}\hfill{}\subfloat[\label{fig:sphere-stanton-number}Stanton number profile.]{\centering{}\includegraphics[clip,width=0.48\textwidth]{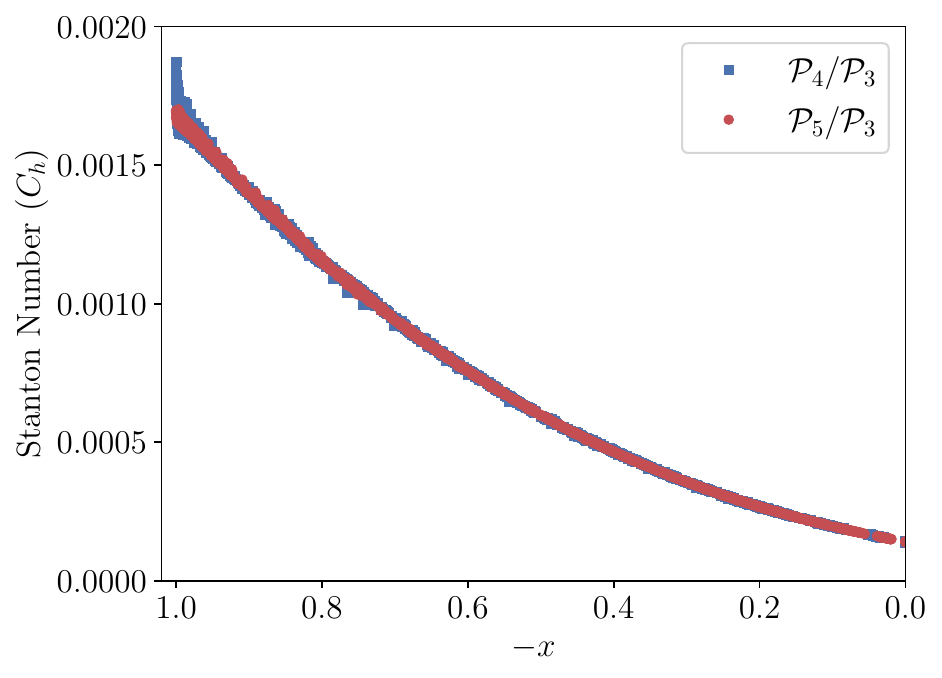}
}

\caption{\label{fig:sphere-surface-quantities} Streamwise variation of surface
pressure and surface heat flux for the subparametric MDG-ICE($\mathcal{P}_{4}/\mathcal{P}_{3}$)
and MDG-ICE($\mathcal{P}_{5}/\mathcal{P}_{3}$) solutions to three-dimensional
Mach 5 flow over a sphere at $\mathrm{Re}=3.775\times10^{6}$. Pressure
and heat-flux values at all auxiliary-variable degrees of freedom
along the sphere wall are shown. The stagnation point is located at
$x=-1$.}
\end{figure}

Figure~\ref{fig:sphere-surface-quantities} presents the streamwise
variation of surface pressure and heat flux evaluated at all degrees
of freedom corresponding to $\Sigma_{h}$ for subparametric MDG-ICE($\mathcal{P}_{4}/\mathcal{P}_{3}$)
and MDG-ICE($\mathcal{P}_{5}/\mathcal{P}_{3}$) solutions. The surface
pressure is essentially perfectly symmetric. Noticeable asymmetries
in the heat flux near the stagnation point for the MDG-ICE($\mathcal{P}_{4}/\mathcal{P}_{3}$)
solution are observed. These asymmetries are significantly reduced
in the MDG-ICE($\mathcal{P}_{5}/\mathcal{P}_{3}$) solution, resulting
in a highly symmetric heat-flux profile. Note that interaction between
the symmetry boundaries near the stagnation point may be a partial
source of the small asymmetries. The stagnation-point Stanton number
in the MDG-ICE($\mathcal{P}_{5}/\mathcal{P}_{3}$) solution is approximately
0.00167, which agrees well with the value of 0.00159 obtained using
the correlation by Fay and Riddell~\citep{Fay58}. These results
demonstrate the ability of MDG-ICE, combined with the enhanced optimization
solver introduced in this work, to produce very accurate surface heating
predictions even in the presence of considerable misalignment between
the grid and the high-gradient features.

\section{Conclusions and future work}

We introduced an enhanced optimization solver for the moving discontinuous
Galerkin method with interface condition enforcement (MDG-ICE), which
automatically fits discontinuous and high-gradient features without
a priori information via curvilinear $r$-adaptivity in both space
and time. Specifically, to overcome the major bottleneck of frequent
cell degeneration, we developed an anisotropic, locally adaptive penalty
technique for the Levenberg-Marquardt method previously employed in~\citep{Ker20}.
The proposed MDG-ICE formulation, without any explicit stabilization
mechanisms, was applied to three test cases involving sharp yet smooth
gradients: Burgers viscous shock formation in space-time, Mach 17.6
viscous flow over a circular half-cylinder in two and three dimensions,
and Mach 5 flow over a three-dimensional sphere. We used simplicial
grids in order to evaluate the ability of the developed MDG-ICE formulation
to obtain symmetric surface heat-flux profiles in the presence of
strong misalignment between the grid and both the shock and boundary
layer, which is considerably challenging for conventional numerical
methods. Oscillation-free solutions and highly symmetric heating predictions
were achieved.

Although the results presented in this work are very promising and
demonstrate how MDG-ICE can significantly alleviate the burden of
mesh generation on the user, additional advancements are needed in
order to more fully realize its potential. Specifically, we plan to
incorporate localized artificial dissipation (either only during intermediate
iterations or in minimal amounts in the final solution), time marching
via space-time MDG-ICE, local $p$-adaptivity, automatic selection
and adjustment of regularization and other solver parameters~\citep{Tra12,Zah20},
and further improvements to the optimization solver that were originally
proposed for the baseline Levenberg-Marquardt method~\citep{Tra12}.
In addition, we will integrate metric-based mesh regeneration/adaptation
strategies into the solver, potentially utilizing the mesh-implied
metric naturally produced by an MDG-ICE solution, which should already
yield small element length scales at high-gradient features. These
developments will likely accelerate convergence, circumvent the need
for continuation in Mach number and Reynolds number, and allow for
even coarser initial grids. We will also explore approximating strong
viscous shocks as inviscid discontinuities as an intrinsic feature
of the solver (i.e., without the use of ad hoc strategies); layers
of high-aspect-ratio cells would then no longer be required since
discontinuous shocks can simply be fit along grid interfaces, as in
the inviscid setting~\citep{Cor18}. Finally, the efficiency of solving
the linear system will be improved.

\section*{Acknowledgments}

This work is sponsored by the Office of Naval Research through the
Naval Research Laboratory 6.1 Computational Physics Task Area and
by Dr. Eric Marineau of the Hypersonic Aerothermodynamics, High-Speed
Propulsion and Materials Program of ONR Code 35.

\bibliographystyle{elsarticle-num}
\bibliography{citations}

\begin{thebibliography}{10}
\expandafter\ifx\csname url\endcsname\relax
  \def\url#1{\texttt{#1}}\fi
\expandafter\ifx\csname urlprefix\endcsname\relax\def\urlprefix{URL }\fi
\expandafter\ifx\csname href\endcsname\relax
  \def\href#1#2{#2} \def\path#1{#1}\fi

\bibitem{Cor18}
A.~Corrigan, A.~Kercher, D.~Kessler, A moving discontinuous {G}alerkin finite
  element method for flows with interfaces, International Journal for Numerical
  Methods in Fluids 89~(9) (2019) 362--406.
\newblock \href {https://doi.org/10.1002/fld.4697}
  {\path{doi:10.1002/fld.4697}}.

\bibitem{Ker20}
A.~Kercher, A.~Corrigan, D.~Kessler, The moving discontinuous {G}alerkin finite
  element method with interface condition enforcement for compressible viscous
  flows, International Journal for Numerical Methods in Fluids 93~(5) (2021)
  1490--1519.
\newblock \href {https://doi.org/10.1002/fld.4939}
  {\path{doi:10.1002/fld.4939}}.

\bibitem{Ker20_LS}
A.~Kercher, A.~Corrigan, A least-squares formulation of the moving
  discontinuous {G}alerkin finite element method with interface condition
  enforcement, Computers \& Mathematics with Applications 95 (2021) 143--171.
\newblock \href {https://doi.org/10.1016/j.camwa.2020.09.012}
  {\path{doi:10.1016/j.camwa.2020.09.012}}.

\bibitem{Ree73}
W.~H. Reed, T.~Hill, Triangular mesh methods for the neutron transport
  equation, Tech. rep., Los Alamos Scientific Lab., N. Mex.(USA) (1973).

\bibitem{Coc00}
B.~Cockburn, G.~Karniadakis, C.-W. Shu, The development of discontinuous
  {G}alerkin methods, in: Discontinuous {G}alerkin Methods, Springer, 2000, pp.
  3--50.

\bibitem{Maj12}
A.~Majda, Compressible fluid flow and systems of conservation laws in several
  space variables, Springer Science \& Business Media, 2012.
\newblock \href {https://doi.org/10.1007/978-1-4612-1116-7}
  {\path{doi:10.1007/978-1-4612-1116-7}}.

\bibitem{Chi18}
E.~Ching, Y.~Lv, P.~Gnoffo, M.~Barnhardt, M.~Ihme, Shock capturing for
  discontinuous {G}alerkin methods with application to predicting heat transfer
  in hypersonic flows, Journal of Computational Physics 376 (2018) 54--75.
\newblock \href {https://doi.org/10.1016/j.jcp.2018.09.016}
  {\path{doi:10.1016/j.jcp.2018.09.016}}.

\bibitem{Mor02}
G.~Moretti, Thirty-six years of shock fitting, Computers \& Fluids 31~(4)
  (2002) 719--723.
\newblock \href {https://doi.org/10.1016/S0045-7930(01)00072-X}
  {\path{doi:10.1016/S0045-7930(01)00072-X}}.

\bibitem{Sal09}
M.~Salas, A shock-fitting primer, CRC Press, 2009.

\bibitem{Sal11}
M.~Salas, A brief history of shock-fitting, in: Computational Fluid Dynamics
  2010, Springer, 2011, pp. 37--53.

\bibitem{Zah18}
M.~Zahr, P.-O. Persson, An optimization-based approach for high-order accurate
  discretization of conservation laws with discontinuous solutions, Journal of
  Computational Physics 365 (2018) 105--134.
\newblock \href {https://doi.org/10.1016/j.jcp.2018.03.029}
  {\path{doi:10.1016/j.jcp.2018.03.029}}.

\bibitem{Zah20}
M.~Zahr, A.~Shi, P.-O. Persson, Implicit shock tracking using an
  optimization-based high-order discontinuous {G}alerkin method, Journal of
  Computational Physics 410 (2020) 109385.
\newblock \href {https://doi.org/10.1016/j.jcp.2020.109385}
  {\path{doi:10.1016/j.jcp.2020.109385}}.

\bibitem{Shi22}
A.~Shi, P.-O. Persson, M.~J. Zahr, Implicit shock tracking for unsteady flows
  by the method of lines, Journal of Computational Physics 454 (2022) 110906.

\bibitem{Hua22}
T.~Huang, M.~Zahr, A robust, high-order implicit shock tracking method for
  simulation of complex, high-speed flows, Journal of Computational Physics 454
  (2022) 110981.

\bibitem{Hua23}
T.~Huang, C.~J. Naudet, M.~J. Zahr, High-order implicit shock tracking boundary
  conditions for flows with parametrized shocks, Journal of Computational
  Physics (2023) 112517.

\bibitem{Luo21}
H.~Luo, G.~Absillis, R.~Nourgaliev, A moving discontinuous {G}alerkin finite
  element method with interface condition enforcement for compressible flows,
  Journal of Computational Physics 445 (2021) 110618.

\bibitem{Nom04}
I.~Nompelis, T.~Drayna, G.~Candler, Development of a hybrid unstructured
  implicit solver for the simulation of reacting flows over complex geometries,
  in: 34th AIAA Fluid Dynamics Conference and Exhibit, 2004, p. 2227,
  {AIAA}-2004-2227.
\newblock \href {https://doi.org/10.2514/6.2004-2227}
  {\path{doi:10.2514/6.2004-2227}}.

\bibitem{Gno04}
P.~Gnoffo, J.~White, Computational aerothermodynamic simulation issues on
  unstructured grids, in: 37th AIAA Thermophysics Conference, 2004, p. 2371,
  {AIAA}-2004-2371.
\newblock \href {https://doi.org/10.2514/6.2004-2371}
  {\path{doi:10.2514/6.2004-2371}}.

\bibitem{Can09}
G.~Candler, D.~Mavriplis, L.~Trevino, Current status and future prospects for
  the numerical simulation of hypersonic flows, in: AIAA (Ed.), 47th AIAA
  Aerospace Sciences Meeting including The New Horizons Forum and Aerospace
  Exposition, 2009, {AIAA}-2009-0153.
\newblock \href {https://doi.org/10.2514/6.2009-0153}
  {\path{doi:10.2514/6.2009-0153}}.

\bibitem{Fid07}
K.~J. Fidkowski, A simplex cut-cell adaptive method for high-order
  discretizations of the compressible {N}avier-{S}tokes equations, Ph.D.
  thesis, Massachusetts Institute of Technology (2007).

\bibitem{Oli08}
T.~A. Oliver, A high-order, adaptive, discontinuous {G}alerkin finite element
  method for the {Reynolds-Averaged Navier-Stokes} equations, Ph.D. thesis,
  Massachusetts Institute of Technology (2008).

\bibitem{Cez15}
M.~Ceze, K.~J. Fidkowski, Constrained pseudo-transient continuation,
  International Journal for Numerical Methods in Engineering 102~(11) (2015)
  1683--1703.

\bibitem{Gno99}
P.~A. Gnoffo, Planetary-entry gas dynamics, Annual Review of Fluid Mechanics
  31~(1) (1999) 459--494.

\bibitem{Lev44}
K.~Levenberg, A method for the solution of certain non-linear problems in least
  squares, Quarterly of applied mathematics 2~(2) (1944) 164--168.
\newblock \href {https://doi.org/10.1090/qam/10666}
  {\path{doi:10.1090/qam/10666}}.

\bibitem{Mar63}
D.~Marquardt, An algorithm for least-squares estimation of nonlinear
  parameters, Journal of the society for Industrial and Applied Mathematics
  11~(2) (1963) 431--441.
\newblock \href {https://doi.org/10.1137/0111030} {\path{doi:10.1137/0111030}}.

\bibitem{Chi21}
E.~Ching, M.~Ihme, Efficient projection kernels for discontinuous {G}alerkin
  simulations of disperse multiphase flows on arbitrary curved elements,
  Journal of Computational Physics 435 (2021) 110266.
\newblock \href {https://doi.org/10.1016/j.jcp.2021.110266}
  {\path{doi:10.1016/j.jcp.2021.110266}}.

\bibitem{Tou13}
T.~Toulorge, C.~Geuzaine, J.~Remacle, J.~Lambrechts, Robust untangling of
  curvilinear meshes, Journal of Computational Physics 254 (2013) 8--26.

\bibitem{Ala14}
F.~Alauzet, A changing-topology moving mesh technique for large displacements,
  Engineering with Computers 30~(2) (2014) 175--200.

\bibitem{Loh08}
R.~L{\"o}hner, Applied {CFD} Techniques, J. Wiley \& Sons, 2008.

\bibitem{Xie13}
Z.~Xie, R.~Sevilla, O.~Hassan, K.~Morgan, The generation of arbitrary order
  curved meshes for 3{D} finite element analysis, Computational Mechanics 51
  (2013) 361--374.

\bibitem{Mox16}
D.~Moxey, D.~Ekelschot, {\"U}.~Keskin, S.~Sherwin, J.~Peir{\'o}, High-order
  curvilinear meshing using a thermo-elastic analogy, Computer-Aided Design 72
  (2016) 130--139.

\bibitem{Mar21}
J.~Marcon, A.~Garai, M.~Denison, S.~Murman, An adjoint elasticity solver for
  high-order mesh deformation, in: AIAA Scitech 2021 Forum, 2021, p. 1238.

\bibitem{Gar15}
A.~Gargallo-Peir{\'o}, X.~Roca, J.~Peraire, J.~Sarrate, Distortion and quality
  measures for validating and generating high-order tetrahedral meshes,
  Engineering with Computers 31~(3) (2015) 423--437.

\bibitem{Tra12}
M.~K. Transtrum, J.~P. Sethna, Improvements to the {L}evenberg-{M}arquardt
  algorithm for nonlinear least-squares minimization, arXiv preprint
  arXiv:1201.5885 (2012).

\bibitem{Geu09}
C.~Geuzaine, J.-F. Remacle, Gmsh: A three-dimensional finite element mesh
  generator with built-in pre- and post-processing facilities, International
  Journal for Numerical Methods in Engineering~(79(11)) (2009) 1310--1331.

\bibitem{Ame01}
P.~Amestoy, I.~S. Duff, J.~Koster, J.-Y. L'Excellent, A fully asynchronous
  multifrontal solver using distributed dynamic scheduling, SIAM Journal on
  Matrix Analysis and Applications 23~(1) (2001) 15--41.

\bibitem{Ame19}
P.~Amestoy, A.~Buttari, J.-Y. L'Excellent, T.~Mary, Performance and scalability
  of the block low-rank multifrontal factorization on multicore architectures,
  ACM Transactions on Mathematical Software 45 (2019) 2:1--2:26.

\bibitem{Bal97}
S.~Balay, W.~D. Gropp, L.~C. McInnes, B.~F. Smith, Efficient management of
  parallelism in object oriented numerical software libraries, in: E.~Arge,
  A.~M. Bruaset, H.~P. Langtangen (Eds.), Modern Software Tools in Scientific
  Computing, Birkh{\"{a}}user Press, 1997, pp. 163--202.

\bibitem{Bal23}
S.~Balay, S.~Abhyankar, M.~F. Adams, S.~Benson, J.~Brown, P.~Brune,
  K.~Buschelman, E.~Constantinescu, L.~Dalcin, A.~Dener, V.~Eijkhout,
  J.~Faibussowitsch, W.~D. Gropp, V.~Hapla, T.~Isaac, P.~Jolivet, D.~Karpeev,
  D.~Kaushik, M.~G. Knepley, F.~Kong, S.~Kruger, D.~A. May, L.~C. McInnes,
  R.~T. Mills, L.~Mitchell, T.~Munson, J.~E. Roman, K.~Rupp, P.~Sanan,
  J.~Sarich, B.~F. Smith, S.~Zampini, H.~Zhang, H.~Zhang, J.~Zhang, {PETSc/TAO}
  users manual, Tech. Rep. ANL-21/39 - Revision 3.20, Argonne National
  Laboratory (2023).
\newblock \href {https://doi.org/10.2172/1968587} {\path{doi:10.2172/1968587}}.

\bibitem{Bar10}
G.~Barter, D.~Darmofal, Shock capturing with {PDE}-based artificial viscosity
  for {DGFEM}: Part {I}. formulation, Journal of Computational Physics 229~(5)
  (2010) 1810--1827.
\newblock \href {https://doi.org/10.1016/j.jcp.2009.11.010}
  {\path{doi:10.1016/j.jcp.2009.11.010}}.

\bibitem{Joh20}
R.~F. Johnson, A.~D. Kercher, A conservative discontinuous {G}alerkin
  discretization for the chemically reacting {N}avier-{S}tokes equations,
  Journal of Computational Physics 423 (2020) 109826.
\newblock \href {https://doi.org/10.1016/j.jcp.2020.109826}
  {\path{doi:10.1016/j.jcp.2020.109826}}.

\bibitem{Chi22_ICCFD}
E.~Ching, A.~Kercher, A.~Corrigan, Anisotropic mesh modifications for the
  moving discontinuous {G}alerkin method with interface condition enforcement
  for robust simulations of high-speed viscous flows, in: Eleventh
  International Conference on Computational Fluid Dynamics (ICCFD11), Maui,
  Hawaii, 2022, {ICCFD}11-0305.

\bibitem{Kit13_2}
K.~Kitamura, E.~Shima, Towards shock-stable and accurate hypersonic heating
  computations: A new pressure flux for {AUSM}-family schemes, Journal of
  Computational Physics 245 (2013) 62--83.

\bibitem{Lv15}
Y.~Lv, M.~Ihme, Entropy-bounded discontinuous {G}alerkin scheme for {E}uler
  equations, Journal of Computational Physics 295 (2015) 715--739.

\bibitem{Jia18}
Y.~Jiang, H.~Liu, Invariant-region-preserving {DG} methods for
  multi-dimensional hyperbolic conservation law systems, with an application to
  compressible {E}uler equations, Journal of Computational Physics 373 (2018)
  385--409.

\bibitem{Chi22}
E.~Ching, R.~Johnson, A.~Kercher, Positivity-preserving and entropy-bounded
  discontinuous {G}alerkin method for the chemically reacting, compressible
  {E}uler equations. {P}art {I}: {T}he one-dimensional case, arXiv preprint
  arXiv:2211.16254~\url{https://arxiv.org/abs/2211.16254} (2022).

\bibitem{Dem10}
L.~Demkowicz, J.~Gopalakrishnan, A class of discontinuous {P}etrov--{G}alerkin
  methods. {P}art {I}: The transport equation, Computer Methods in Applied
  Mechanics and Engineering 199~(23-24) (2010) 1558--1572.
\newblock \href {https://doi.org/10.1016/j.cma.2010.01.003}
  {\path{doi:10.1016/j.cma.2010.01.003}}.

\bibitem{Dem11}
L.~Demkowicz, J.~Gopalakrishnan, A class of discontinuous {P}etrov--{G}alerkin
  methods. {II}. {O}ptimal test functions, Numerical Methods for Partial
  Differential Equations 27~(1) (2011) 70--105.
\newblock \href {https://doi.org/10.1002/num.20640}
  {\path{doi:10.1002/num.20640}}.

\bibitem{Dem15_20}
L.~Demkowicz, J.~Gopalakrishnan, Discontinuous {P}etrov-{G}alerkin ({DPG})
  method, Tech. Rep. 15-20, ICES, retrieved from
  \url{https://www.oden.utexas.edu/media/reports/2015/1520.pdf} (October 2015).

\bibitem{Blo90}
F.~Blottner, Accurate {N}avier-{S}tokes results for the hypersonic flow over a
  spherical nosetip, Journal of spacecraft and Rockets 27~(2) (1990) 113--122.
\newblock \href {https://doi.org/10.2514/3.26115} {\path{doi:10.2514/3.26115}}.

\bibitem{Fis21}
T.~Fisher, High fidelity {CFD} workshop 2021: High speed steady advanced case:
  Blottner sphere, NASA Langley Research Center Turbulence Modeling
  Resource~\url{https://turbmodels.larc.nasa.gov/highfidelitycfd{\_}workshop2022.html}
  (2021).

\bibitem{Mur23}
A.~Murphy, R.~Agarwal, Computational analysis of laminar steady hypersonic flow
  past {B}lottner sphere using {ANSYS} {F}luent, in: AIAA AVIATION 2023 Forum,
  2023, {AIAA}-2023-3847.

\bibitem{Kir10}
B.~Kirk, S.~Bova, R.~Bond, The influence of stabilization parameters in the
  {SUPG} finite element method for hypersonic flows, in: 48th AIAA Aerospace
  Sciences Meeting Including the New Horizons Forum and Aerospace Exposition,
  2010, {AIAA}-2010-1183.

\bibitem{Cou18}
B.~L.~S. Couchman, Comparison of heat flux predictions over the {B}lottner
  cylinder and sphere., Tech. rep., Sandia National Lab. (SNL-NM), Albuquerque,
  NM (United States) (2018).

\bibitem{Fay58}
J.~Fay, F.~Riddell, Theory of stagnation point heat transfer in dissociated
  air, Journal of the Aerospace Sciences 25~(2) (1958) 73--85.

\end{thebibliography}

\end{document}